\pdfoutput=1

\documentclass[a4paper,usenatbib]{mn2e}
\usepackage{graphicx}
\usepackage{float}
\usepackage{subfigure}
\usepackage{amssymb}
\usepackage{amsmath}
\usepackage{afterpage}
\usepackage[usenames]{color}
\usepackage[totalwidth=480pt, totalheight=630pt]{geometry}

\newcommand{\SRCC}{$\rho$}

\title{The Tully--Fisher and mass--size relations from halo abundance matching}
\author[H.~Desmond and R.~Wechsler]{Harry Desmond\thanks{E-mail: harryd2@stanford.edu}$^{1,2}$ and Risa H.~Wechsler$^{1,2}$ \\
$^{1}${Kavli Institute for Particle Astrophysics and Cosmology, Physics Department, Stanford University, Stanford, CA 94305, USA} \\
$^{2}${SLAC National Accelerator Laboratory, Menlo Park, CA 94025, USA}\\
}

\begin{document}

\maketitle

\begin{abstract} The Tully--Fisher relation (TFR) expresses the connection between rotating galaxies and the dark matter haloes they inhabit, and therefore contains a wealth of information about galaxy formation. We construct a general framework to investigate whether models based on halo abundance matching are able to reproduce the observed stellar mass TFR and mass--size relation (MSR), and use the data to constrain galaxy formation parameters. Our model tests a range of plausible scenarios, differing in the response of haloes to disc formation, the relative angular momentum of baryons and dark matter, the impact of selection effects, and the abundance matching parameters. We show that agreement with the observed TFR puts an upper limit on the scatter between galaxy and halo properties, requires weak or reversed halo contraction, and favours selection effects that preferentially eliminate fast-rotating galaxies.  The MSR constrains the ratio of the disc to halo specific angular momentum to be approximately in the range $0.6-1.2$.  We identify and quantify two problems that models of this nature face. (1) They predict too large an intrinsic scatter for the MSR, and (2) they predict too strong an anticorrelation between the TFR and MSR residuals. We argue that resolving these problems requires introducing a correlation between stellar surface density and enclosed dark matter mass. Finally, we explore the expected difference between the TFRs of central and satellite galaxies, finding that in the favoured models this difference should be detectable in a sample of $\sim$ 700 galaxies.  \end{abstract}


\section{INTRODUCTION}
\label{sec:intro}

The complex process of galaxy formation produces several regular scaling relations, which indicate a tight connection between galaxies and their host dark matter haloes. One approach towards modelling this connection uses the technique of `(sub)halo abundance matching' (AM), which sets galaxies' stellar masses based on the virial mass or rotation velocity of their haloes~\citep{Kravtsov,Conroy,Behroozi_2010,Guo,Moster}. Further, galaxy sizes may be predicted from halo spin and concentration under the assumption that galaxy and halo angular momentum are related~\citep[e.g.][]{Fall,MMW}. Models with these basic ingredients have been shown to successfully reproduce observed satellite fractions and galaxy two-point statistics~\citep[e.g.][]{Conroy,Reddick, CAM}, to roughly match the normalization and slope of the stellar mass--size relation (MSR;~\citealt*{MMW, Kravtsov_Radius}), and to provide broad-brush agreement with the luminosity~\citep{TG}, stellar mass~\citep{D10, D11} and baryonic~\citep{Desmond} Tully--Fisher relations (TFRs).

In this work, we push models of this type further by asking whether they are capable of reproducing the \textit{detailed} properties of the stellar mass TFR and MSR for a homogeneously selected sample of nearby spiral galaxies.  Our goal is to assess whether and under which conditions a wide range of galaxy formation models are consistent with observed galaxy sizes and dynamics, and further to understand to what extent these observables bring additional constraining power to this class of models. We examine whether AM models can match the intrinsic TFR and MSR scatter in addition to slope and normalization, compare the predicted and observed correlations between the residuals of the two relations, and explore the question of whether satellite and central galaxies should be expected to lie on different TFRs. Our analysis elucidates and quantifies several issues which have been regarded in some studies as fundamentally problematic for all $\Lambda$ Cold Dark Matter ($\Lambda$CDM)-based TFR analyses (e.g.~\citealt{McGaugh_Main, McGaugh_Complete}).

Most previous work on modelling the TFR has employed one of three broad methodologies: semi-analytic modelling, AM, or hydrodynamical simulation. In semi-analytic models (e.g.~\citealt*{Fall, White, Kauffman93, MMW, Baugh, Somerville, Guo, Yu_Bayes}), the relations between galaxy and halo properties are determined observationally or given simple parametrized forms in accordance with analytic calculations or the results of $N$-body simulations. AM instead calibrates the stellar mass--halo mass relation directly using the observed stellar mass function (SMF) and the halo mass function found in simulations. Assuming that the true galaxy--halo connection lies within the space of models accessible to this methodology, AM greatly reduces the theoretical parameter space and hence increases the constraining power of any set of observations (see also Section~\ref{sec:method}). Finally, in hydrodynamical galaxy simulations one attempts to account for the effects of baryonic physics (such as feedback from star formation and AGN) in a cosmological context (e.g.~\citealt{Abadi, Governato, Tissera, Piontek, Illustris, EAGLE_1}). In principle, this removes the need for a priori assumptions concerning the relationship between galaxies and haloes. However, simulations of this kind are computationally expensive, with the result that it is not feasible to model at sufficient resolution a large number of galaxies spanning a range of values of the parameters relevant for the TFR. In addition to prohibiting statistical tests of goodness of fit, this compromises one's ability to pin down the simulated relation's intrinsic scatter.

Our analysis expands in several ways on similar AM-based TFR studies in the literature~\citep[e.g.][]{D10,TG}. First, we consider a very general set of AM models, including those that have been shown to be consistent with galaxy clustering measurements~\citep{Reddick}. This allows us to test the AM framework itself, not just specific AM models. Secondly, we include in our framework a prescription for taking account of the impact of selection criteria in the observational studies, which to first order eliminate non-rotating early-type galaxies. Thus we intend our theoretical TFR and MSR populations to contain only galaxies that would pass the cuts associated with the data. Thirdly, in addition to considering the slope, intercept, and scatter of the TFR, we also investigate the shape of the stellar MSR and the correlation between the residuals of these two relations. This information provides enhanced constraining power over the TFR alone. Finally, we evaluate our models in a statistically rigorous way by means of a full likelihood formalism.

The structure of this paper is as follows. Section~\ref{sec:obs_data} describes the observational data with which we compare our model, and Section~\ref{sec:sim_data} documents the $N$-body simulations from which halo properties are drawn. Section~\ref{sec:method} details our methodology. In Section~\ref{sec:results} we present our results for the TFR, MSR, correlation between the residuals of the TFR and MSR, and environment dependence of the TFR. In Section~\ref{sec:discussion} we discuss the general significance and broader ramifications of our results.  Section~\ref{sec:suggestions} proposes avenues for future research, and Section~\ref{sec:conc} concludes. Appendix~\ref{sec:app_plots} provides more specific information on the effect of the individual model parameters on the TFR and MSR. In Appendix~\ref{sec:assumptions}, we discuss in detail the assumptions in our model and the systematic errors that may result from their failure, and document the tests we have performed to assess them quantitatively. We provide additional technical implementation details in Appendix~\ref{sec:technical}.

\section{Observational Data}
\label{sec:obs_data}

The observational TFR to which we compare our model is that of~\citet[][hereafter P07]{Pizagno}. This is a sample of 162 galaxies with H$\alpha$ rotation curves obtained via long-slit spectroscopy from the Calar Alto and MDM observatories, and luminosities measured by the Sloan Digital Sky Survey (SDSS;~\citealt{SDSS}). The $r$-band magnitude distribution is roughly uniform between --22 and --18.5, corresponding to stellar masses roughly in the range $10^9-10^{11.2} M_{\odot}$. The sample was selected from SDSS DR2 according to four criteria:

\begin{enumerate}

\item{} $cz > 5000\:\mathrm{km\:s^{-1}}$. This ensures that peculiar velocities are small relative to the Hubble flow.

\item{} $cz < 9000\:\mathrm{km\:s^{-1}}$ for $-18.5 > M_r > -20$, $11000 \mathrm{km\:s^{-1}}$ for $-20 > M_r > -21$, and $15,000\: \mathrm{km\:s^{-1}}$ for $M_r < -21$. This eliminates distant galaxies with more uncertain velocity measurements.

\item{} An isophotal axis ratio cut $b/a < 0.6$, where $b$ is the apparent minor axis of the galaxy and $a$ the apparent major axis, as measured in the $r$ band by the SDSS pipeline. Focusing on edge-on galaxies in this way reduces uncertainties in inclination measurements and increases the likelihood that a given galaxy is a rotation-dominated spiral.

\item{} The rotation curve is well fit by an arctan function, as determined by visual inspection. This eliminates morphologically unusual galaxies and those without significant rotation.

\end{enumerate}

As we describe further in Section~\ref{sec:method}, we take the axis ratio cut (which we can systematically investigate in the SDSS) to be the selection criterion that most strongly influences the position of the accepted galaxies on the TFR plane, and assume that the arctan cut eliminates a random subsample of these galaxies (i.e. whether or not a galaxy passes the latter cut is independent of the properties of its host halo that influence its position on the TFR). Further discussion of this assumption can be found in Appendix~\ref{sec:assumptions}.

The P07 study itself uses luminosity; the conversion to stellar mass requires a mass-to-light ratio for each galaxy, in addition to an extrapolation of the observed surface brightness profile to find the total luminosity. Several different prescriptions for these mappings exist in the literature, sometimes based on quite different assumptions and underlying stellar population models (e.g.~\citealt{Bell, Kauffman03, McGaugh_STFR, Bernardi}), making this step a significant source of systematic uncertainty in most studies of the stellar mass TFR. In this work we take the stellar masses of the P07 galaxies from the NASA Sloan Atlas (NSA). This data set uses improved sky subtraction optimized for low-redshift galaxies, and is in good agreement with recent estimates of the SMF from ~\cite*{Bernardi} and \cite{Kravtsov_BrightEnd}.\footnote{We remove five P07 galaxies (J13213.08+002032.7, J012223.78--005230.7, J124545.20+535702.0, J170310.47+653417.6, J205307.50--002407.0) whose masses are not recorded in the NSA, and another (J215652.70+121857.5) with stellar mass significantly lower than the rest. This leaves 156 galaxies in our TFR sample.}

The advantages of the P07 sample for the purposes of our study are threefold. First, the selection criteria are relatively loose and simple to model theoretically, which is important because selection effects are typically one of the largest (and least discussed) sources of uncertainty in comparison of data with models that make predictions for the full galaxy population. As described in Section~\ref{sec:method}, we model the impact of selection by postulating and then marginalizing over a correlation between a galaxy's axis ratio and rotation velocity. Secondly, the fact that the galaxies were selected for spectroscopic follow-up from a well-understood parent population (those in the SDSS) allows us to achieve consistency between the simulated and observed galaxies using correlations from the SDSS. Finally, the P07 study was done with the aim of theory comparison explicitly in mind. Thus, for example, they use a velocity measure that is easy to model robustly (the velocity at the radius enclosing 80 per cent of the $i$-band light, $V_{80}$), and expend considerable effort on pinning down the relation's intrinsic scatter.

For completeness, we note that a study similar to P07 was more recently performed by~\citet{Reyes}. We choose to use the former because its selection criteria are simpler and hence easier to model within our framework. The two data sets are sufficiently similar that our conclusions would unlikely change, were the~\citet{Reyes} data used instead. Other sizeable TFR samples in the literature -- which do not possess the homogeneity or simple selection required for our detailed modelling -- include those of \citet{Courteau07},~\citet{McGaugh_Complete}, and~\citet{Mocz}.

Although the baryonic TFR (which includes cold gas mass) appears to have smaller scatter and curvature than the stellar mass TFR~\citep{McGaugh_BTFR}, and hence to be more fundamental, practical considerations force us to restrict our attention to the latter. Homogeneous data sets that both contain cold gas mass measurements and are derived from a thoroughly characterized parent population do not yet exist. Stellar masses can be estimated from the information acquired by large-scale galaxy surveys such as the SDSS, whilst cold gas mass measurements (requiring H\textsc{i} observations) have been done for relatively few systems not systematically selected according to a fixed set of criteria. A further consequence of this is that the SMF, required for AM, is known with significantly better accuracy than the corresponding baryon mass function. An analysis at our level of detail is not yet feasible for the baryonic TFR, but will be worth performing when it becomes so.

\section{Simulation Data}
\label{sec:sim_data}

The choice of $N$-body simulation with which to create our theoretical galaxy population is driven by two competing criteria. On the one hand, we require the simulation to contain a statistically representative sample of haloes hosting galaxies with stellar masses at the upper end of the range covered by the P07 data set [$\log(M_\star/M_\odot)\approx11.2$]. On the other, we require haloes at the low mass end [$\log(M_\star/M_\odot)\approx9.0$] to be sufficiently well resolved for their concentrations to be reliable.~\citet{Diemer_Kravtsov} propose three criteria for a halo to be well resolved.

\begin{enumerate}

\item{} At least 1000 particles within $R_\mathrm{200c}$, the halo radius within which the mean density is 200 times the critical density of the Universe.

\item{} At least 200 particles within $r_\mathrm{s}$, the scalelength of the halo.

\item{} $r_\mathrm{s}$ at least 6 times the force softening length, $\epsilon$, used in the simulation.

\end{enumerate}

Many $\log(M_\star/M_\odot)\approx9.0$ haloes (using a fiducial AM prescription) fail at least one of these cuts in a $(250\:\mathrm{Mpc}\:h^{-1})^3$ simulation with $2048^3$ particles and $\epsilon\approx1\:h^{-1}\:\mathrm{kpc}$.  We find that this failure is manifest in a $>10$ per cent discrepancy between median concentrations from the simulation and those predicted by the model of~\citet{Diemer_Kravtsov}, and $>10$ per cent differences between alternative methods of determining concentration. Thus we must use a higher resolution simulation to accurately model low-mass galaxies. To this end, we use the {\tt c-125} simulation, a $(125\:\mathrm{Mpc}\:h^{-1})^3$ box with $2048^3$ particles and comoving softening length $0.5\:h^{-1}$ kpc, simulated using \textsc{l-gadget} (based on \textsc{gadget-2};~\citealt{Gadget1},~\citealt{Gadget2}). This simulation assumes a flat $\Lambda$CDM cosmology with $h=0.7$, $\Omega_\mathrm{m}=0.29$, $\Omega_\mathrm{b}=0.047$, $\sigma_\mathrm{8}=0.82$ and $n_\mathrm{s}=0.96$. The initial conditions were generated by 2\textsc{lptic}~\citep{2LPT} at $z$ = 199, and the power spectrum was generated by \textsc{camb}\footnote{http://camb.info/}.

The {\tt c-125} box, however, contains few high-mass haloes due to its relatively small volume. For example, there are only $\sim$ 220 haloes in the simulation that would be assigned galaxies with $11.15 < \log(M_\star/M_\odot) < 11.2$ by a standard abundance match -- too few to justify statistical inference concerning the high-mass end of the TFR. Modelling this regime accurately therefore requires a larger box.  We use a larger simulation run with the same code and cosmological parameters for this purpose.  The {\tt c-250} box contains $2048^3$ particles in a volume of $(250\:\mathrm{Mpc}\:h^{-1})^3$, yielding a particle mass of $1.46 \times 10^8 M_\odot\:h^{-1}$, and has a comoving softening length of $1\:h^{-1}\:\mathrm{kpc}$. This box contains $\sim$ 1800 haloes with $11.15 < \log(M_\star/M_\odot) < 11.2$.
Both boxes were provided by M. Becker (Becker et al., in preparation).  We identify haloes using the \textsc{rockstar} halo finder~\citep*{Rockstar_1}, and generate merger trees using the \textsc{consistent trees} algorithm~\citep{Rockstar_2}.

To create a final galaxy--halo catalogue with which to model the TFR, we proceed as follows. We first perform AM separately on the haloes generated by the {\tt c-125} and {\tt c-250} simulations (see also Section~\ref{sec:method}).
We then splice together the $\log(M_\star/M_\odot) < 10$ part of the  {\tt c-125} catalogue and the $\log(M_\star/M_\odot) > 10$ part of the {\tt c-250}  catalogue. We have verified that the median halo concentrations, virial masses, and spins of the two simulations differ by $<5$ per cent for $9 < \log(M_\star/M_\odot) < 11.2$, so the discontinuity at $\log(M_\star/M_\odot) = 10$ is not significant. Our final catalogue contains at least 1800 objects in all stellar mass bins, and all haloes are fully resolved.

In order to depend as little as possible on the assumption that haloes have an NFW density profile, we use $M_\mathrm{vir}$ values determined directly by the halo finder (rather than calculating them from $r_\mathrm{vir}$), and determine concentration from the \textsc{rockstar} output `$r_\mathrm{s,klypin}$' (derived from the ratio of $v_\mathrm{max}$ to $v_\mathrm{vir}$) rather than $r_\mathrm{s}$.

\section{Method}
\label{sec:method}

We begin with an executive summary of our method for comparing mock and observed TFRs and MSRs. Each step is expanded upon in the remainder of this section.

\begin{enumerate}

\item{} Use AM to associate a stellar mass with each halo, using one of six halo properties as the AM parameter and allowing for the possibility of non-zero scatter.

\item{} Assume the stellar mass is distributed in a thin exponential disc and find the scalelength by requiring the specific angular momentum, $J$, of the disc to equal some fraction of the specific angular momentum of the host halo.

\item{} Model the effect of disc formation on the density profile of the halo, allowing for standard adiabatic contraction, an expansion of the same magnitude using the same functional form, or anything in between.

\item{} Add a bulge to each model galaxy by bootstrap resampling from the bulge-to-disc ratios of the P07 galaxies, in five bins of stellar mass.

\item{} Model the impact of the selection effects in the P07 data set by imposing a correlation between apparent axis ratio and $V_{80}$ at fixed stellar mass. The strength of this correlation is controlled by a free parameter that interpolates between no correlation and a near-monotonic relationship.

\item{} Generate 200 mock data sets containing as many galaxies as are in the P07 sample, and with the same stellar masses and velocity uncertainties, but with velocities chosen randomly from the theoretical parent population. Fit each data set with a power-law TFR to determine the best-fitting slope, intercept, and scatter values, and compare the distributions thereby obtained to the values for the real data in order to assess goodness of fit via a likelihood.

\item{} Use an MCMC (Markov Chain Monte Carlo) algorithm to explore the parameter space systematically and obtain posterior probability distributions for the model parameters.

\item{} Repeat the above set of steps replacing the TFR by the MSR or the correlation of the residuals of the two relations.

\end{enumerate}

\vspace{5 mm}

\noindent Our model is built on the technique of AM~\citep*{Kravtsov,Conroy,Behroozi_2010,Guo,Moster,Reddick}. This enables the association of a galaxy with each dark matter halo in an $N$-body simulation by hypothesizing a one-to-one relation between stellar mass and some halo property. One begins by rank-ordering the haloes in the simulation at $z=0$ by that property, and then uses a SMF (in our case that of~\citealt{Bernardi}) to ascertain the abundances of galaxies as a function of stellar mass that one would expect to see in the volume of the simulation. One then rank-orders these galaxies by stellar mass and enforces the galaxy--halo connection by associating the galaxy with $n$th largest stellar mass to the halo with the $n$th largest value of the matching property. AM has been shown to produce galaxy catalogues that are consistent with a large number of observed statistics, including galaxy clustering, galaxy kinematics, galaxy--galaxy lensing, galaxy group catalogues, and galaxy void statistics~\citep{Conroy,Vale_Ostriker,Moster,Reddick}. It has also been shown to yield a stellar mass--halo mass relation similar to that obtained directly from studies of weak lensing and satellite kinematics~\citep{Behroozi_2010}.

The simplest halo property to use for AM is the current virial mass of the halo, $M_\mathrm{now}$. Much effort has recently been expended, however, on generalizing the procedure by considering alternative halo properties. For example, it might be expected that the process of galaxy formation would be more closely correlated with the depth of the potential well in which the galaxy is located, for which a better proxy than $M_\mathrm{now}$ is the present-day maximum rotation velocity of the halo, $V_\mathrm{max}$. However, galaxies did not form at $z=0$; the stellar mass that a given halo acquires may in fact be better traced by the peak value of the halo's maximum rotation velocity over its merger history, $V_\mathrm{peak}$. Satellite galaxies, on the other hand, likely form when the subhaloes that harbour them first cross the virial radius of the host halo; thus another popular choice ($V_\mathrm{acc}$) is to use the maximum rotation velocity at the time of accretion for subhaloes, and the current maximum rotation velocity for distinct haloes. Along with the analogues of $V_\mathrm{acc}$ and $V_\mathrm{peak}$ obtained by replacing the halo's velocity by its mass ($M_\mathrm{acc}$, $M_\mathrm{peak}$), these are the models that we investigate in parallel in this work.\footnote{We have also tried abundance matching to the maximum rotation velocity of a halo at the time it attains its maximum mass, $V_\mathrm{mpeak}$. This generates a very similar galaxy--halo connection to the $V_\mathrm{peak}$ model, and hence an almost identical TFR.}

It is also possible to loosen the galaxy--halo connection within the AM framework. This is done with a parameter known as `AM scatter', measured in dex, which sets the standard deviation of a normal distribution. After completing AM with zero scatter, the stellar mass of each galaxy is perturbed by an amount drawn from this distribution, and then the abundance of galaxies as a function of stellar mass is modified to restore consistency with the true SMF (see~\citealt{Behroozi_2010} for details). The larger the AM scatter, the larger the scatter in the relationship between stellar mass and halo mass or velocity. We take the AM scatter to be a free parameter with uniform prior in the range 0--0.5 dex, and assume for simplicity that it is mass independent.

The analysis of~\citet{Reddick} suggests that only by matching to $V_\mathrm{peak}$ with a scatter of $0.2 \pm 0.03$ dex can one simultaneously fit observed satellite fractions and galaxy two-point correlation functions. Similar results have been found by other authors using satellite velocity dispersions~\citep[e.g.][]{More09}. We investigate a range of AM models rather than restricting ourselves to those deemed acceptable by~\citet{Reddick} in order to explore in full generality the aptitude of the AM ansatz for reproducing the observed TFR and MSR, and to bring out important physical differences between the models in terms of the internal galaxy dynamics they imply. While the clustering analysis performed by~\citet{Reddick} is optimal for distinguishing the models on the basis of their assembly bias and satellite fractions, the TFR is most suited to bringing out differences in the stellar mass--concentration relations. We use the $V_\mathrm{peak}$ model most often for illustrating our results in recognition of the fact that it is currently the most favoured observationally, but highlight where alternative proxies differ.

After setting the stellar mass of the galaxy within each halo in this way, we model its structure. We assume each galaxy to be a thin exponential disc, a good approximation for the late-type spirals under investigation. We set the disc's scalelength by the requirement that its specific angular momentum be some fixed fraction of that of its host halo (as determined by the halo spin parameter, $\lambda$), a common assumption since the work of~\citet{MMW} and still popular today. The ratio of the disc and halo specific angular momentum is the third free parameter in the model: $j \equiv \frac{J_\mathrm{disc}}{J_\mathrm{halo}}$. We assume this to be a constant for all haloes.\footnote{It is possible that $j$ is mass dependent. However, we show below that no mass dependence is required in order to fit the slope and intercept of the TFR or MSR.  A mass-dependence term included in the model would be poorly constrained and consistent with zero, and would be unlikely to alter any of our conclusions.} Since a priori expectations for $j$ from hydrodynamical simulations range from $j \ll 1$ (e.g.~\citealt{LowJ_2, LowJ_1, Scannapieco}) to $j>1$ (e.g.~\citealt{Kimm, Stewart2, Danovich}), we begin with a uniform prior over the range (0,5]. We will find in Section~\ref{sec:MSR_Results} that $j$ is constrained by the MSR to be $<1.4$ at $3\sigma$ for all AM proxies, while it is poorly constrained by the TFR. To prevent a broad $j$ posterior from degrading the constraints on the other model parameters with which it is degenerate, we restrict $j$ to the range (0,1.4] for our primary TFR analysis. We discuss the effect of the $j$ prior further in Section~\ref{sec:TFR_Results}.

We next add a bulge to our model galaxies to further augment consistency with the P07 data set. To avoid assumptions about the distribution of bulge-to-disc ratios in our parent population, and their correlation with other galaxy properties, we do bootstrap sampling of their values for the galaxies in the P07 sample. We partition the P07 galaxies into five bins of $\log(M_{\star}/M_{\odot})$ (9--9.5, 9.5--10, 10--10.5, 10.5--11, 11--11.5), and then select for each model galaxy a bulge-to-disc ratio from the corresponding bin.\footnote{This procedure coarsely incorporates the correlation found in the data between bulge mass fraction and stellar mass. Correlations between bulge mass fraction and other properties of the P07 galaxies are found to be negligible.} We assume that the bulge-to-disc $i$-band flux fractions measured by P07 equal the corresponding mass fractions, and defer to Appendix~\ref{sec:assumptions} a discussion of the (small) magnitude of the systematic error that may thereby be induced. The bulge and disc masses sum to the galaxy stellar mass assigned by AM.

The disc scalelength, $R_\mathrm{d}$, that gives the galaxy the correct angular momentum is determined by the density profile of the halo. Although we assume the pure dark matter haloes to have NFW density profiles~\citep*{NFW}, we allow for the possibility that these are modified by disc formation. In particular, the collapse of the baryons into a relatively small galaxy at the centre is expected to draw in the surrounding dark matter in a process known as `adiabatic contraction'~\citep{Blumenthal, Gnedin_2004, Gnedin_2011}. Our fiducial model of adiabatic contraction in this study is that of~\citet{Gnedin_2011}, which builds upon earlier work by accounting for non-circular dark matter orbits.  In this model 

\begin{equation}
M_\mathrm{f}(\bar{r}_\mathrm{f})\;r_\mathrm{f} = M_\mathrm{i}(\bar{r}_\mathrm{i})\;r_\mathrm{i},
\end{equation}

\noindent where subscripts i and f denote before and after disc formation, respectively,

\begin{equation}
M_\mathrm{i}(r) = M_\mathrm{i,\mathrm{halo}}(r)
\end{equation}

\noindent is the initial mass enclosed within radius $r$ (with the halo described by an NFW profile),

\begin{equation}
M_\mathrm{f}(r) = M_\mathrm{disc}(r)\;+\;M_\mathrm{bulge}\;+\;(1-M_\star/M_\mathrm{vir})\:M_\mathrm{f,halo}(r)
\end{equation}

\noindent is the final enclosed mass, and

\begin{equation}
\bar{r} = A_0\;r_0\;(r/r_0)^w,
\end{equation}

\noindent where $A_0 = 1.6$, $w=0.8$, and $r_0 = 0.03\:R_\mathrm{vir}$ (see~\citealt{Gnedin_2011}, eq. 4). For a given halo mass, concentration, and spin, and galaxy disc and bulge mass, these equations may be solved iteratively with the requirement that $J_\mathrm{disc} = j\:J_\mathrm{halo}$ 
to determine the final halo mass within any given radius, and the disc scalelength. This gives the `standard' adiabatic contraction solution. However, it is commonly argued (largely motivated by a desire to solve various small-scale problems with $\Lambda$CDM such as the `cusp/core' -- \citealt{cusp-core} -- and `too big to fail' -- \citealt*{TBTF} -- problems) that processes such as baryon ejection due to stellar winds and energy injection by supernovae can in fact generate an expansion of the halo that erases the impact of the earlier stage of contraction. We remain agnostic about the detailed effect that baryonic feedback has on halo density profiles and choose to employ a simple generalization of the adiabatic contraction formalism created by~\citet{D07}. In particular, we define 

\begin{equation}
\Gamma(\bar{r}_\mathrm{i}) \equiv \bar{r}_\mathrm{f}/\bar{r}_\mathrm{i},
\end{equation}

\noindent and then take the true final radius enclosing a given amount of mass to be given by

\begin{equation}
\bar{r}_{f,\mathrm{true}} = \Gamma^\nu \bar{r}_i.
\end{equation}

\noindent The free parameter $\nu$ interpolates between standard adiabatic contraction ($\nu=1$), no effect of baryonic collapse on the dark matter halo ($\nu=0$), and halo \emph{expansion} by the same factor as the standard case ($\nu=-1$).\footnote{There is no a priori theoretical reason why $\nu$ must lie in the range [--1,1]. However, we show below that the TFR constrains $\nu$ to lie comfortably within these bounds, and hence a larger range need not be considered.} Different values of $\nu$ result in different disc scalelengths and halo density profiles for given input parameters.

We now have all the information needed to calculate the full rotation curve of all model galaxies, and hence locate them on the TFR. To do so, we record the velocity at the radius which encloses 80 per cent of the $i$-band light (converting from stellar disc radius to $i$-band flux radius using a correction factor of 1.13; ~\citealt{D07}). Under the assumption of a constant stellar mass-to-light ratio over the bulge and disc (discussed further in Appendix~\ref{sec:assumptions}), this velocity is the model analogue of P07's $V_{80}$.

Finally, we model the potential impact of the selection criteria employed by P07. Since we cannot determine which of our model galaxies would have properties allowing them to pass the P07 selection cuts (that is to say, we do not know how to map selection criteria on to halo properties), we attempt merely to bracket their effect on the TFR. Thus we assume that there exists a correlation between apparent axis ratio and the $V_{80}$ value of a galaxy, with larger axis ratio implying larger $V_{80}$.\footnote{An alternative would be to use the age matching model of~\citet{Hearin_Watson} to assign a colour to each model galaxy, and then assume a correlation between apparent axis ratio and colour. We have found that the correlations between halo age and the dynamical properties which affect $V_{80}$ are too small for selection effects to have a significant impact on the TFR, under this assumption.} There is weak evidence for such a correlation -- galaxies with larger apparent axis ratios are more likely to be early type, which have been found to lie preferentially in more massive or concentrated haloes at fixed galaxy mass (e.g.~\citealt{Puebla, Wojtak}). However, we do not motivate this model on the strength of this evidence -- we wish only to assess the extent to which selection effects could shape the TFR, and provide a framework for taking these into account when comparing theory with observation. Since galaxies are typically predicted by AM to rotate too fast for their stellar mass at $\log(M_\star/M_\odot) \gtrsim 10$~\citep[e.g.][]{TG, Desmond}, removing the galaxies with high $V_{80}$ in this way is expected to improve agreement with the observations. Our formalism allows us to determine how significant such selection effects must be.

The details of the implementation of this correlation in our model are as follows. We begin by determining the fraction of galaxies in the local Universe, as a function of stellar mass, which pass the P07 axis ratio cut ($b/a < 0.6$; see Section~\ref{sec:obs_data}). This is done using the SDSS Value Added Galaxy Catalog (VAGC;~\citealt{VAGC}), and determines the fraction of galaxies that should be included in the theoretical parent population under construction. The acceptance fraction falls from around 40 per cent at $M_\star = 10^9 M_\odot$ to around 15 per cent at $M_\star = 10^{11.2} M_\odot$. Next, we establish at each stellar mass a probability density function on the set of $V_{80}$ values output by our model at that mass, which by construction describes the likelihood of a galaxy with that $V_{80}$ value having $b/a < 0.6$, and thus the likelihood that it should be included in our catalogue.  This function is made linear, and the ratio of the probability density at the smallest $V_{80}$ value produced by the model at that mass to the largest is taken to be another free parameter in the model, with a value in the range [0.5,$\infty$).\footnote{The lower bound of 0.5 is arbitrary, and simply limits the possibility for strong preferential selection of \textit{high} $V_{80}$ galaxies at fixed stellar mass. Such selection would conflict with the findings of~\citet{Puebla} and~\citet{Wojtak}, and will be shown in Section~\ref{sec:TFR_Results} to be disfavoured by the TFR.} To compactify this semi-infinite parameter range into a finite one, we perform the re-parametrization $y = \arctan(x-0.5)$ where $x$ is the original parameter and $y$ the new one. Thus $y$ (which we refer to hereafter as the `selection factor') lies in the range [0,$\pi/2$).

In equations, we calculate the probability of accepting a model galaxy with $V_{80}=V$ (that is, the probability of assuming it has $b/a<0.6$) as

\begin{equation}
P_\mathrm{accept}(V) = A+B\:V,
\end{equation}

\noindent where $A$ and $B$ are constants set by

\begin{equation}
\int^{V_\mathrm{max}}_{V_\mathrm{min}} P_\mathrm{accept}(V_{80})\:\mathrm{d}V_{80} = 1 \;\;\;\;\; \mathrm{and} \;\;\;\;\; \frac{P(V_\mathrm{min})}{P(V_\mathrm{max})} = x.
\end{equation}

Finally, we select the number of galaxies at each stellar mass corresponding to the fraction determined from the VAGC, according to this probability density function. The larger the value of the selection factor, the more likely galaxies are to be selected from the low end of the $V_{80}$ function at fixed stellar mass, and hence the stronger the preference in the model for TFR selection effects to eliminate high-velocity galaxies. This completes the creation of a full theoretical parent population of galaxies, from which, absent systematic errors, the P07 sample should have been randomly drawn. Each set of values for the five free parameters in our model (summarized in Table~\ref{tab:Parameters}) generates a different parent population.

\begin{table*}
  \begin{center}
    \begin{tabular}{l|l|c|}
      \hline
      &Description &Allowed values\\ 
      \hline
      AM parameter    &    Halo property to which galaxy stellar masses are matched   & $V_\mathrm{peak}$, $V_\mathrm{acc}$, $V_\mathrm{max}$, $M_\mathrm{peak}$, $M_\mathrm{acc}$, $M_\mathrm{now}$\\
      AM scatter    & Universal Gaussian scatter in stellar mass at fixed halo property & [0, 0.5] dex\\
      $j$    &  Ratio of specific angular momentum of disc and halo & MSR: (0, 5] $\:$ TFR: (0, 1.4]\\
      $\nu$    &  Controls degree of halo expansion (--) or contraction (+) & [--1, 1]\\
      Selection factor    & Governs impact of P07 selection effects on theoretical TFR & [0, $\pi/2$)\\
      \hline
    \end{tabular}
  \caption{Table of model parameters and their allowed values.}
  \label{tab:Parameters}
  \end{center}
\end{table*}

It now remains to assess the extent of agreement between the theory and P07 data. To do this, we generate a large number (200) of mock data sets from the theoretical parent population, each one containing the same number of galaxies as the P07 sample, and with the same masses and velocity uncertainties, but with velocities randomly selected from the model galaxies. To simulate the effect of measurement error, we perturb each model velocity by a random number drawn from a normal distribution with mean zero and standard deviation equal to the measurement uncertainty. We then fit a power law to the $\log(V_{80}/\mathrm{km\:s^{-1}})$--$\log(M_{\star}/M_{\odot})$ relation of each mock data set (using a Gaussian likelihood model) and record the best-fitting values of the slope, intercept and scatter, for comparison with the analogous quantities derived from the observational data. We pivot the fit on the median $\log(M_{\star}/M_{\odot})$ of the sample, 10.47, to eliminate the degeneracy between the slope and intercept, and take stellar mass to be the independent variable throughout. To condense this information into a single measure of goodness of fit, we locate each mock data set in the space of \{slope, intercept, scatter\} values and fit a 3D Gaussian to the 200 points thereby obtained. This describes the theoretical TFR probability density function in this space. The value of this function evaluated at the \{slope, intercept, scatter\} coordinate of the P07 data (0.264, 2.24, 0.0780) is taken to be the likelihood of the model under consideration: the further the P07 coordinate from the centroid of the theoretical distribution, the worse the agreement between theory and observation. A goodness of fit measure using mock data sets with characteriztics identical to the real data treats simulated galaxies in the same way as observed galaxies, and is therefore a more sound basis for statistical inference than a measure based on the entire model population.

Thus we obtain the likelihood of the observational data given a particular set of model parameter values. We now use Bayes' theorem to convert this into a probability of the parameters given the data (using uniform priors over the allowed ranges):

\begin{equation}
P(\boldsymbol{p}|\boldsymbol{d}) \propto \frac{1}{\sqrt{|\boldsymbol{\mathsf{C}}|}} \mathrm{e}^{-\frac{1}{2} (\boldsymbol{x} - \boldsymbol{x}_\mathrm{d}) \boldsymbol{\mathsf{C}^{-1}} (\boldsymbol{x} - \boldsymbol{x}_\mathrm{d})^\mathrm{T}}
\end{equation}

\noindent where $\boldsymbol{p} = \{\mathrm{AM\;parameter}, \mathrm{AM\;scatter}, j,\nu, \mathrm{selection\;factor}\}$ is the vector of model parameters, $\boldsymbol{d} = \{M_\star,V_{80}\}$ is the P07 data vector, $\boldsymbol{\mathsf{C}}$ is the $3\times3$ covariance matrix of \{slope, intercept, scatter\} values of fits to the mock data sets, $\boldsymbol{x}$ is a $1\times3$ vector of the mean \{slope, intercept, scatter\} values of the mock data, and $\boldsymbol{x}_\mathrm{d} = (0.264, 2.24, 0.0780)$. Finally, we calculate the posterior distributions of the model parameters with the MCMC algorithm. We employ the sampler created by~\citet{Yu_Bayes_Real},\footnote{https://github.com/ylu2010/PSEtoolbox} using 32 walkers and at least 4000 steps after burn-in, and check convergence by eye. Since the AM parameter is a discrete variable, we perform the MCMC analysis separately for each of its six possible values.

To further assess the agreement of our model with the P07 data, and to further constrain our model parameters, we examine the relation between $\log(M_\star/M_\odot)$ and $\log(R_\mathrm{d}/\mathrm{kpc})$ (MSR) using the same formalism. Since the disc scalelengths were measured by P07 (and were calculated in our model en route to $V_{80}$), this simply requires replacing $V_{80}$ by $R_\mathrm{d}$ in the final step and then repeating the MCMC.  The best-fitting \{slope, intercept, scatter\} of the P07 MSR is \{0.233, 0.486, 0.170\}. We also investigate the cross-correlation of the TFR and MSR by generating combined radius and velocity mock data, and establish goodness of fit using a likelihood based on the strength of the correlation of the radius and velocity residuals. Further discussion of these analyses and the results they yield may be found in Sections~\ref{sec:MSR_Results} and~\ref{sec:residuals}.

More technical details of our method (in particular to make each likelihood evaluation quick enough for an MCMC analysis to be feasible) are given in Appendix~\ref{sec:technical}.

\section{Results}
\label{sec:results}

\subsection{Tully--Fisher relation}
\label{sec:TFR_Results}

\begin{figure*}
  \subfigure[$V_\mathrm{peak}$]
  {
    \includegraphics[width=0.403\textwidth]{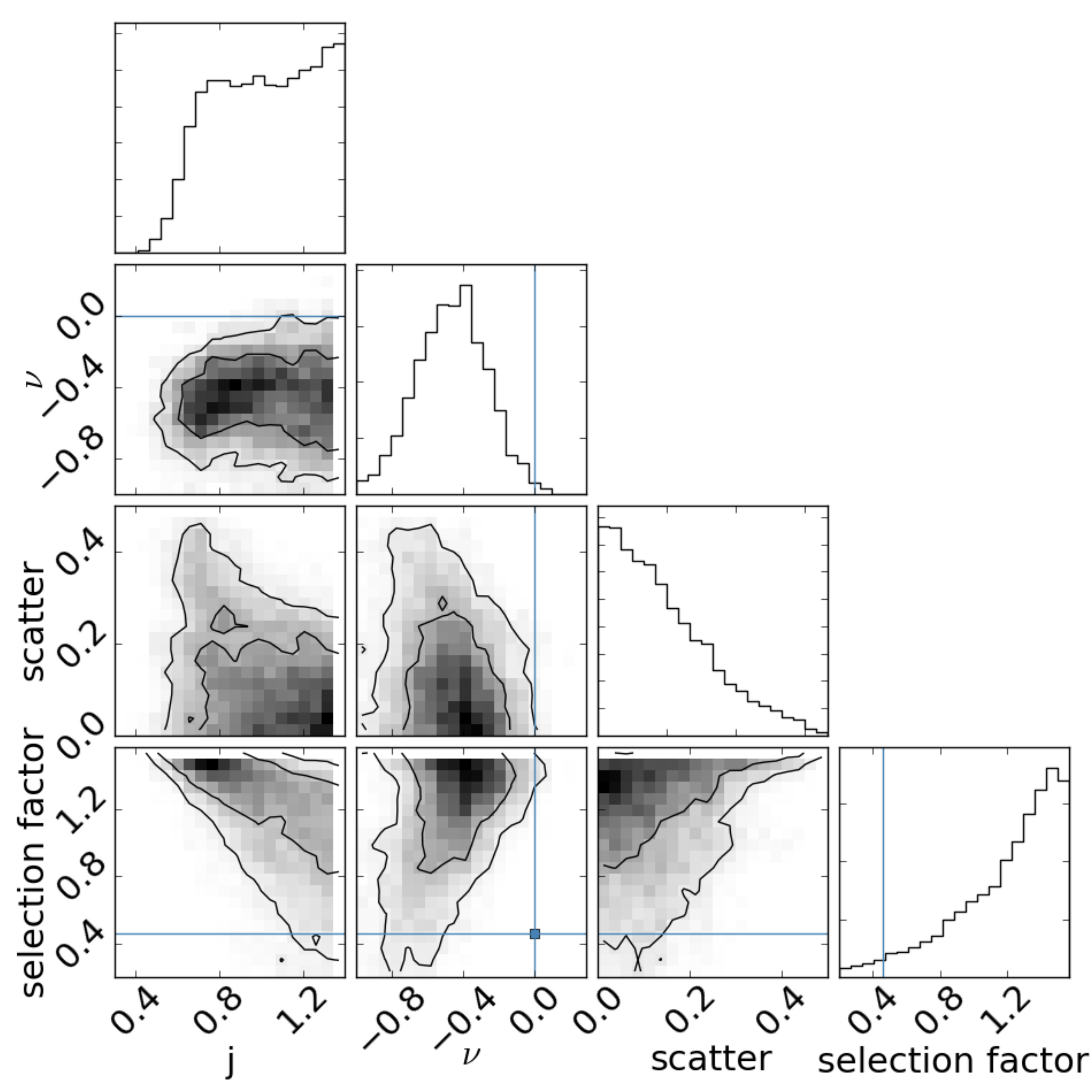}
    \label{fig:Vpeak}
  }
  \subfigure[$M_\mathrm{peak}$]
  {
    \includegraphics[width=0.403\textwidth]{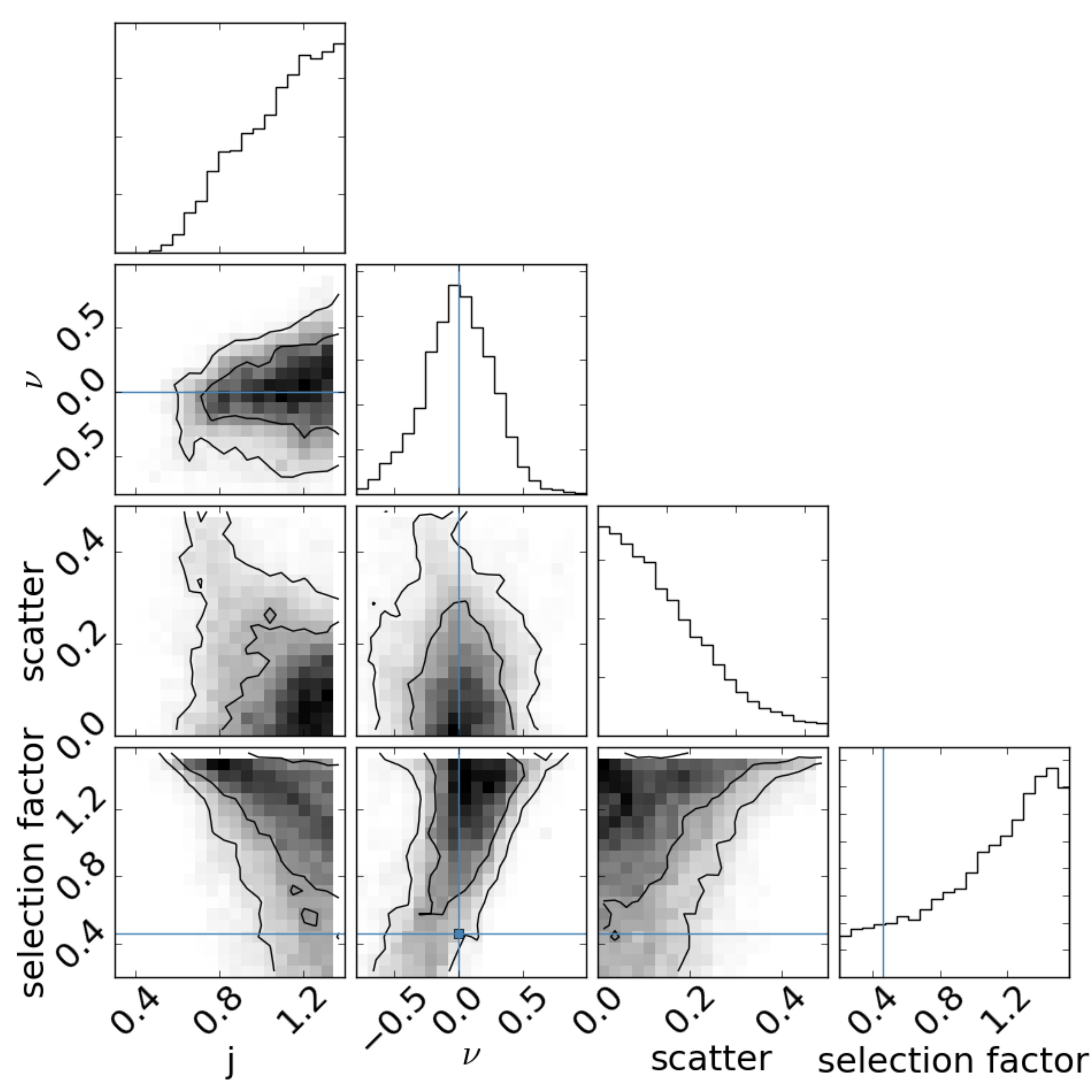}
    \label{fig:Mpeak}
  }
  \subfigure[$V_\mathrm{acc}$]
  {
    \includegraphics[width=0.403\textwidth]{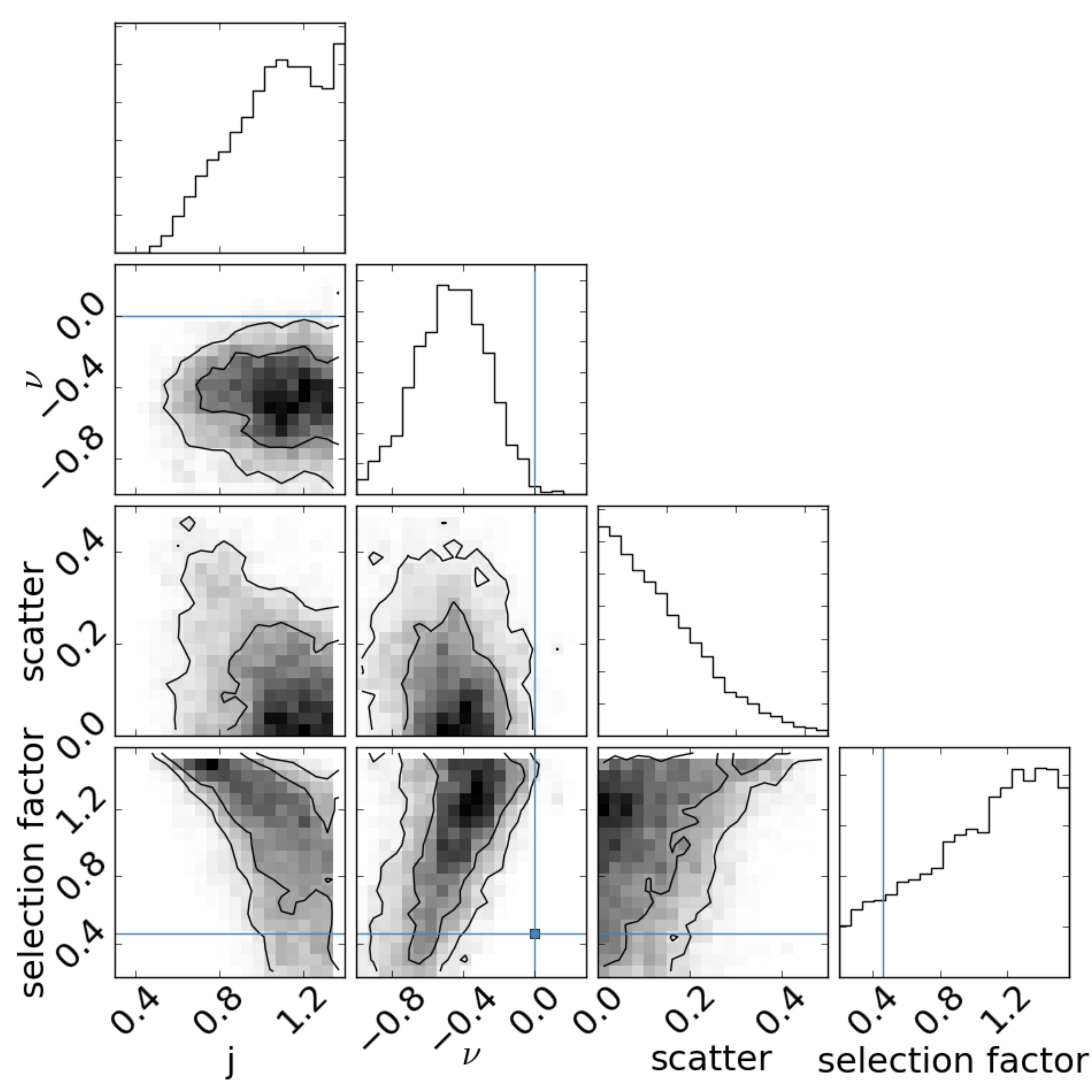}
    \label{fig:Vacc}
  }
  \subfigure[$M_\mathrm{acc}$]
  {
    \includegraphics[width=0.403\textwidth]{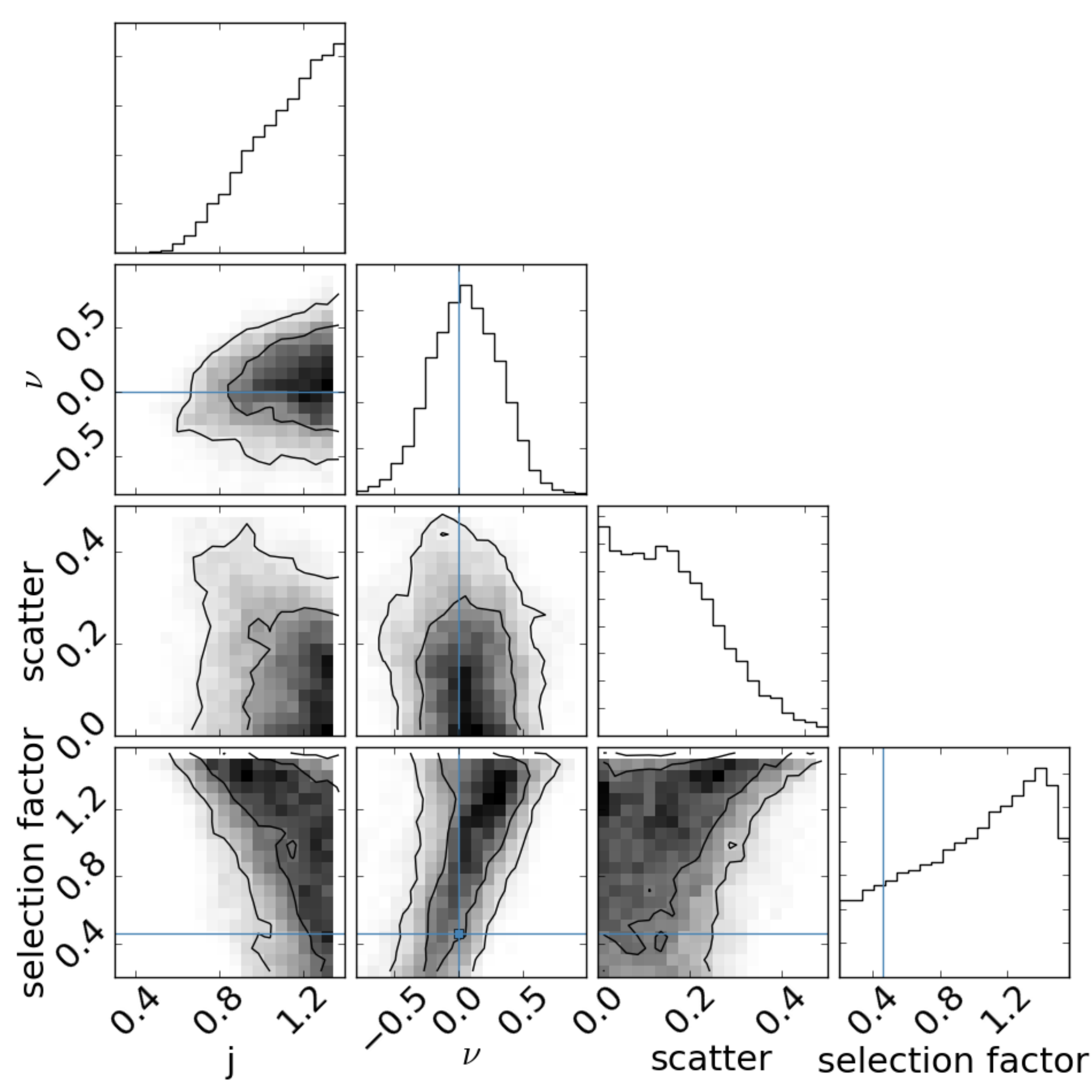}
    \label{fig:Macc}
  }
  \subfigure[$V_\mathrm{max}$]
  {
    \includegraphics[width=0.403\textwidth]{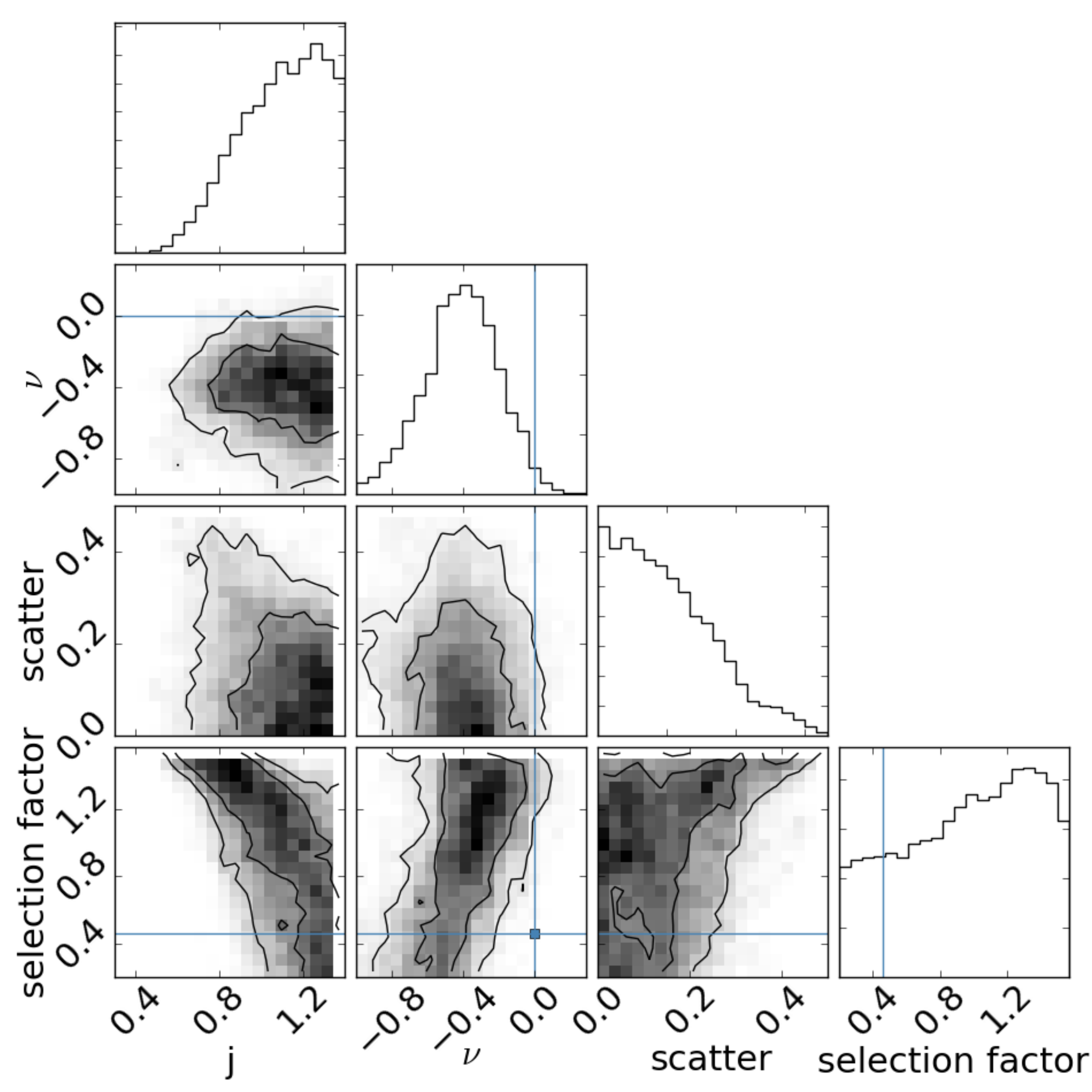}
    \label{fig:Vmax}
  }
  \subfigure[$M_\mathrm{now}$]
  {
    \includegraphics[width=0.403\textwidth]{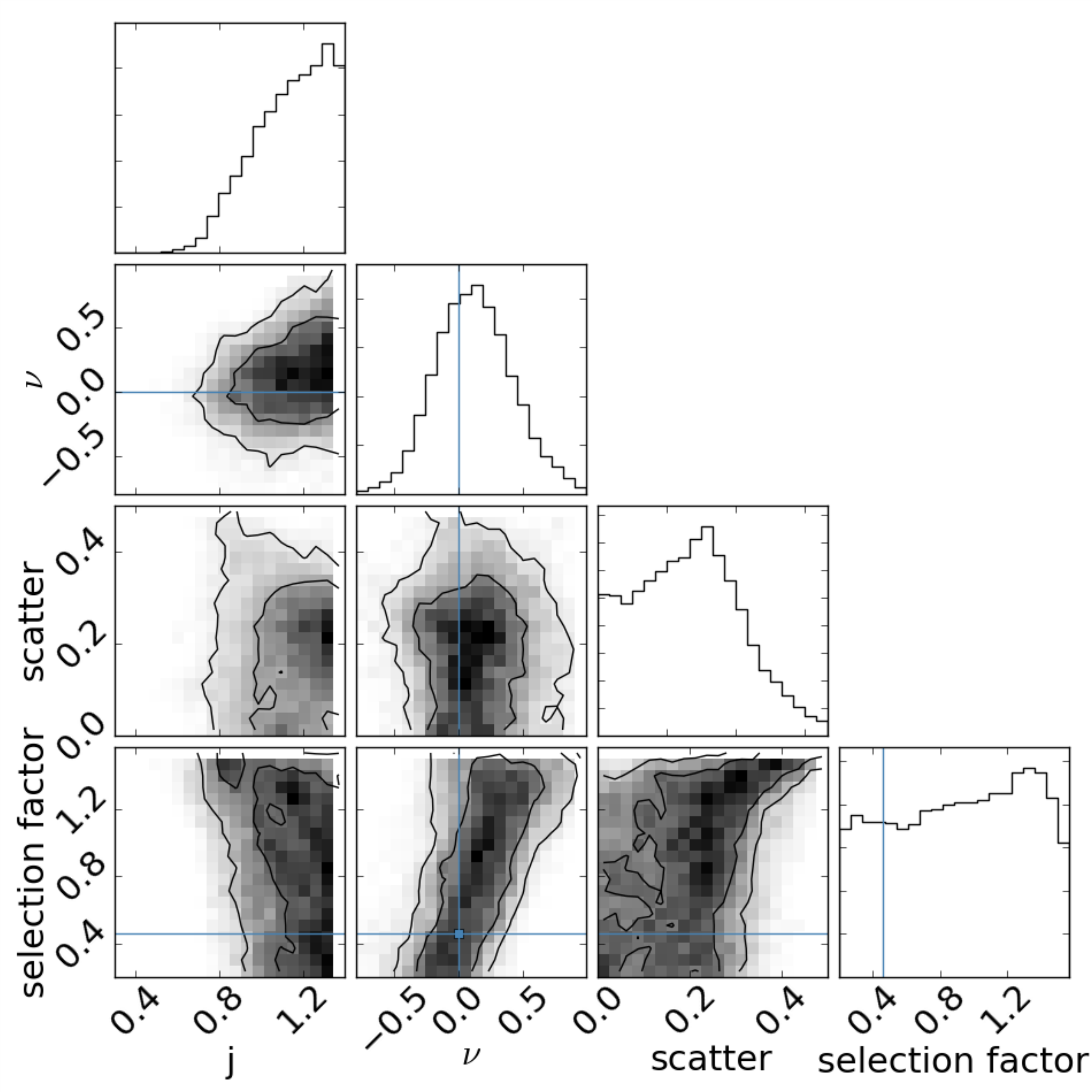}
    \label{fig:Mnow}
  }
  \caption{Posterior probability distributions of model parameters obtained from the TFR (diagonal panels), and correlations among the parameters (off-diagonal panels). Contours indicate the 68 and 95 per cent confidence levels. Each subfigure shows the result for a different AM model; note the change of range on the $\nu$ axis between the $M$ and $V$ models. Light blue lines show the case of no effect of disc formation on halo density profile ($\nu=0$) and random selection [selection factor = $\arctan(0.5) = 0.464$]. In several places in the paper we compare the $V_\mathrm{peak}$ (top left) and $M_\mathrm{now}$ (bottom right) models, which span the range of possible effects of AM proxy on the model TFR.}
  \label{fig:Posteriors}
\end{figure*}

\begin{figure}
  \centering
  \subfigure[Slope]
  {
    \includegraphics[width=0.366\textwidth]{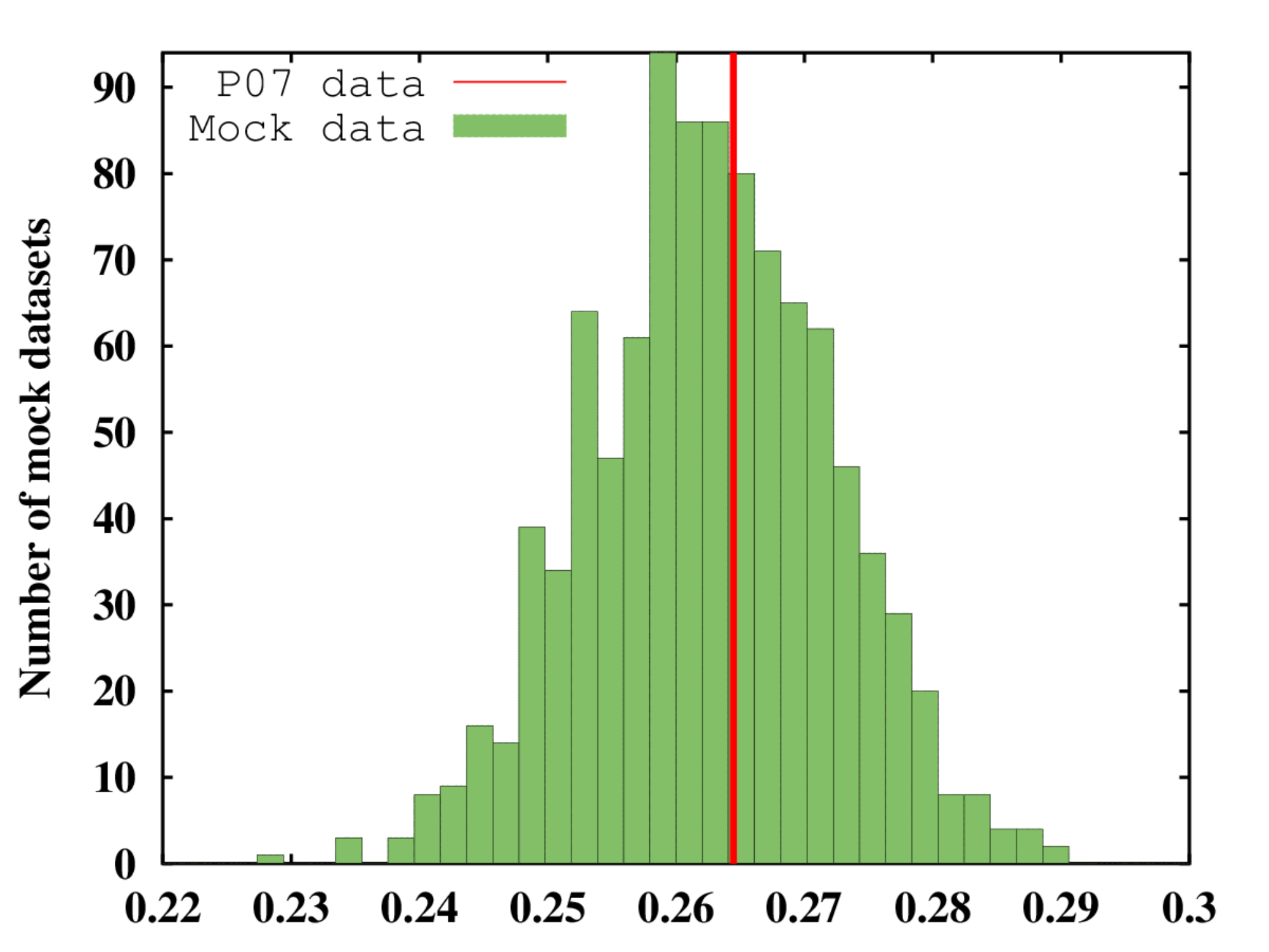}
    \label{fig:slope}
  }
  \subfigure[Intercept]
  {
    \includegraphics[width=0.366\textwidth]{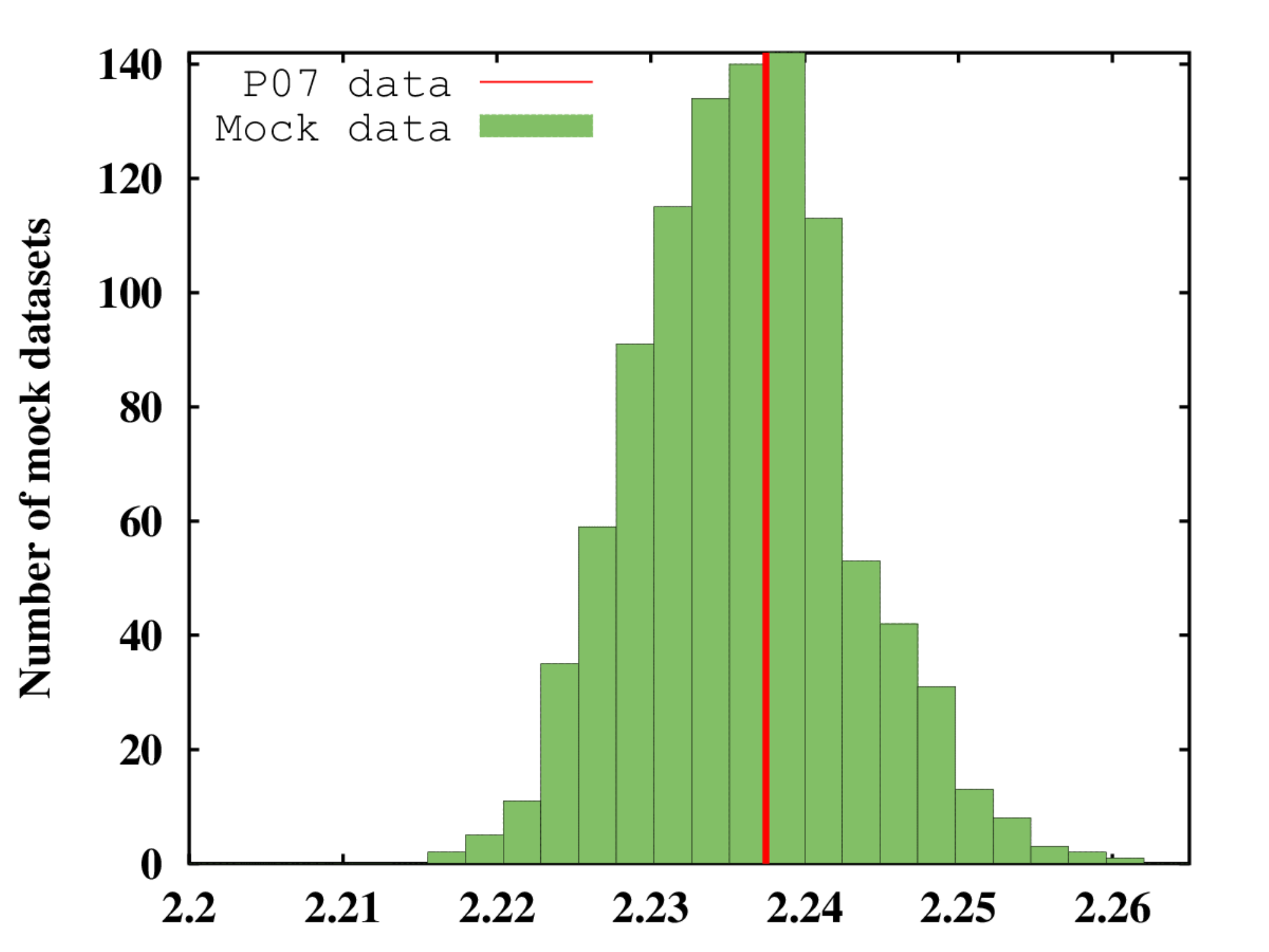}
    \label{fig:intercept}
  }
  \subfigure[Intrinsic Scatter]
  {
    \includegraphics[width=0.366\textwidth]{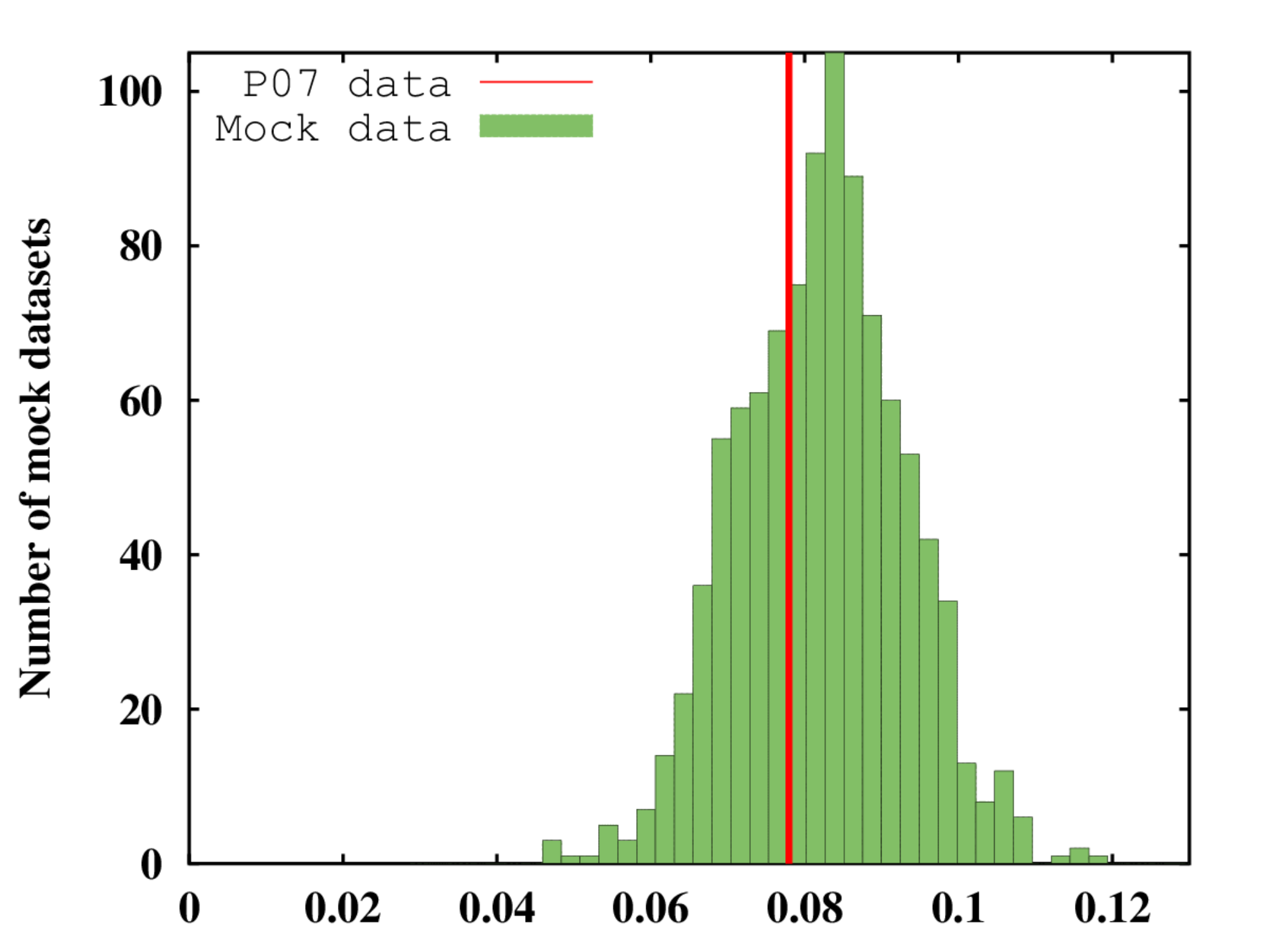}
    \label{fig:sigma}
  }
  \caption{The distribution of slope, intercept and scatter values of 1000 mock TFR data sets, generated using the maximum-likelihood parameters of the $V_\mathrm{peak}$ model (Table~\ref{tab:ML_Parameters}), compared to the values for the observational data (red lines). There is good agreement for each power-law parameter.}
  \label{fig:Histograms}
\end{figure}

Table~\ref{tab:Constraints} lists the $1\sigma$ and $2\sigma$ constraints on our model parameters after comparison to the P07 TFR, and Table~\ref{tab:ML_Parameters} gives the maximum-likelihood parameter values and their associated goodness of fit. Fig.~\ref{fig:Posteriors} shows the posterior probability distributions obtained for the parameters, and Fig.~\ref{fig:Histograms} compares the distribution of slope, intercept and scatter values obtained from our mock data sets using the maximum-likelihood parameter values to those of the data. In Appendix~\ref{sec:app_plots} (Fig.~\ref{fig:PointsPlots}), we show and explain the effect of each of the model parameters on the predicted TFR. The most significant findings are the following.

\begin{figure}
  \centering
  \includegraphics[width=0.5\textwidth]{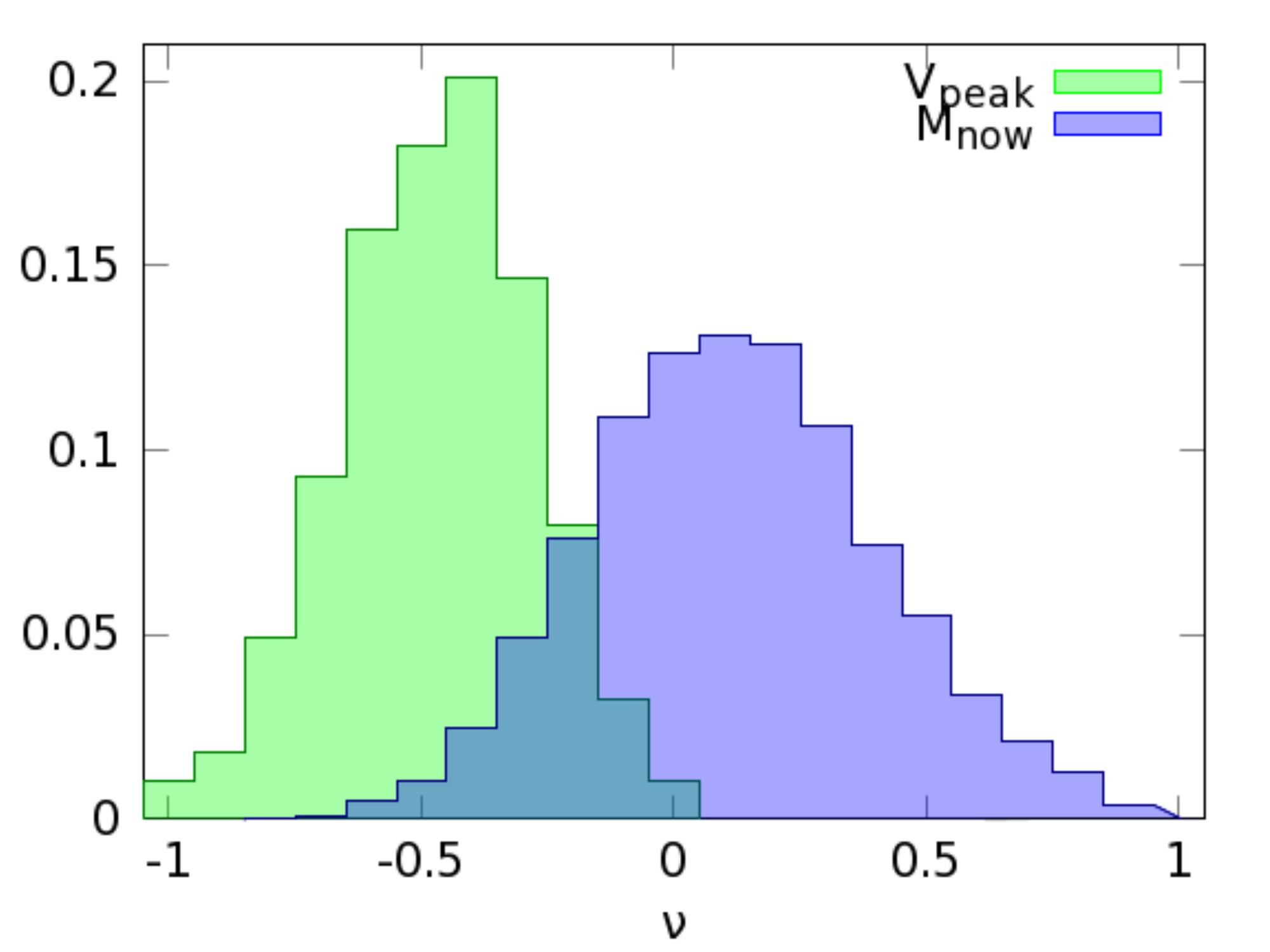}
  \caption{Comparison of the $\nu$ posterior probability distributions (determining the halo response to disc formation) derived from the TFR for the $V_\mathrm{peak}$ and $M_\mathrm{now}$ models. Models that assign galaxies to haloes based on mass are consistent with mild halo contraction ($\nu > 0$) or expansion ($\nu < 0$); models that assign galaxies to haloes based on velocity require expansion.}
  \label{fig:nu_comp}
\end{figure}

\begin{itemize}

\item{} In all cases, reasonably low AM scatter (e.g. $< 0.36$ dex at $2\sigma$ for the $V_\mathrm{peak}$ model\footnote{Technically, each AM scatter value presented here is the quadrature sum of the uncertainty in the input SMF and the intrinsic scatter associated with the AM procedure itself. So, for example, if the stellar masses used to construct the SMF have an uncertainty of 0.15 dex, then our $V_\mathrm{peak}$ constraint implies that the pure AM scatter must be $< 0.33$ dex at $2\sigma$ or $< 0.10$ dex at $1\sigma$. See Appendix~\ref{sec:assumptions} for further discussion of stellar mass uncertainties.}) is required in order to limit the scatter in the model TFRs to the small value observed in the data. This constraint is currently weaker than that afforded by clustering and satellite fraction measurements~\citep[e.g.][]{More10, Reddick}, but can be strengthened when larger homogeneous TFR data sets become available. In addition, the TFR has the potential to constrain the scatter at lower mass (see also Section~\ref{sec:constraints}).

\item{} The $V_\mathrm{peak}$ model requires at the $2\sigma$ level that galaxies passing the P07 selection cuts have systematically low rotation velocities for their stellar mass (see final column of Table~\ref{tab:Constraints} and bottom-right panel of Fig.~\ref{fig:Vpeak}). This further reduces the predicted scatter and also improves agreement with the data by shifting the TFR towards lower $V_{80}$ at fixed stellar mass. All other AM models are consistent at $2\sigma$ with random selection [i.e. independent of $V_{80}$, corresponding to selection factor $=\arctan(0.5)=0.464$]. It is important to note, however, that models with weak selection require high $j$ to be consistent with the relatively low normalization of the TFR, and hence that lowering the upper limit of the prior on $j$ would tighten the selection requirements. For example, if $j$ is capped at 1 instead of 1.4, only the $M_\mathrm{now}$ and $M_\mathrm{acc}$ models would be consistent with random selection at the $2\sigma$ level.

\item{} For all AM models, standard adiabatic contraction would make galaxies rotate too fast at fixed stellar mass and is ruled out at the $\gtrsim 4\sigma$ level. For mass-based AM models, agreement with the data is maximized by switching contraction off altogether. In contrast, for models that match to halo velocity, expansion of the halo in response to galaxy formation is in fact required at the $2\sigma$ level. (However, we show in Appendix~\ref{sec:assumptions} that a small amount of contraction is permitted if $\Omega_\mathrm{m}$ and $\sigma_8$ take their lowest values allowed by the Planck uncertainties.) We compare the $\nu$ posterior probability distributions for the $V_\mathrm{peak}$ and $M_\mathrm{now}$ models in Fig.~\ref{fig:nu_comp}.

\item{} Low $j$ values ($j \lesssim 0.7$, depending on the model) are excluded. Lowering $j$ reduces the angular momentum of all galaxies, tending to decrease their size and increase their rotation velocity. Our main analysis uses the prior $0 < j \le 1.4$, based on constraints from the MSR discussed in the following section. If this prior is relaxed, the upper bound placed on $j$ by the TFR ranges from $\sim 2.5$ for the $M_\mathrm{now}$ model to $\sim 5$ for the $V_\mathrm{peak}$ model. In this case, high $\nu$ values ($\nu > 1$) are allowed, and no lower limit may be placed on the selection factor for any AM model.

\item{} The trends in the constraints derived using the different AM models may be understood by means of the trend in average halo concentration as a function of stellar mass. The models in Table~\ref{tab:Constraints} are listed in decreasing order of average halo concentration over the stellar mass range of interest (with the largest difference being between $V_\mathrm{max}$ and $M_\mathrm{peak}$), and hence in decreasing order of TFR normalization for fixed values of the other parameters. Thus models higher in the table require stronger selection and lower $\nu$ to achieve consistency with the data.

\item{} There are moderate degeneracies between the selection factor and AM scatter (which influence the TFR scatter) and between $\nu$, $j$ and the selection factor (which influence the normalization). The slope is the least constraining of the three TFR characteriztics; it is satisfactorily reproduced in a large proportion of parameter space because the stellar mass range probed by the P07 sample is relatively small (2.2 decades).

\item{} The maximum-likelihood values generated by each of the models are very similar, and correspond to a very good fit to the P07 data. This demonstrates that our framework has sufficient flexibility to match the observed stellar mass TFR regardless of the AM scheme employed.

\end{itemize}

\bgroup
\def\arraystretch{1.7}
\begin{table*}
  \begin{center}
    \begin{tabular}{l|c|c|c|c|}
      \hline
      AM parameter &AM scatter / dex &$j$ &$\nu$ &Selection factor\\ 
      \hline
      $V_\mathrm{peak}$    &    $<$ 0.18 (0.36) &  $>$ 0.87 (0.63) & $-0.4^{+0.1 (+0.3)}_{-0.2 (-0.4)}$ & $>$ 1.09 (0.476)\\
      $V_\mathrm{acc}$    &     $<$ 0.16 (0.34) & $>$ 0.97 (0.67) & $-0.5^{+0.2 (+0.4)}_{-0.1 (-0.3)}$ & --\\
      $V_\mathrm{max}$    &     $<$ 0.18 (0.36) & $>$ 1.0 (0.73) & $-0.4^{+0.2 (+0.4)}_{-0.2 (-0.4)}$ & --\\
      $M_\mathrm{peak}$    &    $<$ 0.18 (0.36) & $>$ 1.0 (0.72) & $0.0^{+0.3 (+0.5)}_{-0.2 (-0.5)}$ & --\\
      $M_\mathrm{acc}$    &     $<$ 0.20 (0.36) & $>$ 1.0 (0.77) & $0.0^{+0.3 (+0.6)}_{-0.2 (-0.4)}$ & --\\
      $M_\mathrm{now}$    &     $<$ 0.24 (0.40) & $>$ 1.1 (0.81) & $0.1^{+0.2 (+0.6)}_{-0.3 (-0.5)}$ & --\\
      \hline
    \end{tabular}
  \caption{Table of parameter constraints derived by comparing the model to the P07 TFR. Numbers outside of brackets are $1\sigma$ limits and numbers within brackets are $2\sigma$ limits. Lower rows correspond to lower average halo concentration at given stellar mass. The constraints on both $j$ and the selection factor depend sensitively on the prior range allowed for $j$ (here $0<j<1.4$); when $j$ is allowed to take larger values the selection factor may be smaller. We quote limits on the selection factor only in cases where random selection (selection factor = 0.464) is ruled out at the $>2\sigma$ level.}
  \label{tab:Constraints}
  \end{center}
\end{table*}
\egroup

\begin{table*}
  \begin{center}
    \begin{tabular}{l|c|c|c|c|c|}
      \hline
      AM parameter &AM scatter / dex &$j$ &$\nu$ &Selection factor &$P$ value\\ 
      \hline
      $V_\mathrm{peak}$   & 0.0 & 0.86 & -0.4 & 1.34 & $\sim0.9$\\
      $M_\mathrm{now}$    & 0.4 & 0.92 & 0.2 & 1.51 & $\sim0.9$\\
      \hline
    \end{tabular}
  \caption{Maximum-likelihood parameter values for the TFR and associated goodness of fit, for the $V_\mathrm{peak}$ and $M_\mathrm{now}$ models. `$P$ value' is roughly the fraction of mock data sets with \{slope, intercept, scatter\} values at least as far from the centroid as those of the data. All AM matching proxies are capable of generating good fits to the observed TFR.}
  \label{tab:ML_Parameters}
  \end{center}
\end{table*}

\subsection{Mass--size relation}
\label{sec:MSR_Results}

On constructing mock MSTs in our framework, we immediately find that the model is not capable of reproducing the observed intrinsic scatter, $0.17 \pm 0.01$ dex. In particular, the predicted scatter is always significantly larger (at least 0.28 dex), even when AM scatter is switched off and the selection factor maximized. The reason for this discrepancy is easily identified: in our model, in which the specific angular momentum of a galaxy is proportional to that of the halo in which it lives, disc size is proportional to halo spin. Thus the scatter in radius at fixed stellar mass receives a contribution equal to the scatter in halo spin at fixed AM parameter, which is roughly 0.27 dex and therefore by itself exceeds the measured value.\footnote{We use the `Peebles' spin~\citep{Peebles} for compatibility with previous work such as~\citet{MMW}. The alternative `Bullock' spin~\citep{Bullock_Spin} has a typical scatter of around 0.29 dex and so would slightly exacerbate the discrepancy if used instead.} We postpone further discussion of this issue to Section~\ref{sec:MSR_Problem}, and in the remainder of this section investigate the extent to which our model parameters can be constrained using only the slope and normalization of the MSR.

\begin{figure}
  \centering
  \includegraphics[width=0.5\textwidth]{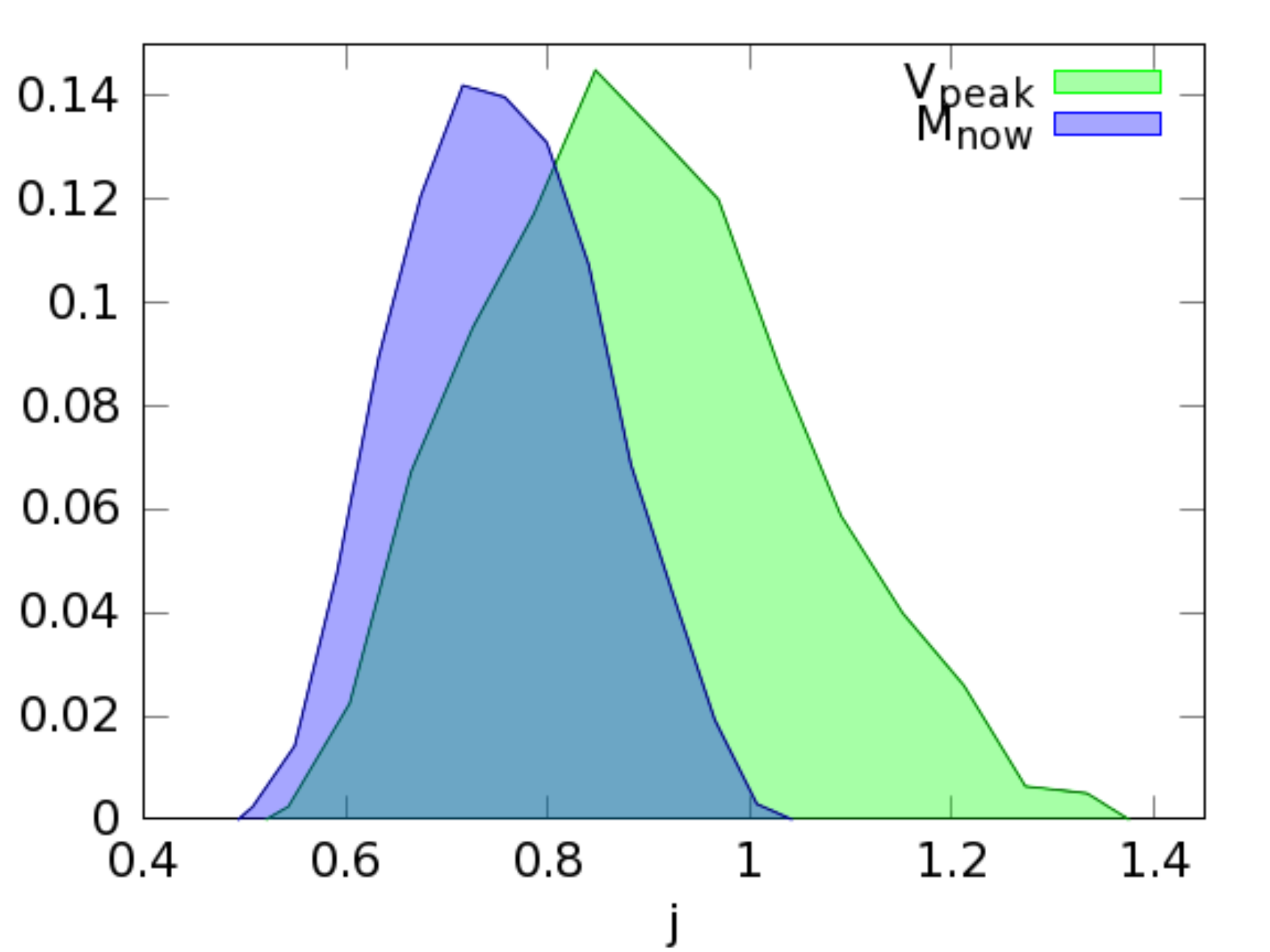}
  \caption{Comparison of the posterior probability distributions of $j$ (ratio of specific angular momentum of galaxy and halo) derived from the MSR, for the $V_\mathrm{peak}$ and $M_\mathrm{now}$ models. Velocity-based AM models allow somewhat higher $j$ than mass-based models.}
  \label{fig:j_comp}
\end{figure}

\begin{figure}
  \centering
  \subfigure[Slope]
  {
    \includegraphics[width=0.366\textwidth]{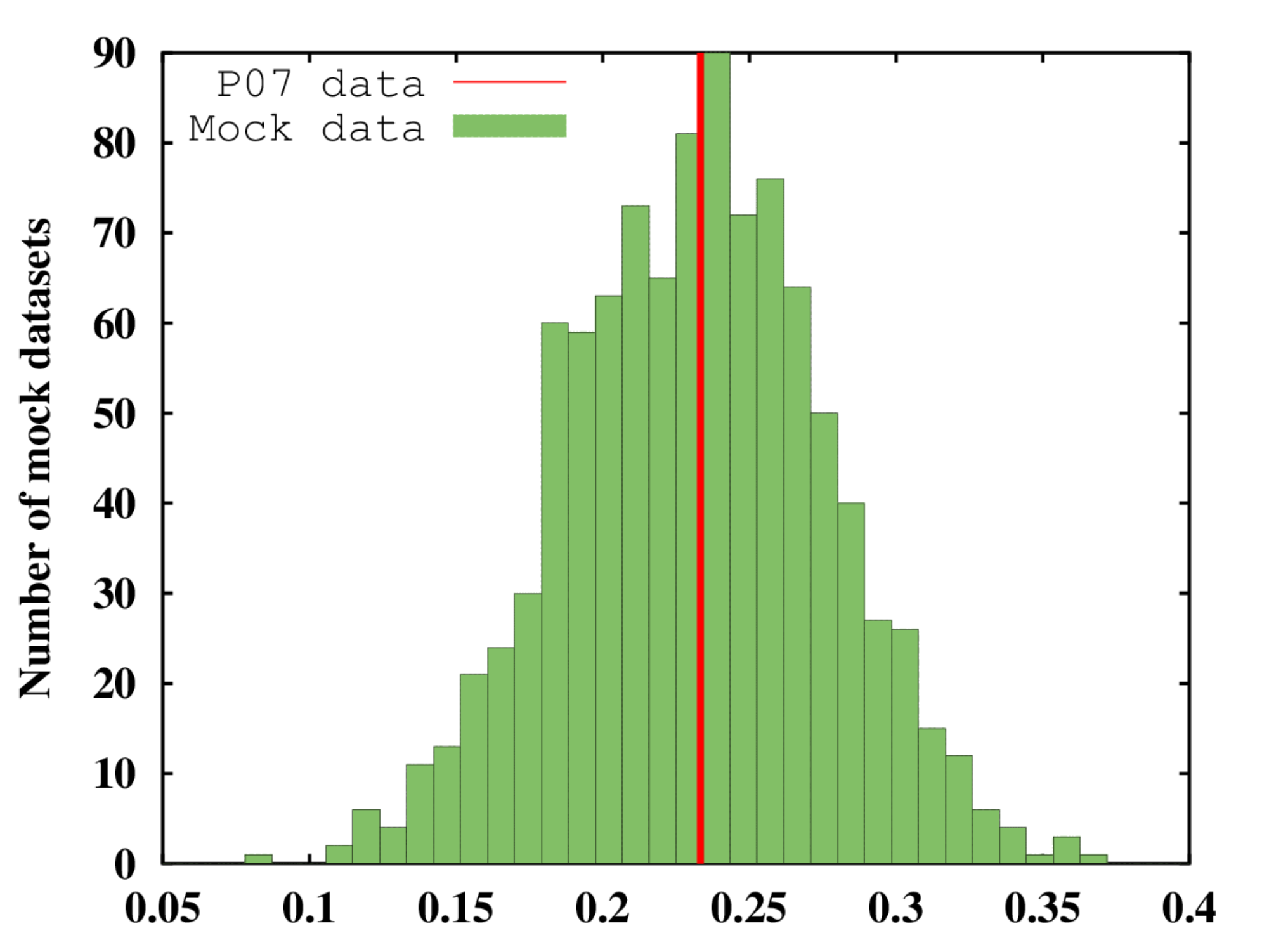}
    \label{fig:slope_Rad}
  }
  \subfigure[Intercept]
  {
    \includegraphics[width=0.366\textwidth]{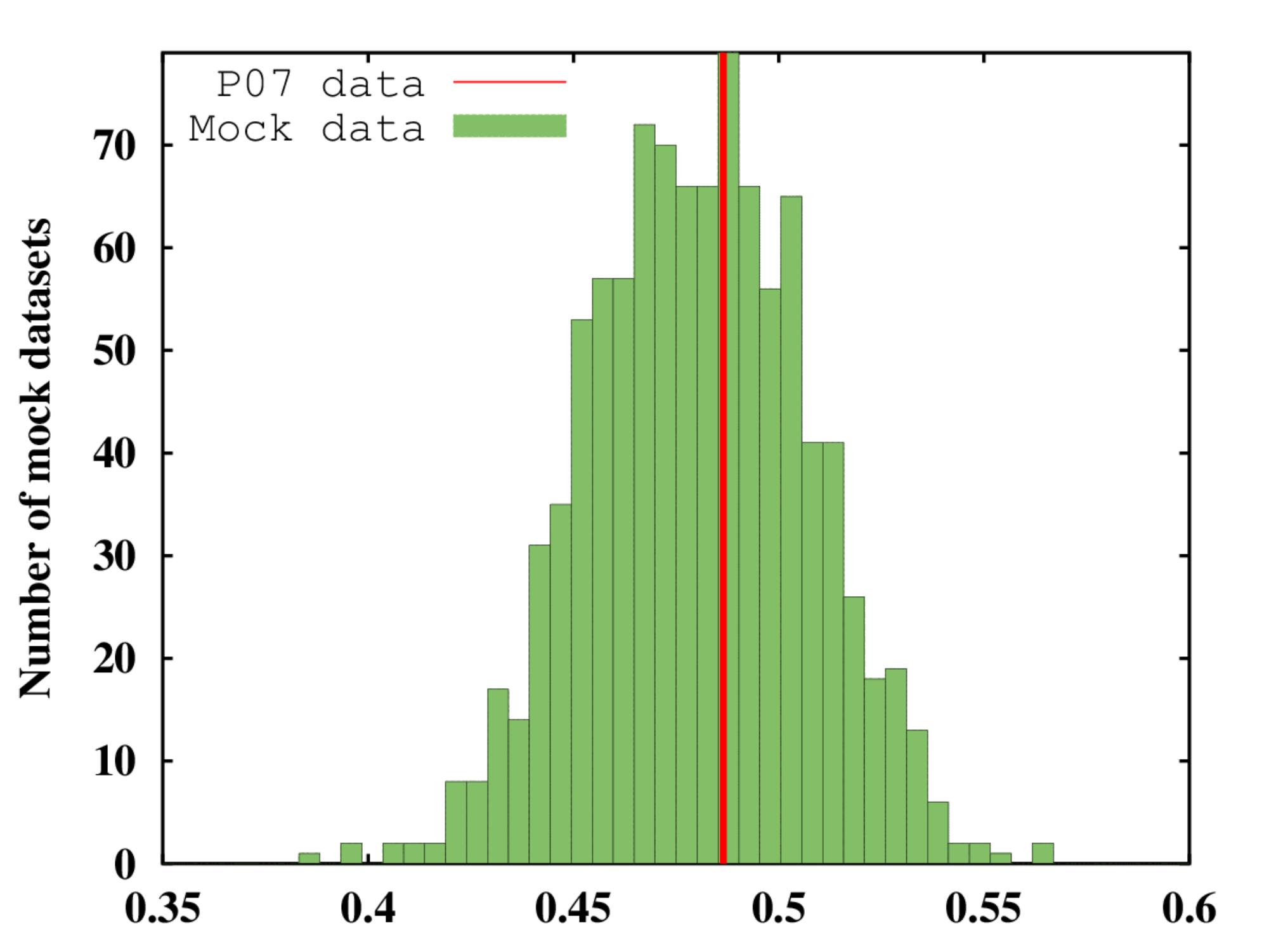}
    \label{fig:intercept_Rad}
  }
  \subfigure[Intrinsic Scatter]
  {
    \includegraphics[width=0.366\textwidth]{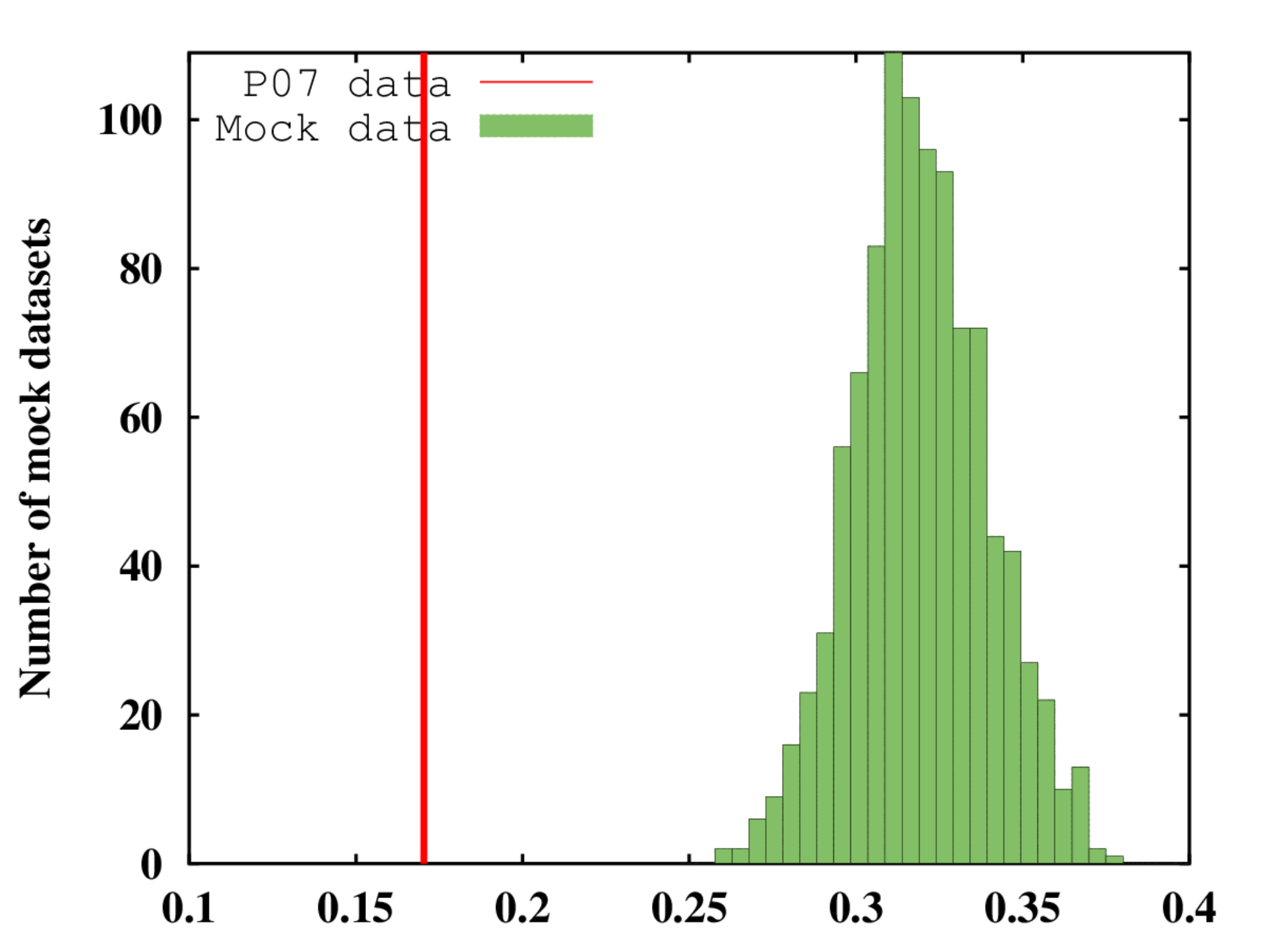}
    \label{fig:sigma_Rad}
  }
  \caption{Same as Fig.~\ref{fig:Histograms}, but for the MSR instead of the TFR (see Table~\ref{tab:ML_Parameters_Rad}). The slope and intercept agree well, but the predicted scatter is significantly larger than that of the data due to the relatively large scatter in halo spin.}
  \label{fig:Histograms_Rad}
\end{figure}

To this end, we repeat the MCMC analysis described for the TFR in Section~\ref{sec:method} but replacing a galaxy's $V_{80}$ value by its scalelength, $R_\mathrm{d}$, and calculating the likelihood for a particular set of parameter values by comparing the slopes and intercepts of power-law fits to mock data generated using the model to those of the data. We find that only $j$, the ratio of the specific angular momentum of the disc and halo, can be significantly constrained using this information, since this is mainly responsible for setting median disc size at fixed stellar mass. $j$ values around 0.8 are favoured\footnote{Differences between the Peebles and Bullock spin parameters are no more than $\sim10$ per cent over the mass range of interest, so the specific definition of halo spin affects our constraints on $j$ at the $\lesssim10$ per cent level.} (see Table~\ref{tab:Constraints_Rad} and Fig.~\ref{fig:j_comp}), suggesting that a small fraction of discs' angular momentum may be transferred to their haloes. (Note that these constraints on $j$ include full marginalization over the other parameters in the model.) Slightly higher $j$ values are preferred for velocity-based AM models than mass-based ones since the former tend to put relevant galaxies in more concentrated haloes, making those galaxies smaller at fixed angular momentum. In Appendix~\ref{sec:app_plots} (Fig.~\ref{fig:PointsPlots_Rad}), we show and explain the effect of each of the model parameters on the predicted MSR. Fig.~\ref{fig:Histograms_Rad} demonstrates that the model can fit well the slope and intercept of the relation (using the best-fitting parameter values listed in the first row of Table~\ref{tab:ML_Parameters_Rad}), and illustrates its failure at reproducing the intrinsic scatter.

Our choice to ignore the MSR scatter when deriving constraints on the model parameters corresponds to the assumption that a mechanism employed to resolve the discrepancy (see Section~\ref{sec:MSR_Problem}) would not significantly affect the median relation. We caution, however, that the constraints on $j$ described above may be subject to systematic uncertainties associated with our use of a model that does not adequately capture all aspects of the MSR. Furthermore, the constraint on $j$ depends on the priors placed on the other model parameters. For example, restricting $\nu$ and the AM scatter to small values and the selection factor to large values -- in order to minimize the MSR scatter as much as the framework allows -- tends to lower the preferred value of $j$, and reduce its statistical uncertainty.

\bgroup
\def\arraystretch{1.7}
\begin{table}
  \begin{center}
    \begin{tabular}{l|c|}
      \hline
      AM parameter &$j$\\
      \hline
      $V_\mathrm{peak}$ &$0.84^{+0.19 (+0.38)}_{-0.14 (-0.24)}$\\
      $V_\mathrm{acc}$ &$0.86^{+0.16 (+0.27)}_{-0.15 (-0.28)}$\\
      $V_\mathrm{max}$ &$0.87^{+0.12 (+0.25)}_{-0.17 (-0.26)}$\\
      $M_\mathrm{peak}$ &$0.74^{+0.18 (+0.25)}_{-0.07 (-0.16)}$\\
      $M_\mathrm{acc}$ &$0.77^{+0.14 (+0.22)}_{-0.10 (-0.19)}$\\
      $M_\mathrm{now}$ &$0.76^{+0.09 (+0.19)}_{-0.12 (-0.19)}$\\
      \hline
    \end{tabular}
  \caption{Table of constraints on $j$ derived by comparing the model to the slope and intercept of the P07 MSR. Numbers outside of brackets are $1\sigma$ limits and numbers within brackets are $2\sigma$ limits.}
  \label{tab:Constraints_Rad}
  \end{center}
\end{table}
\egroup

\begin{table*}
  \begin{center}
    \begin{tabular}{l|c|c|c|c|c|c|}
      \hline
      &AM scatter / dex &$j$ &$\nu$ &Selection factor &$P$ value (w/o scatter) &$P$ value (with scatter)\\ 
      \hline
      $V_\mathrm{peak}$    &     0.34 & 0.60 & --0.2 & 1.55 & $>0.9$ & $\sim10^{-13}$\\
      $M_\mathrm{now}$    &      0.24 & 0.57 & --0.4 & 1.55 & $>0.9$ & $\sim10^{-10}$\\
      \hline
    \end{tabular}
  \caption{Maximum-likelihood parameter values for the MSR and associated goodness of fit, for the $V_\mathrm{peak}$ and $M_\mathrm{now}$ models. `$P$ Value' is roughly the fraction of mock data sets with \{slope, intercept\} (`w/o scatter') or \{slope, intercept, scatter\} (`with scatter') values at least as extreme as those of the data. Our models are capable of reproducing the slope and intercept of the observed MSR, but not its small scatter.}
  \label{tab:ML_Parameters_Rad}
  \end{center}
\end{table*}

\subsection{Correlation of Velocity and Radius Residuals}
\label{sec:residuals}

If a galaxy's stellar mass contributes significantly to its rotation velocity, one would expect smaller galaxies at fixed stellar mass to have larger $V_{80}$ values, and hence that galaxies with different surface brightnesses would lie on systematically offset TFRs. In other words, there would be a significant anticorrelation between velocity and radius residuals at fixed stellar mass. This expectation is not confirmed by the data~\citep{Zwaan, McGaugh_SB, McGaugh_Residuals, Courteau07}. Here we explore this relation in the context of our model to elucidate its theoretical significance.

\begin{figure}
  \subfigure[$V_\mathrm{peak}$; TFR parameter values]
  {
    \includegraphics[width=0.5\textwidth]{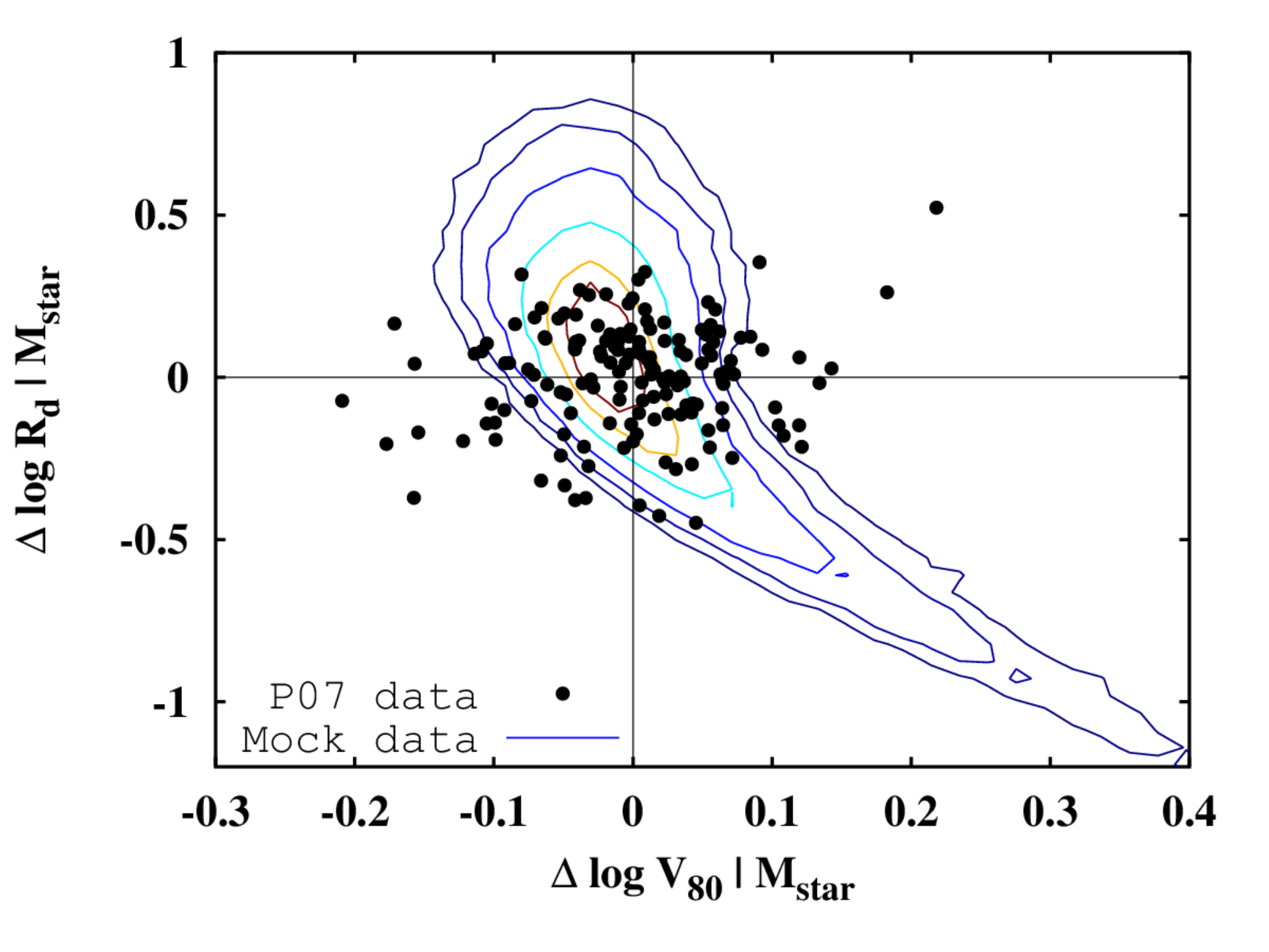}
    \label{fig:Res_Comp}
  }
  \subfigure[$V_\mathrm{peak}$; optimum parameter values]
  {
    \includegraphics[width=0.5\textwidth]{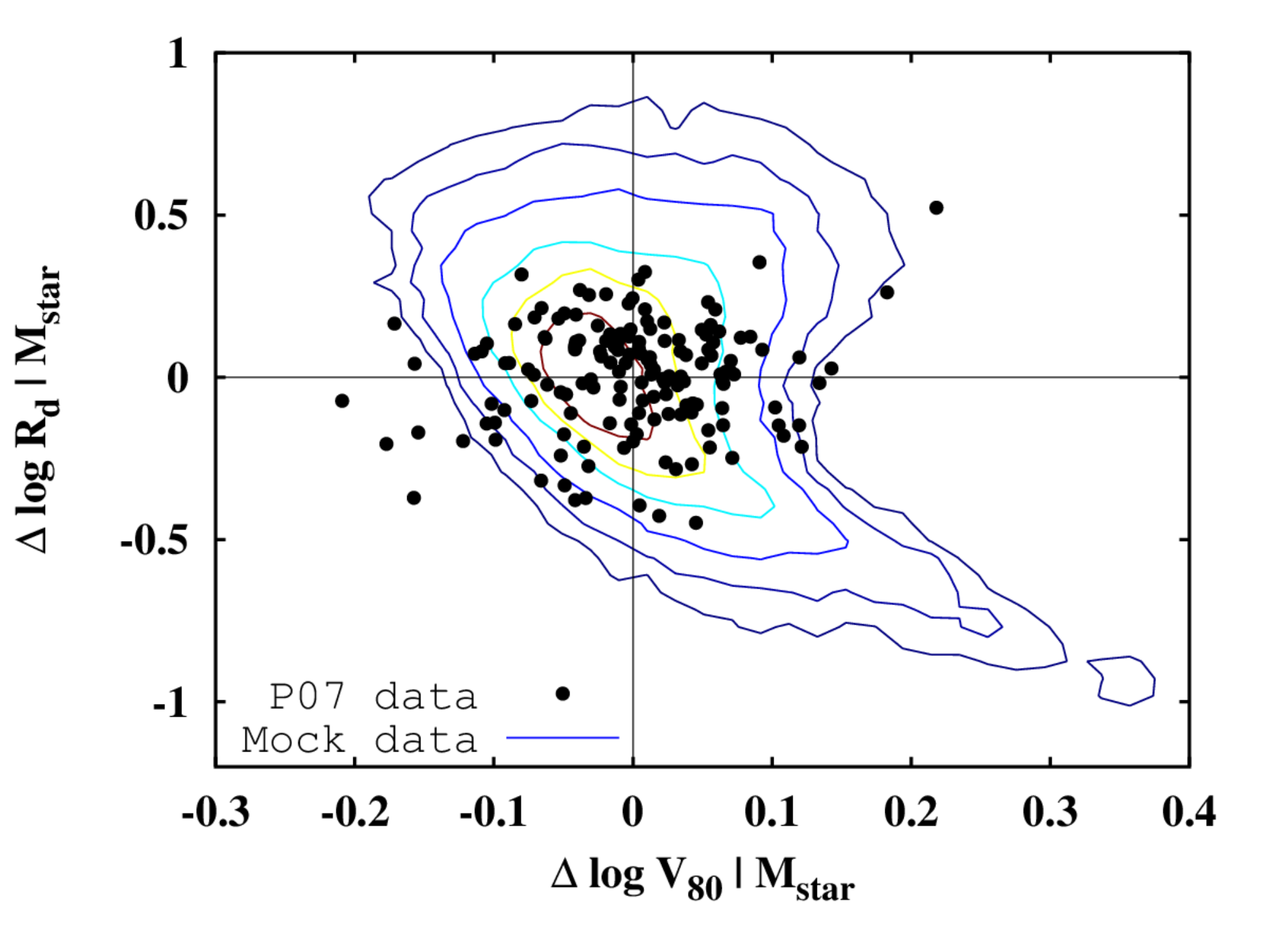}
    \label{fig:Res_Comp_ML}
  }
  \subfigure[]
  {
    \includegraphics[width=0.5\textwidth]{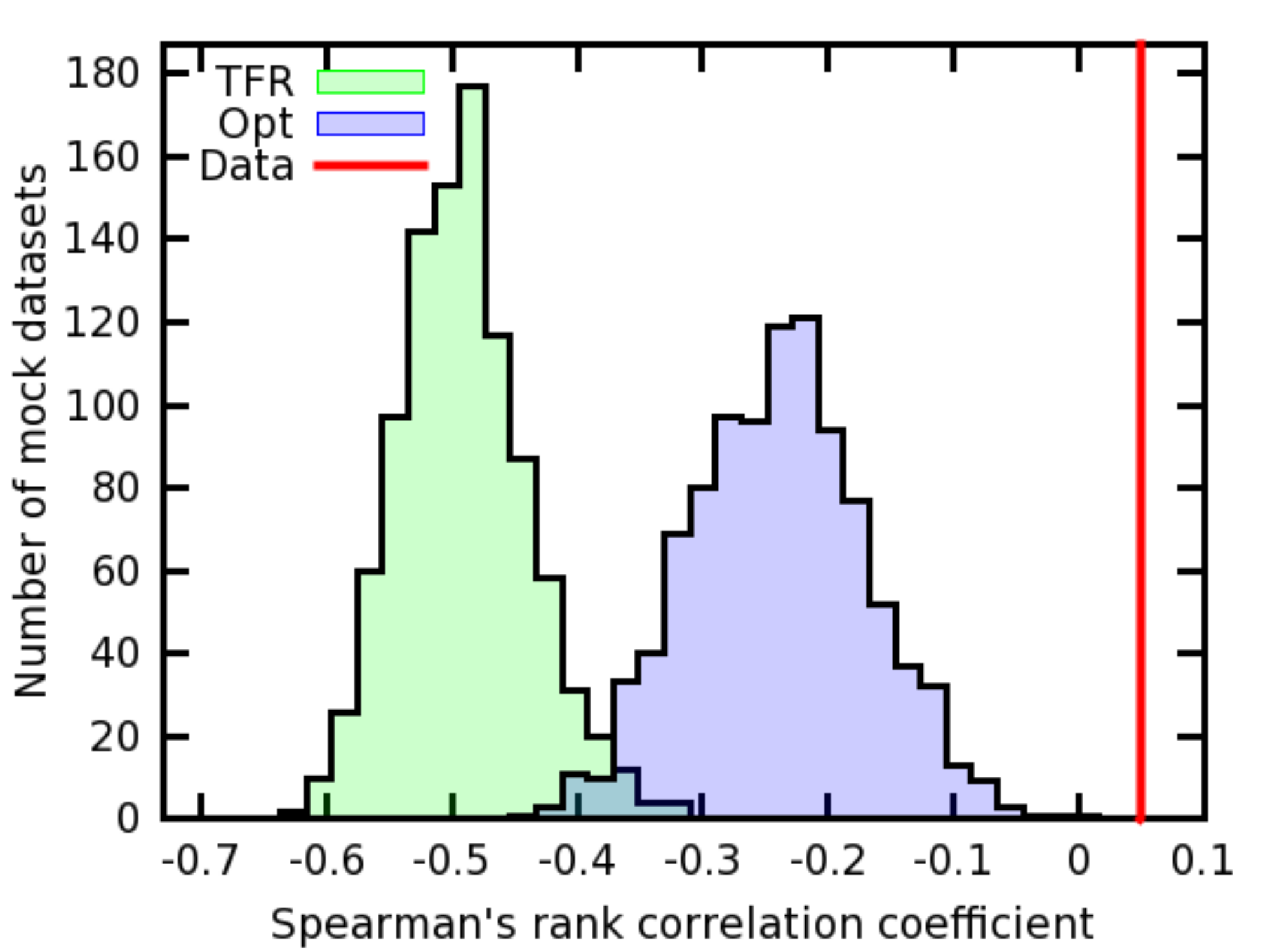}
    \label{fig:Res_Hist}
  }
  \caption{Comparison of the Spearman's rank correlation coefficient ($\rho$) between the radius and velocity residuals in the data and in 1000 mock data sets drawn from the $V_\mathrm{peak}$ model. In Figs~\ref{fig:Res_Comp} and~\ref{fig:Res_Comp_ML}, the black points are the P07 galaxies and the contours enclose 20, 40, 60, 80, 90, and 95 per cent of the model galaxies. Fig.~\ref{fig:Res_Comp} uses the parameter values that fit the TFR best; Fig.~\ref{fig:Res_Comp_ML} uses the optimum parameter values that maximize the \SRCC\ values of the mock data. Fig.~\ref{fig:Res_Hist} shows the distribution of $\rho$ across the mock data sets using the TFR (green) and optimum (blue) parameter values, and compares them to the $\rho$ value of the real data (0.0485; red line). The data exhibit a significantly weaker correlation than is predicted by the model. }
  \label{fig:Residual_Total}
\end{figure}

\begin{figure}
  \subfigure[$M_\mathrm{now}$; TFR parameter values]
  {
    \includegraphics[width=0.5\textwidth]{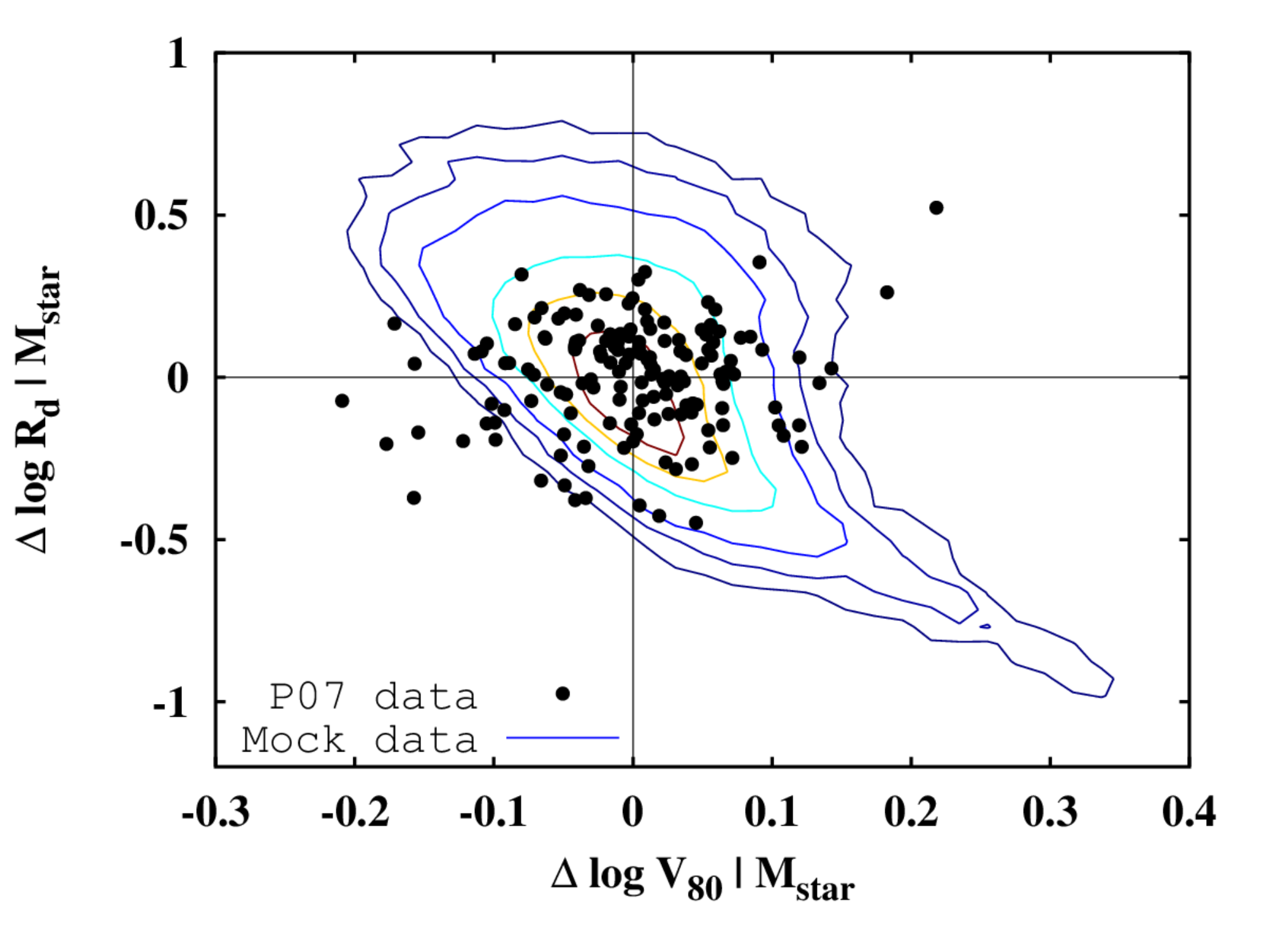}
    \label{fig:Res_Comp_Mnow}
  }
  \subfigure[$M_\mathrm{now}$; optimum parameter values]
  {
    \includegraphics[width=0.5\textwidth]{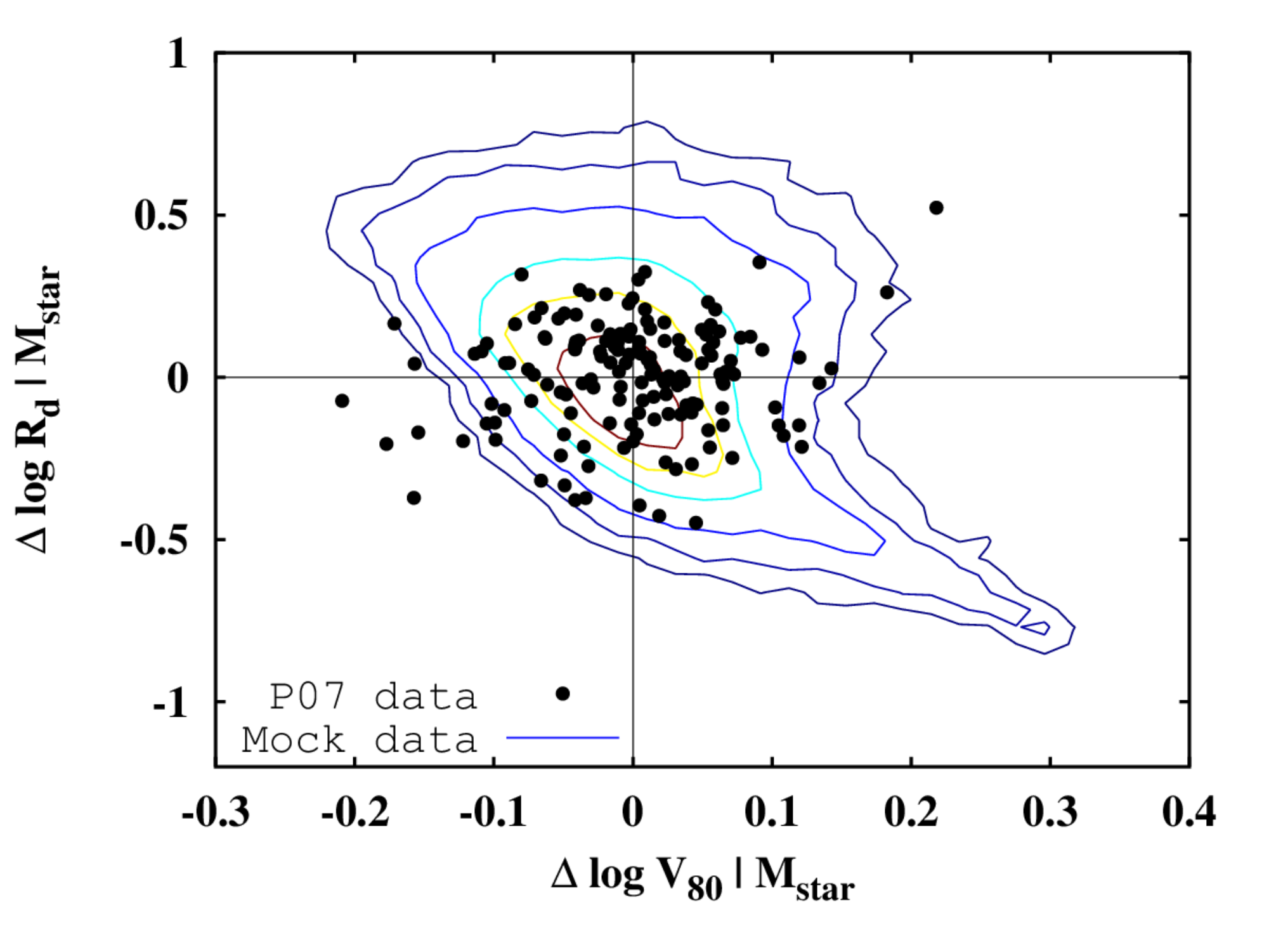}
    \label{fig:Res_Comp_Mnow_ML}
  }
  \subfigure[]
  {
    \includegraphics[width=0.5\textwidth]{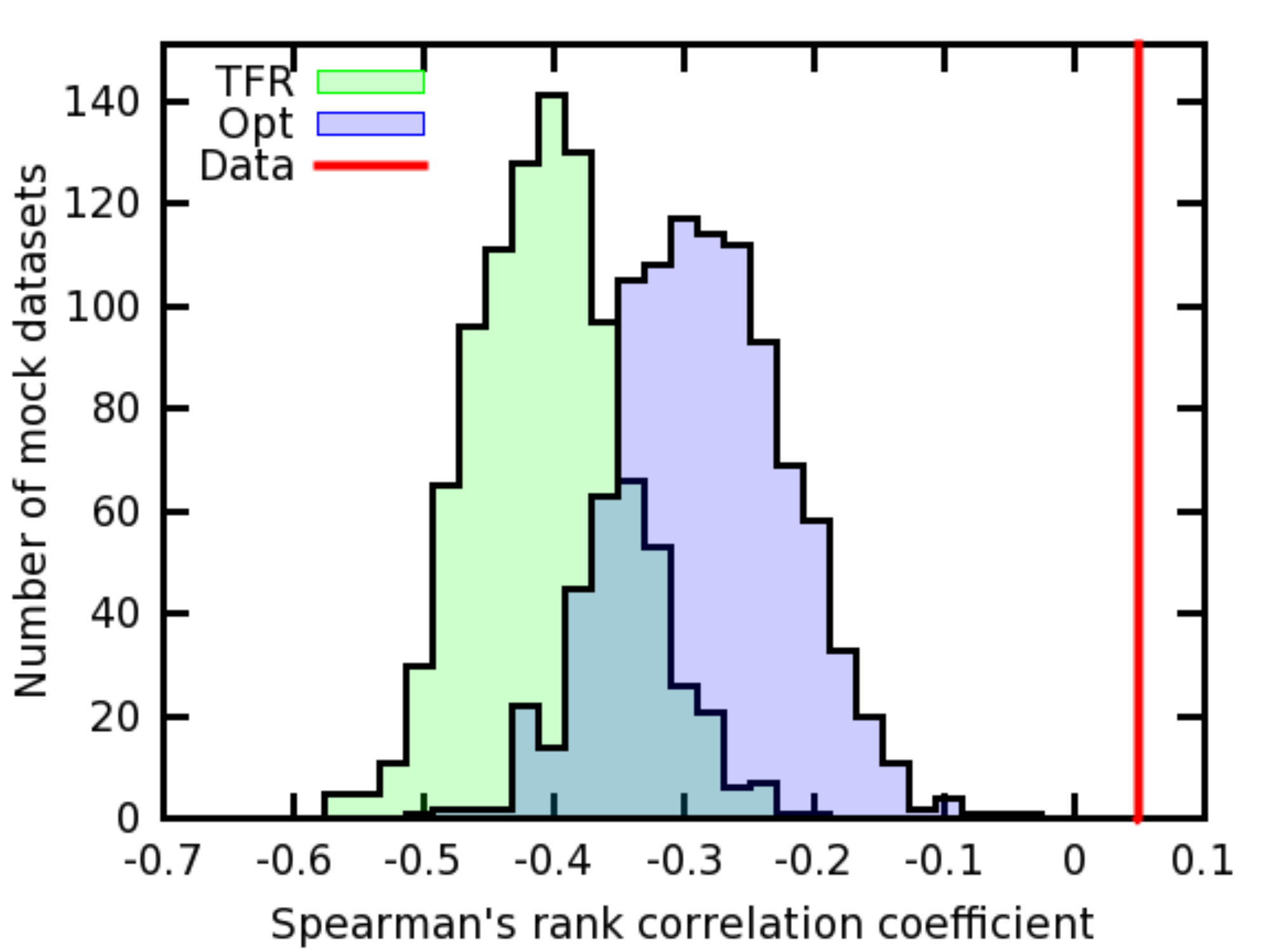}
    \label{fig:Res_Hist_Mnow}
  }
  \caption{Same as Fig.~\ref{fig:Residual_Total}, but for the $M_\mathrm{now}$ model.}
  \label{fig:Residual_Mnow_Total}
\end{figure}

We begin by taking the maximum-likelihood parameter values from Section~\ref{sec:TFR_Results} for a given AM proxy and generating 1000 mock data sets comprising radii and velocities of randomly chosen model galaxies with the same masses and velocity and radius uncertainties as the P07 sample. We then fit the TFR and MSR of each mock data set with separate power laws, and calculate the residual of each galaxy as the difference between its $\log(V_{80}/\mathrm{km\:s^{-1}})$ or $\log(R_\mathrm{d}/\mathrm{kpc})$ value and the expected value for this quantity given its mass and the power-law fit. We then plot the residuals against each other and display the resulting contour plot (overlaying the points from all 1000 mock data sets) in Figs~\ref{fig:Res_Comp} and~\ref{fig:Res_Comp_Mnow}, for the $V_\mathrm{peak}$ and $M_\mathrm{now}$ models respectively. We quantify the strength of the correlation by calculating the Spearman's rank correlation coefficient \SRCC\ for each mock data set and for the P07 data, which specifies how close the two variables are to being monotonically related to each other. $\rho=1$ corresponds to a perfectly monotonic correlation, $\rho=0$ to no correlation, and $\rho=-1$ to a perfectly monotonic anticorrelation.

Figs~\ref{fig:Res_Hist} and~\ref{fig:Res_Hist_Mnow} (green histograms) compare the distributions of the 1000 coefficients thereby obtained to the corresponding value for the real data. Table~\ref{tab:SRCC_Significance} (first column) displays the significance levels at which the hypothesis that the P07 data was drawn from the model populations can be rejected, according to this test, for each AM proxy. We see that the parameter values that give the best fit to the TFR generate at least $8\sigma$ tension with the data, showing that our framework fails to capture observed velocity--radius correlations.\footnote{The fact that the predicted radius residuals are significantly larger than those observed is a manifestion of the MSR scatter problem described in Section~\ref{sec:MSR_Results}. Since \SRCC\ is calculated only from the ranks of the variables and not their absolute values, it would be unchanged if all radius discrepancies were scaled down by the same factor.} The velocity-based AM models tend to produce a stronger anticorrelation and are therefore disfavoured by this test. The predicted value of $\rho$ is largely driven by the tail of galaxies assigned to haloes with relatively high concentration (hence large $V_{80}$) and low spin (hence low $R_\mathrm{d}$). This tail contains $\sim 10$ per cent of the model galaxies.

We note parenthetically that the discrepancy is even more pronounced for galaxies near the peak of the stellar mass--halo mass relation ($M_\star \approx 10^{10} M_{\odot}$), because at these masses the galaxy's relative contribution to the rotation velocity is maximized and the predicted anticorrelation of the residuals therefore strongest. Due to the paucity of P07 galaxies near this mass, however, we are not able to strengthen our statistical conclusions by focusing on this regime.

\begin{table}
  \begin{center}
    \begin{tabular}{l|c|c|}
      \hline
      &TFR params &Optimum params\\ 
      \hline
      $V_\mathrm{peak}$    &   12    &  4.7\\
      $V_\mathrm{acc}$    &   14    &  4.8\\
      $V_\mathrm{max}$    &   17    &  5.1\\
      $M_\mathrm{peak}$    &   8.1    &  5.6\\
      $M_\mathrm{acc}$    &   8.2    &  5.6\\
      $M_\mathrm{now}$    &   8.4    &  5.6\\
      \hline
    \end{tabular}
  \caption{Significance level (in $\sigma$) at which the value of the correlation coefficient \SRCC\ between the radius and velocity residuals of the P07 data is inconsistent with the theoretical prediction. The first column uses the parameter values that provide the best fit to the P07 TFR; the second uses the parameter values that yield the weakest anticorrelation between the radius and velocity residuals and hence minimize the significance of the disagreement with the data.}
  \label{tab:SRCC_Significance}
  \end{center}
\end{table}

We now investigate the possibility of resolving this discrepancy within our framework by adopting different parameter values. To this end, we repeat the MCMC analysis of Section~\ref{sec:method} but replace the likelihood function with one determined by the residual correlation as opposed to the TFR or MSR individually. In particular, we fit a normalized 1D Gaussian to the distribution of \SRCC\ values of our mock data sets, and take the likelihood to be the value of this Gaussian at the value of the real data.

\begin{equation}
P(\boldsymbol{p}|d) \propto \frac{1}{\sigma} \mathrm{e}^{-\frac{(\mu-0.0485)^2}{2 \sigma^2}}
\end{equation}

\noindent where $\boldsymbol{p}$ is the vector of model parameters, $d$ is the \SRCC\ value of the P07 data (0.0485), $\sigma$ is the standard deviation of a Gaussian fitted to the \SRCC\ values of the mock data sets, and $\mu$ its mean. As in Section~\ref{sec:TFR_Results}, we cap $j$ at 1.4.

For any AM parameter, we find that the predicted anticorrelation is minimized by maximizing $j$ and the AM scatter, and minimizing $\nu$. The discrepancy can be reduced to the $\sim5\sigma$ level in this way [see Table~\ref{tab:SRCC_Significance} (second column), Figs~\ref{fig:Res_Comp_ML},~\ref{fig:Res_Comp_Mnow_ML}, and Figs~\ref{fig:Res_Hist},~\ref{fig:Res_Hist_Mnow} (blue histograms)] at the price of predicting too low a TFR intercept and slope, and too high a scatter ($P$ value of TFR fit $\lesssim10^{-4}$). In particular, almost any mock data set drawn randomly from our theoretical population would contain at least one galaxy with large positive velocity residual and large negative radius residual (corresponding to low halo spin and high concentration relative to the mean at that stellar mass); that the real data do not do so is therefore highly surprising. Further discussion of this problem, including its relation to previous work and potential solutions, may be found in Section~\ref{sec:Residual_Problem}.

\subsection{Dependence on Environment}
\label{sec:environment}

Haloes in different large-scale environments have different mass accretion histories (e.g.~\citealt*{Merge1, Wechsler06, Merge2}), and galaxies with close neighbours may experience accelerated star formation due to tidal interactions~\citep[e.g.][]{Kennicutt87,MihosHernquist,Henriksen, Alonso04, Barton, Behroozi15}. In addition, the relationship between halo properties and galaxy star formation rates evolves differently for centrals and satellites.  One might therefore expect galaxies in different environments to lie on systematically different TFRs. However, several previous studies have found no evidence for such an effect. For example,~\citet{Masters_Environment} find no difference between the TFRs of galaxies lying in 31 groups and clusters of varying sizes, and identify no correlation between TFR residuals and projected distance from the cluster centre. The analysis of~\citet{Mocz} complements this work by splitting a large sample of SDSS galaxies into four bins of projected surface density and investigating their TFRs separately, finding no statistically significant difference attributable to environmental effects. The data set of~\citet{Courteau07} contains larger samples of both cluster and field spirals, and does not exhibit a significant difference between their respective TFRs.\footnote{However, there is some evidence for a systematic offset in the rotation velocities of satellite dwarf galaxies and isolated dwarf galaxies at low stellar mass~\citep*{Blanton_West_Geha}.}  Here we do not explicitly test for environment dependence in the P07 data but instead ask whether our model predicts a measurable dependence, and hence determine the information this relation could provide.

Rather than investigating environmental trends generically, we use as a measure of a galaxy's environment whether or not it lies within the virial radius of another halo (i.e. whether it is a `satellite' galaxy that resides in a `subhalo', or a `central' galaxy that resides in a distinct halo). For a given matching parameter, the standard AM approach makes a specific assumption about the galaxy growth for centrals versus satellites; we test the extent to which this is measurably manifest in the TFR. We begin by separating our model galaxies into satellites and centrals by splitting the halo population into subhaloes and main haloes as determined by \textsc{rockstar}. We adopt the maximum-likelihood parameter values from Section~\ref{sec:TFR_Results} and generate 1000 mock data sets, each with 156 galaxies at the stellar masses of the P07 galaxies. We then fit a separate power law to the TFRs of the satellite and central galaxies of a given mock data set, and record the distributions of the differences of their slope, intercept and scatter values. Finally, we calculate the significance levels at which no difference between the populations can be excluded by the model; in other words, we calculate the confidence with which one would reject the hypothesis that the real data was drawn from the theoretical populations if one were to perform this analysis on the P07 data and find no difference between the satellite and central TFRs.

\begin{figure}
  \subfigure[$V_\mathrm{peak}$]
  {
    \includegraphics[width=0.5\textwidth]{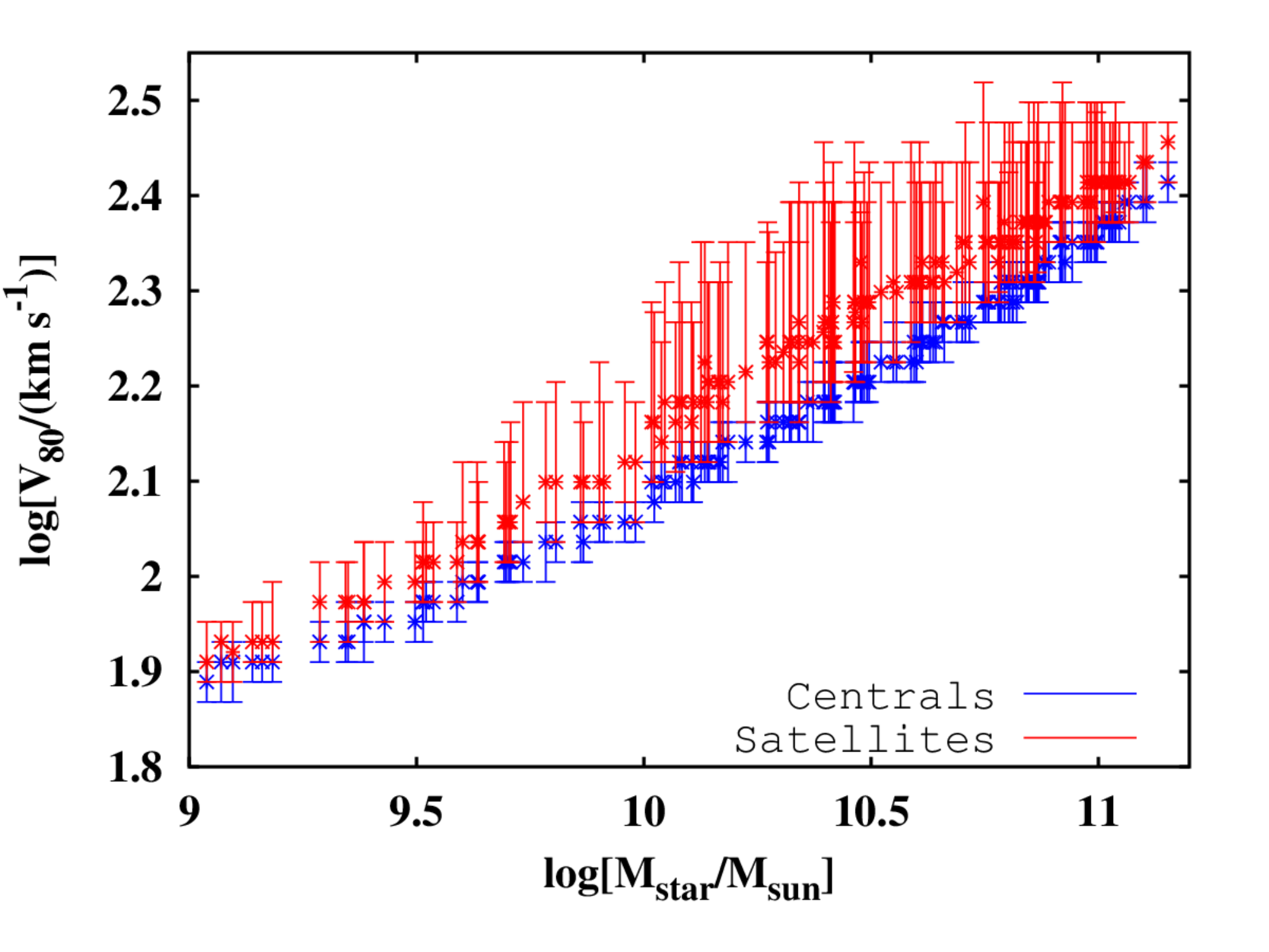}
    \label{fig:Sat_Vpeak}
  }
  \subfigure[$M_\mathrm{now}$]
  {
    \includegraphics[width=0.5\textwidth]{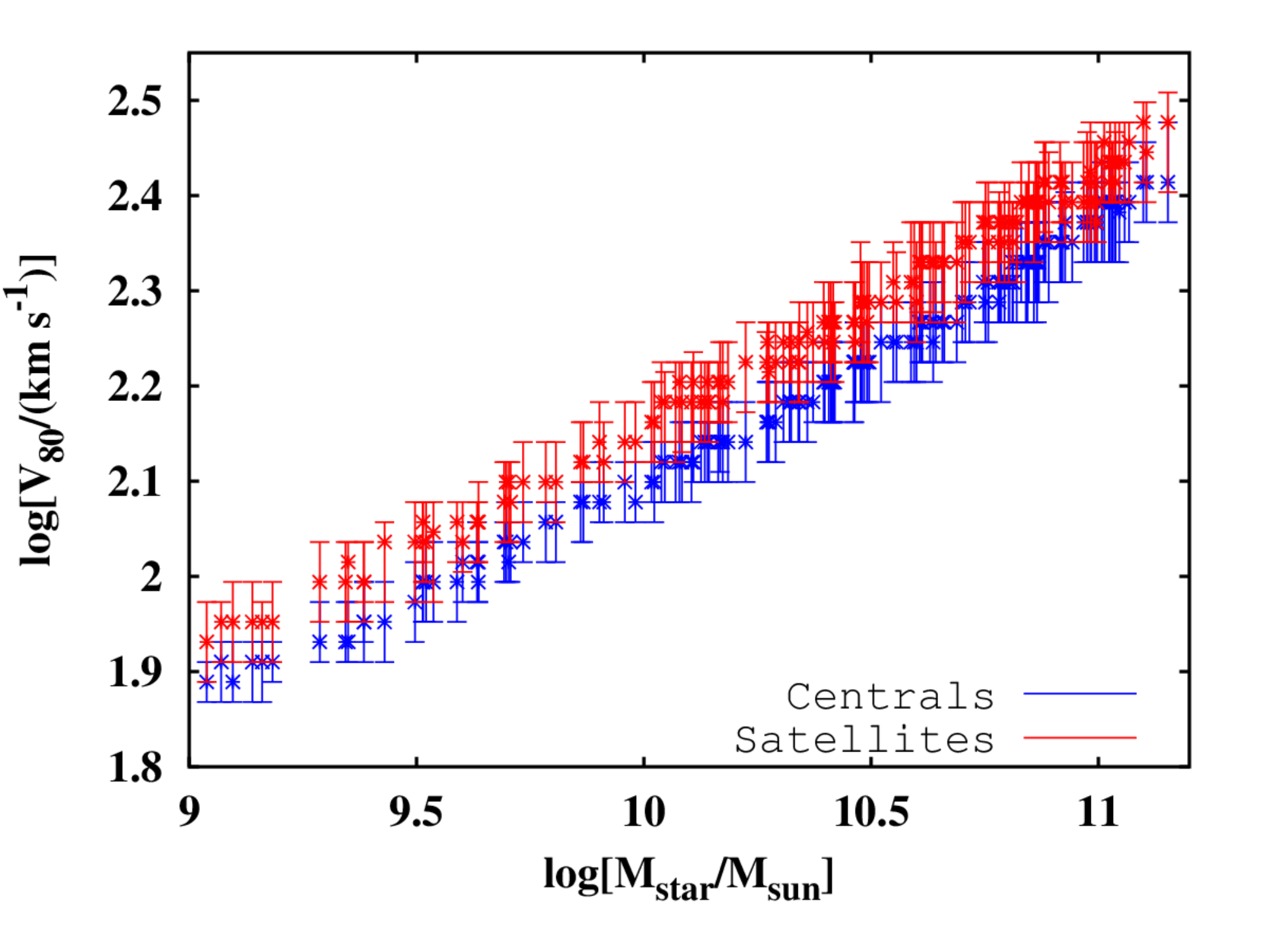}
    \label{fig:Sat_Mnow}
  }
  \caption{Comparison of the predicted satellite-only and central-only TFRs, using the best-fitting parameter values associated with the $V_\mathrm{peak}$ and $M_\mathrm{now}$ models respectively. The points show the median velocities of the model galaxies in 1000 mock data sets, and the error bars show the interquartile range. TFRs for satellite galaxies only are predicted to have somewhat higher normalizations and intrinsic scatters than those for central galaxies only.}
  \label{fig:Sat}
\end{figure}

Fig.~\ref{fig:Sat} shows the separate theoretical TFRs for the best-fitting $V_\mathrm{peak}$ and $M_\mathrm{now}$ models, and Table~\ref{tab:Sat_Significance} quantifies their differences (along with the other AM models). We find that although satellite galaxies have lower halo $V_\mathrm{max}$ than centrals of the same stellar mass (due to stripping of their outer regions during infall), they tend to have slightly higher $V_{80}$ values. This is because the rotation curves of subhaloes peak at smaller radii than distinct haloes (due to their higher concentration), closer to the $R_{80}$ values of the galaxies they host. In addition, satellite galaxies are predicted to exhibit a larger scatter in $V_{80}$, since the strength of stripping varies between subhaloes. This effect is larger for models that match to halo velocity than mass because these cause more satellite galaxies' stellar masses to be set before stripping. However, these differences are not significant: were one to perform this test on the real data and find no difference between the two TFRs (in particular their intrinsic scatters), the most one could say is that the $V_\mathrm{acc}$ and $V_\mathrm{peak}$ models are disfavoured at the $2\sigma$ level.

The main reason for the statistical insignificance of the difference between the theoretical satellite and central TFRs is the size of the sample: with satellite fractions around 10--35 per cent in this stellar mass range, the satellite-only mock data sets are small, resulting in a large variation among the parameters of the power-law fit between them. However, the framework we have set up is easily generalized to permit the analysis of a hypothetical data set with similar characteriztics to the P07 sample but more galaxies, allowing us to estimate the number of galaxies needed to definitively manifest a difference between the satellite and central TFRs. In Table~\ref{tab:Sat_Significance_2} we show the analogue of Table~\ref{tab:Sat_Significance} but for a data set comprising 700 galaxies with $\log(M_\star/M_\odot)$ uniformly distributed in the range of the P07 sample (9--11.2) and with $\log(V_{80}/\mathrm{km\:s^{-1}})$ uncertainties equal to the P07 RMS value (0.0346). In this case, we would expect to measure a significant ($> 3\sigma$) difference between the two TFRs for the velocity-based AM models, but not for the mass-based ones. We note that although there is a difference in the mean relations between the two populations, it is still too small to be robustly distinguished; the most significant difference remains in the scatter. The separation expected for the $V_\mathrm{peak}$ model is over $5\sigma$. Were such a homogeneous data set available which used assumptions and definitions consistent with our model, an analysis of the present type would yield useful information about the general validity of our framework, and might provide another way to distinguish between the various AM schemes.

We caution that our methodology assumes that the AM process itself makes no distinction between the satellite and central populations, so that they can be cleanly separated after the matching has been performed. It is also possible that satellite galaxies should instead be assigned to different haloes at the outset, or that effects not taken into account by AM (e.g. significant continued star formation after accretion, accelerated star formation due to galaxy--galaxy interactions, or systematic variations in halo shape with environment) are significant. In this case our model would need to be generalized for agreement with the data to be expected.

\section{Discussion}
\label{sec:discussion}

We have constructed a general framework for assessing whether AM-based galaxy formation models are capable of reproducing the detailed characteriztics of the observed Tully--Fisher and mass--size relations, and determined what information can be gleaned from them. In Section~\ref{sec:constraints} we compare our parameter constraints to those obtained by studies in the literature. We then provide in Sections~\ref{sec:MSR_Problem} and~\ref{sec:Residual_Problem}, respectively, a detailed discussion of the two major problems that our analysis has revealed: the theoretical MSR scatter is too large, and the velocity and radius residuals are too strongly anticorrelated. In particular, we show that the solution to both of these problems (in addition to several others already known) requires a correlation between stellar surface density and enclosed dark matter mass that is not present in models currently under investigation.

\subsection{Parameter Constraints and Comparison with the Literature}
\label{sec:constraints}

The simplest AM models (which assume adiabatic or near-adiabatic contraction and do not take account of selection effects) give galaxies velocities that are too high for their stellar mass. This result is present in several other TFR studies (e.g.~\citealt{D07, TG, Desmond, McGaugh_Complete}), and its quantitative solution likely requires not only that haloes contract less strongly than prescribed by the adiabatic formalism (in agreement with~\citealt{D07}), but also that observational TFR studies select galaxies with systematically low velocities for their stellar mass. In other words, rotation-supported spirals must live in low-mass or low-concentration haloes at fixed stellar mass. This trend has been suggested by other theoretical considerations~\citep{Puebla} in addition to measurements of satellite kinematics~\citep{Wojtak}, but our model allows us to investigate quantitatively the required magnitude of the effect and its dependence on other galaxy formation parameters. In the context of the TFR, the required magnitude of selection is degenerate with $j$, the ratio of the specific angular momentum of the disc and halo. Improving our knowledge of $j$ would therefore afford a tighter constraint on the selection factor. We have checked that the difference between the haloes of `spirals' and `other types' suggested by our analysis is not in obvious disagreement with the measurements of~\citet{Wojtak}, but it remains to be seen whether the required magnitude of the difference between the selected subsample and the full population is theoretically reasonable and can be corroborated in detail by independent measurements.

Reproducing the small scatter in the stellar mass TFR requires that little scatter is associated with AM itself -- that is, the stellar mass is in near-monotonic correspondence with halo mass or velocity. In particular, the scatter in this relation must be less than around 0.35 dex ($2\sigma$) for all models tested. This is consistent with the results of~\citet{Reddick}, who find that an AM scatter of $\sim$ 0.2 dex is required for agreement with satellite fraction and galaxy clustering measurements. We have also tried focusing attention on the least massive P07 galaxies in order to set a limit on the AM scatter that applies specifically at low stellar mass, where constraints from alternative methods are weak. We found that the $2\sigma$ upper limit on the scatter is around 0.7 dex for a subset of the P07 data with median $\log(M_\star/M_\odot)$ of 9.6. However, this constraint is weaker than that obtained from the full data set only because fewer galaxies are used, and we are unable to probe even lower stellar masses due to the paucity of galaxies in the P07 sample with $\log(M_\star/M_\odot) < 9.5$. Applied to a future TFR data set with more galaxies, however, this method should clearly reveal whether AM scatter values required for clustering and the TFR are consistent, and set robust constraints on the scatter at low stellar mass.  Constraining the scatter at low mass is difficult with clustering because the bias function is not a strong function of halo mass in this regime, and from group statistics due to the scarcity of samples of robustly identified groups with centrals of $M_\mathrm{vir} < 10^{12}\:M_\odot$.

Turning now to the MSR, we see from Table~\ref{tab:Constraints_Rad} that the ratio of the specific angular momentum of a disc and its halo ($j$) must be in the approximate range 0.6--1.2 to ensure the correct normalization of the relation. This result suggests that the baryonic material retains most of its primordial angular momentum when it collapses into a disc, and further reinforces the need to resolve the overcooling problem that has plagued hydrodynamical galaxy simulations~\citep{LowJ_2, LowJ_1, Dekel_Maller, Scannapieco}. Our result is in rough agreement with~\citet{Kravtsov_Radius} who use a simplified version of the~\citet{MMW} method to compare the theoretical and observed sizes of galaxies with a range of morphologies across a wide range of stellar mass. We anticipate that our constraint will provide a useful touchstone for prescriptions of baryonic dissipation and infall implemented in hydrodynamical galaxy simulations.

\begin{table*}
\parbox{.45\linewidth}{
\centering
\begin{tabular}{l|c|c|c|}
      \hline
      &Slope &Intercept &Scatter\\ 
      \hline
      $V_\mathrm{peak}$    &      0.23 & 1.6 & 2.3\\
      $V_\mathrm{acc}$    &      0.29 & 1.5 & 2.0\\
      $V_\mathrm{max}$    &      0.12 & 1.2 & 1.3\\
      $M_\mathrm{peak}$    &      0.48 & 1.1 & 0.95\\
      $M_\mathrm{acc}$    &      0.34 & 1.0 & 0.44\\
      $M_\mathrm{now}$    &      0.10 & 0.82 & 0.61\\
      \hline
\end{tabular}
\caption{Table of significance levels (in $\sigma$) at which the satellite-only and central-only TFRs of the P07 data set are expected to be distinguishable, according to each power-law parameter, using the parameter values that provide the best fit to the P07 TFR. AM models that match to halo velocity predict a larger difference between the two TFRs than those that match to halo mass, although the difference is not significant in any case.}
\label{tab:Sat_Significance}
}
\hfill
\parbox{.45\linewidth}{
\centering
\begin{tabular}{l|c|c|c|}
\hline
      &Slope &Intercept &Scatter\\ 
      \hline
      $V_\mathrm{peak}$    &      1.9 & 1.8 & 5.1\\
      $V_\mathrm{acc}$    &      1.8 & 1.7 & 4.2\\
      $V_\mathrm{max}$    &      1.0 & 1.3 & 3.1\\
      $M_\mathrm{peak}$    &      1.4 & 1.0 & 2.2\\
      $M_\mathrm{acc}$    &      1.3 & 1.1 & 1.4\\
      $M_\mathrm{now}$    &      0.025 & 1.0 & 0.41\\
\hline
\end{tabular}
\caption{Same as Table~\ref{tab:Sat_Significance} but for a hypothetical data set comprising 700 galaxies with stellar masses uniformly distributed in the range $9 < \log(M_\star/M_\odot) < 11.2$. In this case, the satellite and central-only TFRs are expected to be distinguishable on the basis of their intrinsic scatters (at the $>3\sigma$ level) for all of the velocity-based AM models.}
\label{tab:Sat_Significance_2}
}
\end{table*}

\subsection{The Excess Intrinsic Scatter Predicted for the Mass--Size Relation}
\label{sec:MSR_Problem}

A significant problem identified in this work (shown most clearly in Fig.~\ref{fig:sigma_Rad}) is that our model predicts that the MSR scatter should be significantly larger than is observed. The smallest scatter we are able to obtain is around 0.28 dex (by matching to $M_\mathrm{now}$ and setting AM scatter = 0, selection factor = $\pi/2$), while that of the P07 data is $0.17 \pm 0.01$ dex. The theoretical scatter in disc scalelength $R_\mathrm{d}$ at fixed stellar mass is set by five factors, which are assumed to be uncorrelated.

\begin{enumerate}

\item{} The scatter in halo spin $\lambda$ at fixed virial mass (at fixed $j$, $\lambda$ sets a galaxy's angular momentum and hence regulates its size).

\item{} The scatter in halo concentration at fixed virial mass (more concentrated haloes generate larger gravitational potential gradients in the inner regions where the disc is situated, boosting the galaxy's rotation velocity and hence reducing its radius at fixed angular momentum).

\item{} The scatter in present-day halo mass at fixed stellar mass (set by the AM prescription).

\item{} The scatter in bulge mass fractions.

\item{} The impact of selection effects.

\end{enumerate}

Of these, the third can be eliminated by adopting the $M_\mathrm{now}$ AM model with zero scatter, the fourth is small, and the first contributes a fixed 0.27 dex (in our model, $R_\mathrm{d} \propto \lambda$). This is already more than is allowed for by the data. When galaxies are selected on the basis of their $V_{80}$ values, as we have assumed, selection effects are capable of removing around 0.21 dex in quadrature. Scatter in concentration largely makes up the remaining $\sim0.22$ dex. These results agree with the broad-brush analysis of~\citet{Kravtsov_Radius}, who finds the observed relation to have a scatter of 0.2 dex (the difference to P07's 0.17 dex being largely attributable to the subtraction of observational uncertainties in the latter), but does not include the additional theoretical scatter expected from concentration, stellar mass fraction or bulge mass fraction, and does not consider selection effects.\footnote{\citet{Courteau07} find an even smaller MSR scatter of $\sim 0.14$ dex, although the use of a different fitting method, and luminosity as opposed to stellar mass, make this result not directly comparable to ours.}

This result implies that the simplest angular momentum-based methods for assigning disc sizes do not agree with the observed scatter in the MSR. This problem was noted by~\citet{deJong} and further quantified by~\citet{Gnedin_new}, who noted that the width of the $\lambda$ distribution \textit{of discs} can be no larger than $\sim 0.17$ dex. The discrepancy would be exacerbated by introducing scatter into the ratio of the specific angular momentum of the disc and halo (presumably expected at some level due to differences in their formation). We now describe three specific ways in which this problem could be resolved whilst retaining the notion that galaxy sizes are set by angular momentum partition between the disc and halo.

The first is to argue that selection effects in the data play a more significant role than is allowed for in our model. In particular, we could reduce the predicted MSR scatter by more than 0.21 dex if we assume that the quantities by which the P07 galaxies were chosen for inclusion are correlated more strongly with galaxy size or host halo spin than with $V_{80}$. For example, it has been suggested that low angular momentum discs are more prone to the development of a bar instability, and may therefore become early-type galaxies (e.g~\citealt*{deJong_PhD,Dalcanton};~\citealt{MMW}). Three points, however, argue against this as a solution to the problem. First, we have tried constructing a simple toy model in which our model galaxies are selected not on the basis of $V_{80}$ alone but rather a linear combination of $V_{80}$ and $R_\mathrm{d}$, and have not been able to reduce the predicted MSR scatter to the observed value while preserving agreement with the TFR. It is possible to lower the MSR scatter without affecting that of the TFR by preferentially eliminating low-spin haloes at fixed $M_\star$ and $V_{80}$; however, conditional selection of this type is somewhat fine-tuned and we do not investigate it further here.  A more physically motivated prescription that gives similar correlations could be interesting to investigate in future work. Secondly, a resolution of the MSR scatter problem in terms of selection effects would imply that a galaxy sample chosen using a broader set of selection criteria would exhibit significantly larger scatter than the P07 sample. As noted above, however, the~\citet{Kravtsov_Radius} sample includes galaxies with a range of morphologies yet does not find an intrinsic MSR scatter discrepant with that of P07. Finally, reducing the $\sim0.35$ dex predicted scatter to the required $0.17$ dex by selecting only the $\sim50$ per cent with largest radii would impart a strong skew on the resulting distribution, which would therefore have a tail towards high radius. This is not apparent in the observational data.\footnote{This argument applies also to the TFR, and suggests that future data sets with many more galaxies than P07 may be able to constrain the impact of selection by means of the third moments of the velocity and radius distributions in bins of stellar mass.}

Second, one might attempt to resolve this problem by using the fact that star formation proceeds faster in higher-density regions~\citep{Schmidt, Kennicutt}. This suggests that at fixed baryonic mass, larger discs would have smaller stellar masses, which could reduce the width of the size distribution at fixed \textit{stellar} mass relative to the distribution at fixed \textit{total baryonic} mass. This mechanism could therefore be tested by comparing the scatters of the $M_\star$--$R_\mathrm{d}$ and $M_\mathrm{baryon}$--$R_\mathrm{d}$ relations. However, it cannot be accommodated within the basic AM framework where stellar mass is set purely by halo mass or velocity. 

A final way of lowering the predicted scatter in the MSR is to postulate a correlation between a galaxy's stellar surface mass density (or equivalently its size at fixed stellar mass) and the mass or concentration of the halo in which it lives. Suppose that at fixed stellar mass, larger-than-average galaxies are made to live in more massive haloes. The effect of this would be to increase the rotational velocity of such galaxies. At fixed disc angular momentum, this would reduce those galaxies' radii, moving them closer to the average. Thus the magnitude of radius discrepancies would be decreased. An equivalent way to view this is to suppose that at fixed halo mass, larger galaxies are made to have larger stellar mass. This would push a larger-than-average galaxy to the right on the $M_\star$--$R_\mathrm{d}$ plane, reducing the difference between its size and the average for that stellar mass. From either perspective, we see that the intrinsic scatter in the MSR would be reduced.

This is not the only observation which suggests a correlation between surface mass density and enclosed dark matter mass, and evidence for such a correlation exists for early as well as late-type galaxies. It has been shown by~\citet{Sonnenfeld} using weak lensing and kinematical observations of ellipticals that the inferred dark matter mass within the central 5 kpc anticorrelates with surface brightness. In addition, three very massive and high surface brightness ellipticals have been observed to have velocity dispersions that fall off in a Keplerian fashion from the centre of the galaxy, indicating that the enclosed dark matter is dynamically insignificant~\citep{Romanowsky}. The correlation has been documented for a wide range of galaxy types in~\citet{Gentile}, and we show in Section~\ref{sec:Residual_Problem} that the observed relation between the residuals of the TFR and MSR is yet another manifestation of this effect. The most pristine form of the correlation, however, is the `mass discrepancy--acceleration relation' for spirals, which suggests a tight correlation between baryonic surface mass density and enclosed dark matter mass~\citep{Sanders_MDAcc, McGaugh_MDAcc, Tiret_Combes, Famaey_McGaugh}.\footnote{Indeed, it is evidence for such a correlation that gives Modified Newtonian Dynamics (MOND) its phenomenological efficacy.}

We identify two mechanisms whereby a relation of this form could be introduced into our model. The first is to postulate a correlation between the AM scatter (the scatter in stellar masses of galaxies associated with haloes of fixed mass or velocity) and halo spin, which is the main determinant of galaxy size. In particular, we might suppose that the galaxies with the greatest stellar masses for their halo mass live in the most rapidly rotating haloes, tending to make them larger than average. In this case, increasing the AM scatter would \textit{decrease} the scatter in the MSR for the reason described above. We have investigated this possibility using a simple toy model, which suggests that an AM scatter of around 0.3 dex may be required to bring the MSR scatter down to the observed level.\footnote{Alternatively, one could anticorrelate abundance matching scatter with halo concentration. This would require a larger AM scatter to be effective because disc size at fixed stellar mass is affected less strongly by concentration than spin.} This is inconsistent with the conditional SMF measurements of~\citet{Reddick}. Furthermore, a perfect correlation between AM scatter and halo spin is unlikely, and a noisy correlation would require an even larger AM scatter to be effective. Nevertheless, we cannot rule out this scenario as potentially offering a solution to this problem.

An alternative way to enforce a correlation between stellar surface mass density and dark matter mass would be to perform `conditional abundance matching'~\citep{Hearin_Watson, CAM}. In this case, one would no longer assign disc sizes via angular momentum considerations, but instead use directly the observed distribution of disc sizes as a function of stellar mass. The radii of model galaxies would be assigned in monotonic or near-monotonic correspondence with the present-day mass or concentration of the haloes in which they live. The disadvantage of this method is that it loses the theoretical motivation of the~\citet{MMW} model (that the baryons and dark matter share angular momentum), but, if successful, could provide an important clue to understanding the galaxy--halo connection.

\subsection{The Excessive Anticorrelation between Radius and Velocity Residuals}
\label{sec:Residual_Problem}

The second issue illuminated by our work is that the AM plus disc size models we consider predict too strong an anticorrelation between a galaxy's radius discrepancy (the difference between its disc scalelength and that expected given its stellar mass and a power-law fit to the full MSR) and its velocity discrepancy (the analogue for the TFR). This is illustrated by Figs~\ref{fig:Residual_Total} and~\ref{fig:Residual_Mnow_Total}. We have demonstrated that models which match the TFR predict an anticorrelation too strong at the $\gtrsim8\sigma$ level, and the discrepancy can be reduced (though never below $\sim5\sigma$ within the present framework) only by adopting parameter values inconsistent with the TFR alone. This suggests that models of this type, at least with standard assumptions about the galaxy--halo connection, are incapable of capturing in detail the correlations observed between galaxies' sizes and rotation speeds. Here we compare our conclusion to others in the literature, and provide suggestions for how the discrepancy might be resolved.

One of the best-known studies of the correlation of velocity and radius residuals is~\citet{CR}, in which it is argued that the observed level of correlation requires that the disc contributes ($60 \pm 10$) per cent of the total rotation velocity at 2.2 disc scalelengths. In this case, according to their model, the increase in rotation velocity due to the baryons as the disc's size decreases at fixed stellar mass is fully compensated for by a reduction in the amount of dark matter mass within $R_{2.2}$. In our model the discs contribute on average an even smaller fraction of the total rotation velocity (due to the fact that AM induces stellar mass fractions far below the cosmic baryon fraction), yet result in a residual correlation that is in stronger disagreement with the data. We identify three reasons for this difference.

First, the~\citet{CR} model uses the fact that halo spin is uncorrelated with concentration to argue that galaxy size is too. However, halo spin sets only the angular momentum of the disc, and does not fully specify its size; in more concentrated haloes, discs rotate faster and are therefore smaller at fixed angular momentum. This effect intensifies the predicted anticorrelation of the radius and velocity residuals: not only is the galaxy's own contribution to the velocity larger when the galaxy is smaller, the halo's is likely to be too. Secondly, a significant fraction of our model galaxies are larger than the fiducial 3 kpc assumed by~\citet{CR}, and we measure the rotation velocity at $R_{80}$ (about 3$R_\mathrm{d}$) instead of $R_{2.2}$. This means that we are probing a larger physical radius, often around the peak of the halo's rotation curve, and hence the halo does not provide significantly greater rotation velocity for larger discs. Finally, the degree of residual anticorrelation found in the sample to which~\citet{CR} compare their results is significantly larger than ours (the slope is $-0.18 \pm 0.05$ in their `MAT' sample). The statistically insignificant correlation observed in the P07 sample appears more representative of TFR studies in general~\citep{verheijen, McGaugh_Residuals, Courteau07, Reyes}. Furthermore, our analysis improves upon~\citet{CR} by modelling all correlations between disc and halo properties (stellar mass, disc scalelength, halo mass, concentration, and spin), including their scatter, in a fully self-consistent and integrated way that relies only on the AM ansatz and the assumption that halo spin regulates galaxy size. We conclude that the~\citet{CR} solution to the residual correlation problem is not practicable.\footnote{Other studies in the literature attempt to eliminate the residual anticorrelation by postulating and fine-tuning a correlation between stellar mass and baryonic surface density induced by a star formation threshold (e.g.~\citealt{FA,D07}). Such models nevertheless predict a non-zero anticorrelation of the residuals, so it is unclear whether they are in statistical agreement with the data. Furthermore, they predict that \emph{baryonic} TFR residuals \emph{would} be significantly anticorrelated, in conflict with some observations~\citep{McGaugh_Residuals}. In any case, this mechanism is not available within the basic AM framework.}

Several other methods by which one might attempt to resolve the discrepancy can be seen to lead to or exacerbate conflict with other observations. For example, making discs larger at fixed stellar mass would reduce the galaxy's contribution to $V_{80}$ but generate disagreement with the normalization of the MSR. Increasing the dark matter mass in the central regions would aggravate the `cusp/core' and, potentially, the `too big to fail' problems. Finally, reducing stellar mass fractions would intensify the missing baryon problem~\citep{MissingBaryon_2, MissingBaryon_1}. Furthermore, implementing these would likely require a break with the AM ansatz. Alternatively, one might hypothesize that selection criteria in typical TFR surveys conspire to reduce the correlation by eliminating small galaxies with high velocity and large galaxies with low velocity. This is possible but not well motivated, and is not guaranteed to be successful if the scatter in the residual correlation is insufficiently large.

The problem could be alleviated by imposing a correlation between stellar surface density and halo mass or concentration, as described in Section~\ref{sec:MSR_Problem}. As discussed above, in models like ours which set galaxy size by angular momentum considerations, smaller galaxies tend to reside in more concentrated haloes. However, the observed lack of correlation of the residuals indicates that at fixed stellar mass, smaller galaxies must reside in less massive or concentrated haloes so that the increase in rotation velocity due to the galaxy is offset by a decrease in that due to the halo. This conclusion was also reached by~\citet{Gnedin_new}, who found agreement with the observed correlation of the residuals when central disc-to-halo mass fraction $m_\mathrm{d}$ and surface stellar mass density $\Sigma_\star \equiv M_\star\:R_\mathrm{d}^{-2}$ were related by $m_\mathrm{d} \propto \Sigma_\star^{0.65}$ (assuming standard adiabatic contraction). It remains to be seen whether either of the strategies described in Section~\ref{sec:MSR_Problem} for introducing the required correlation within the AM framework (or indeed any other) are capable of simultaneously resolving both problems identified here, reproducing observations of ellipticals, and receiving a priori theoretical support.

\section{Suggestions for Further Work}
\label{sec:suggestions}

We identify several ways in which this study could be usefully extended or its results used to further improve our knowledge of galaxy formation.

\begin{itemize}

\item{} We have used the TFR and MSR to constrain the ratio of disc to halo specific angular momentum, the response of the halo to disc formation, the scatter in the relation between stellar mass and halo mass or rotation velocity, and the impact of selection criteria. These constraints require theoretical motivation from detailed hydrodynamical simulations of galaxy formation, and should be checked for consistency against other observations (e.g. the Faber--Jackson relation and Fundamental Plane of ellipticals) using a similar framework.

\item{} Enlarging the observational TFR data set could significantly improve constraints on the scatter between galaxy and halo properties. At present the AM scatter is better constrained by clustering than by the TFR, but a larger TFR sample would allow a stronger test of consistency between the two methods. Extending the TFR sample to low mass would enable the mass dependence of the scatter to be investigated in a regime that is not accessible to clustering measurements. An analysis of the baryonic TFR along the lines described here would, when feasible, likely provide even more constraining power.

\item{} As described in Section~\ref{sec:environment}, a sample of around 700 galaxies split into satellites and centrals should reveal an significant difference between the TFRs of the two populations if galaxies populate haloes according to a velocity-based AM scheme. Quantifying this difference would provide a direct handle on potential differences between the galaxy--halo connections of centrals and satellites.

\item{} We have argued that several problems associated with galaxy phenomenology require a correlation between galaxy stellar surface mass density and halo mass or concentration that is not induced naturally by simple one-parameter AM models. This could be produced either by correlating AM scatter with halo spin or concentration, or by using conditional abundance matching to correlate galaxy size with halo mass or concentration at fixed stellar mass. An obvious next step is to implement such models to see whether they are capable of resolving the problems described in Section~\ref{sec:discussion}, and, if so, what strength of correlation is required. A correlation between disc scalelength and halo mass would suggest that galaxies of different size but the same stellar mass would cluster differently, which could be tested in SDSS data. A correlation between halo concentration and galaxy size would also likely imply some degree of galactic conformity in the theoretical population, which could be compared to observations~\citep*{Conformity, Hearin_Conformity, Paranjape}. It would then be incumbent upon hydrodynamic simulations or detailed analytic calculations to explain how the correct correlation between stellar surface mass density and dark matter mass comes about.

\end{itemize}

\section{Summary and Conclusions}
\label{sec:conc}

We have explored a range of plausible assumptions about the galaxy--halo connection, implementing AM models that have previously been shown to match a number of observed galaxy statistics in the local Universe. We investigated the extent to which these models are able to reproduce the stellar mass Tully--Fisher and mass--size relations for local galaxies, using a homogeneously selected data sample from the SDSS. Our main findings are as follows.

\begin{itemize}

\item{} A model that matches galaxy stellar mass to maximum halo velocity (either today, at accretion, or at the point in its merger history when this quantity was largest) can fit the normalization of the stellar mass TFR in the range $9 < \log(M_\star/M_\odot) < 11.2$ only if the halo expands in response to disc formation. (A small amount of contraction is permitted if $\Omega_\mathrm{m}$ and $\sigma_\mathrm{8}$ take their lowest values allowed by the Planck uncertainties.) For a model that matches instead to halo mass, no contraction or expansion is favoured. In either case, `standard' adiabatic contraction is ruled out at the $\gtrsim 4\sigma$ level.

\item{} The small intrinsic scatter of the stellar mass TFR places an upper limit on the AM scatter ($\lesssim$ 0.35 dex at 2$\sigma$) and suggests that galaxies selected for observational TFR studies preferentially populate the low end of the rotational velocity function at fixed stellar mass. Once these constraints are satisfied, an AM model that matches to any of the properties considered is able to successfully reproduce the observed TFR.

\item{} The slope and normalization of the MSR can be matched simultaneously with the TFR, and constrain the ratio of the specific angular momenta of the baryonic and dark matter mass to be in the approximate range $0.6-1.2$. However, the intrinsic scatter predicted for the MSR is significantly larger than that in the data even when AM scatter is switched off (at least 0.28 dex, compared to 0.17 dex). This appears to be a fundamental consequence of setting galaxy size proportional to halo spin, which itself has significant scatter.

\item{} The observational data show no statistically significant correlation between the residuals of the MSR and TFR. However, the model predicts an anticorrelation because smaller galaxies at fixed stellar mass generate a larger rotation velocity. The Spearman's rank correlation coefficient is found to be at least $8\sigma$ discrepant with the data for parameter values which generate a good fit to the TFR. This discrepancy can be reduced to around $5\sigma$ only by adopting parameter values that do not generate agreement with the TFR alone.

\item{} Both of the above problems (in addition to several others already known) imply a correlation between stellar surface mass density and enclosed dark matter mass which is stronger than that induced by the current standard range of AM models. This could be introduced by correlating AM scatter ad hoc with halo spin or concentration, or by using `conditional abundance matching' to connect galaxy size to halo mass or concentration at fixed stellar mass. It remains to be seen whether a model along either of these lines can be theoretically motivated and used to successfully resolve the discrepancies described here.

\item{} In the context of our model, splitting the data sample we have investigated into satellite and central galaxies would not be expected to yield distinguishable TFRs. However, we predict that a sample with uniform selection and at least 700 galaxies could be resolved into separate satellite and central TFRs at the $>3\sigma$ level if galaxies populate haloes according to a velocity-based AM scheme. Analysis of a future data set in the way we describe would provide another crucial test of the AM framework, and allow us to further constrain its degrees of freedom.

\end{itemize}

\section*{Acknowledgements}

We are grateful to Matthew Becker, Peter Behroozi, James Bullock, Aaron Dutton, Simon Foreman, Yu Lu, Ari Maller, Stacy McGaugh, Zachary Slepian, and Frank van den Bosch for helpful discussions and comments, and to Matthew Becker for running and making available the simulations used in this work. This work was supported in part by the US Department of Energy contract to SLAC no. DE-AC02- 76SF00515. The simulations used here were run using computational resources at SLAC as well as resources of the National Energy Research Scientific Computing Center, a DOE Office of Science User Facility supported by the Office of Science of the U.S. Department of Energy under contract no. DE-AC02-05CH11231.

\appendix

\section{Model Fits to the TFR and MSR}
\label{sec:app_plots}

We describe and explain here the effect of each of the individual model parameters on the TFR (see Fig.~\ref{fig:PointsPlots}) and MSR (see Fig.~\ref{fig:PointsPlots_Rad}).

\begin{itemize}

\item{} Reducing $j$ increases the normalization and scatter of the TFR, and decreases the normalization of the MSR. Lower $j$ corresponds to smaller disc angular momentum, and hence smaller size for fixed values of other disc and halo parameters. This increases the disc's contribution to the rotation curve, which contributes additional scatter to $V_{80}$ according to the scatter in halo mass, concentration, and spin, and stellar bulge-to-disc ratio, at fixed stellar mass.

\item{} Increasing $\nu$ causes more dark matter mass to be pulled within $R_{80}$, increasing the normalization of the TFR. This lowers slightly the normalization of the MSR, since faster-rotating discs are smaller when other properties are fixed, and increases its scatter.

\item{} Increasing the AM scatter generates additional variation in halo mass and concentration at fixed stellar mass, increasing the scatter in the TFR and MSR. It also reduces the small amount of curvature in the predicted TFR, and decreases the MSR slope.

\item{} Reducing the selection factor weakens the preference for low-$V_{80}$ galaxies to be selected at fixed stellar mass. This increases the normalization of the TFR, reduces the normalization of the MSR (high-$V_{80}$ galaxies tend to be small), and increases the scatter in both.

\end{itemize}

\renewcommand{\thefigure}{A1}

\afterpage{

\begin{figure}
  \centering
  \hspace*{-0.9cm}
  \includegraphics[height=4.3cm, width=0.38\textwidth]{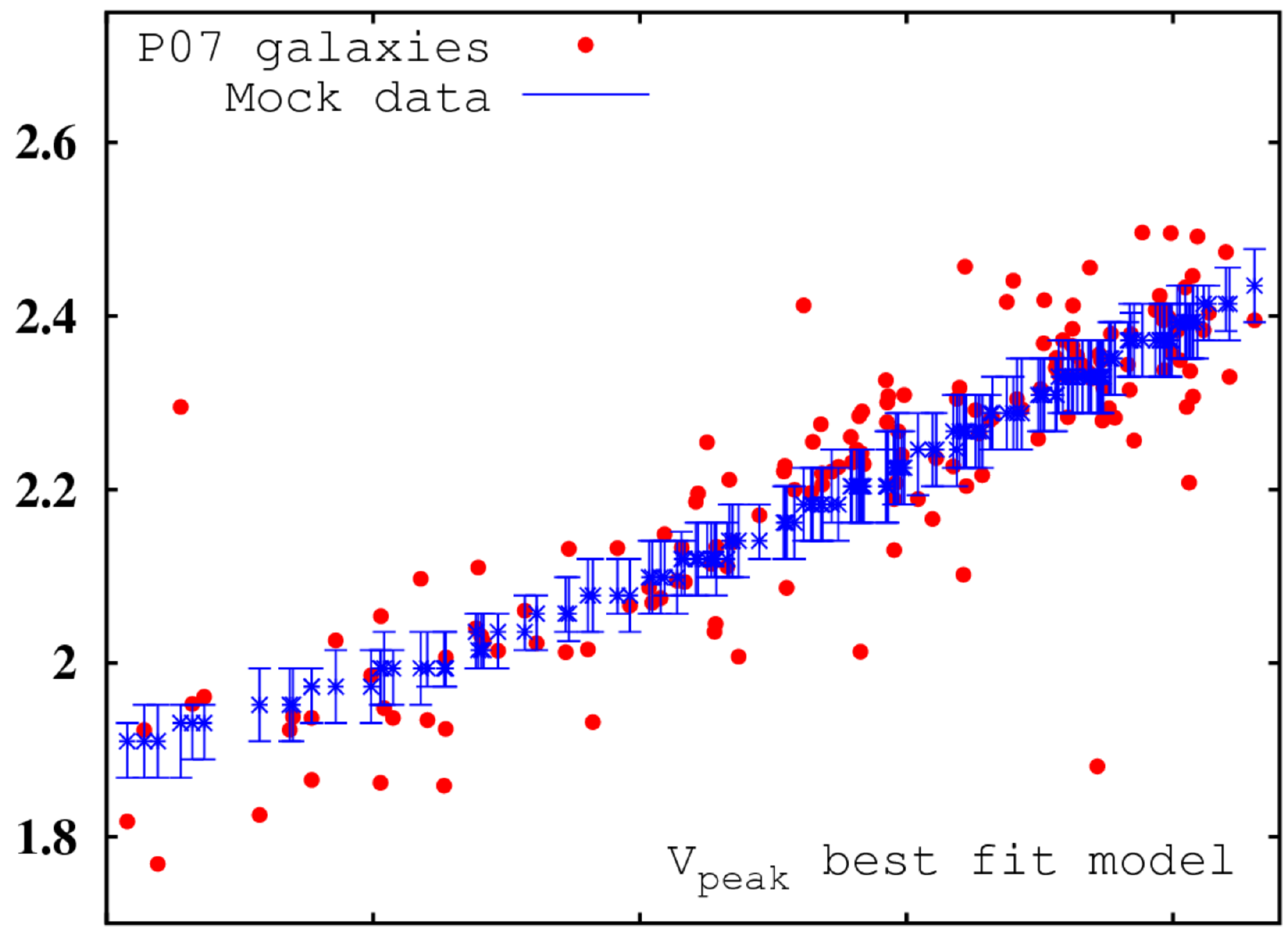}
  \hspace*{-0.9cm}
  \includegraphics[height=4.3cm, width=0.38\textwidth]{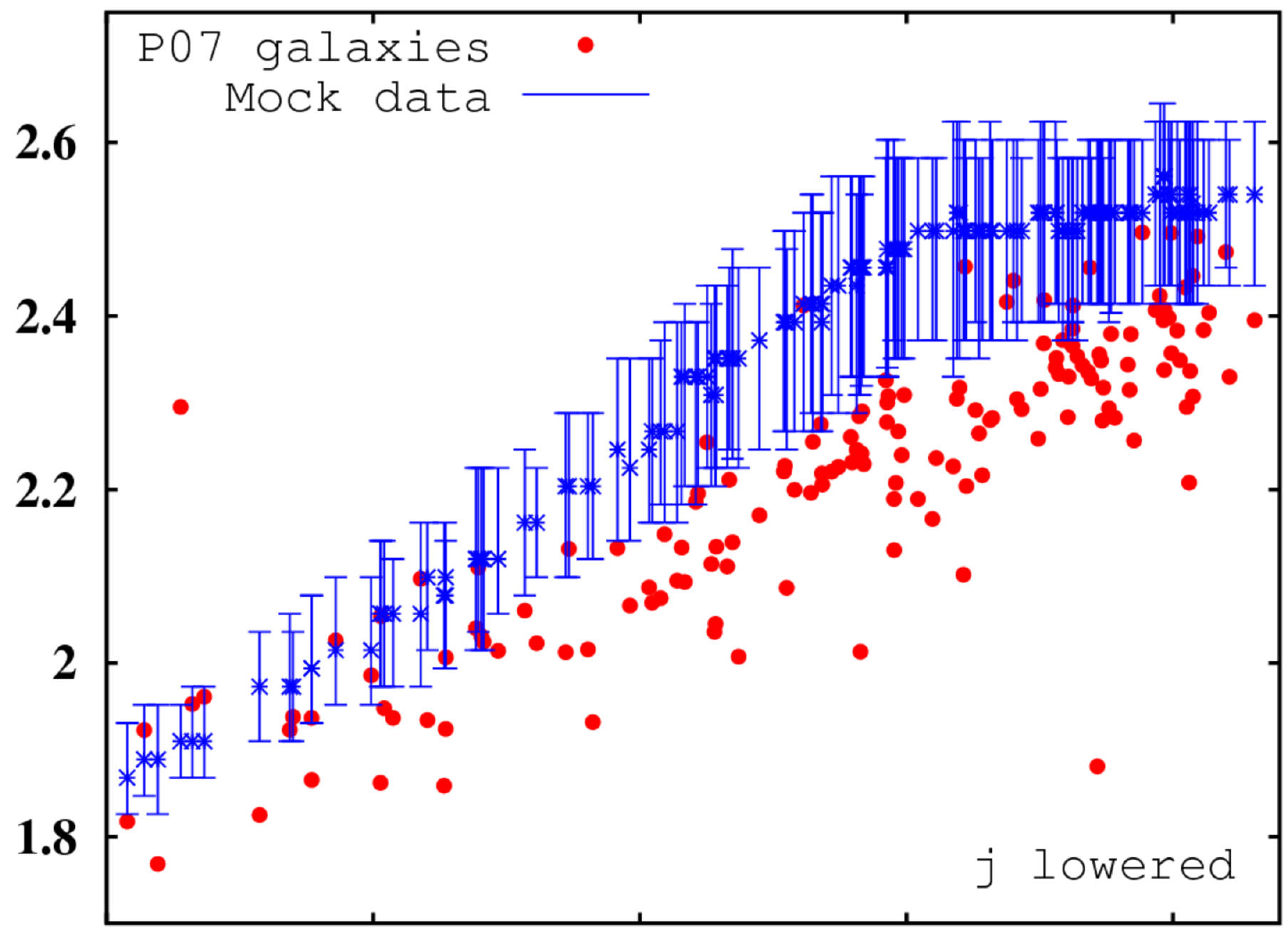}
  \hspace*{-1.47cm}
  \includegraphics[height=4.3cm, width=0.415\textwidth]{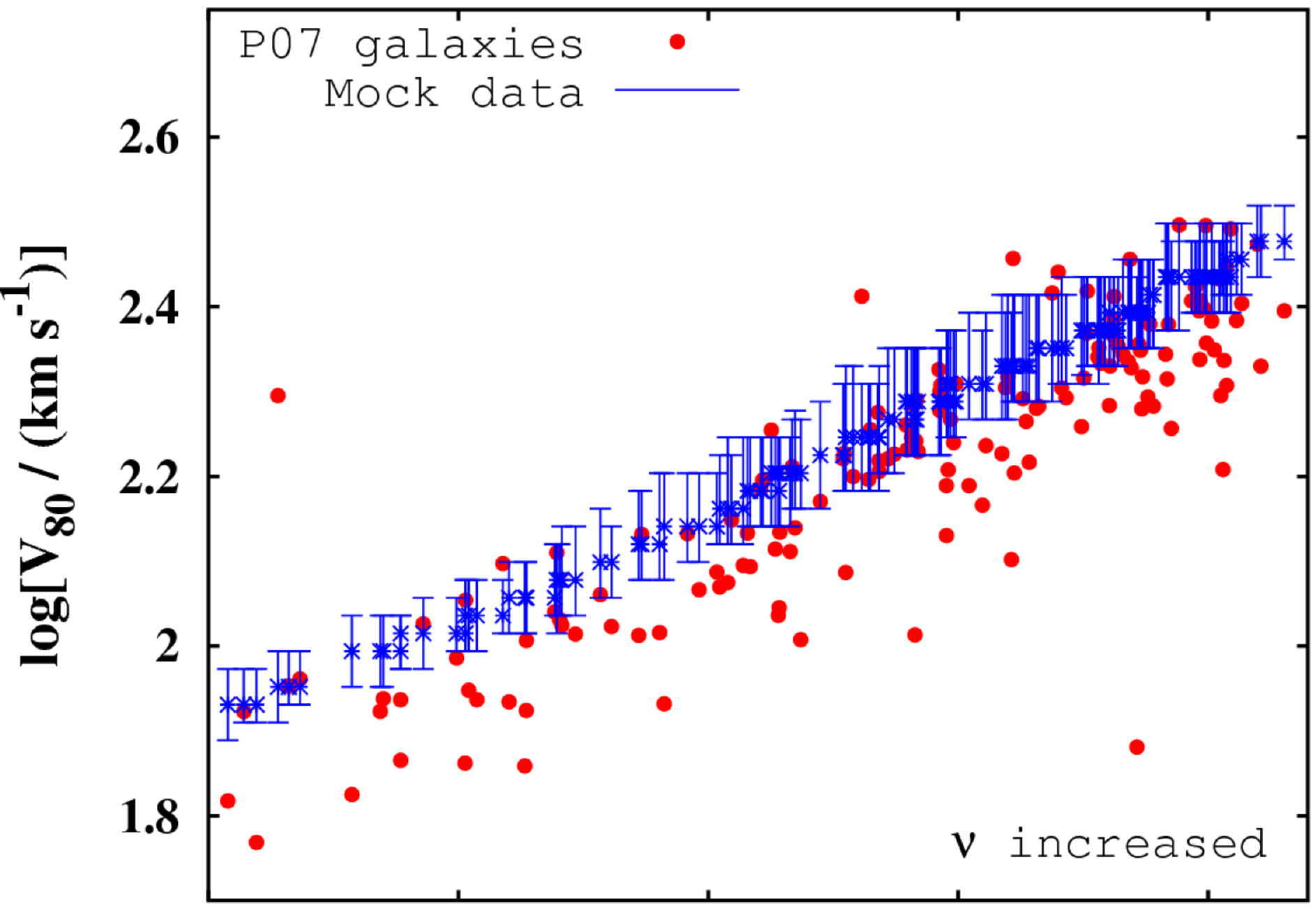}
  \hspace*{-0.9cm}
  \includegraphics[height=4.3cm, width=0.38\textwidth]{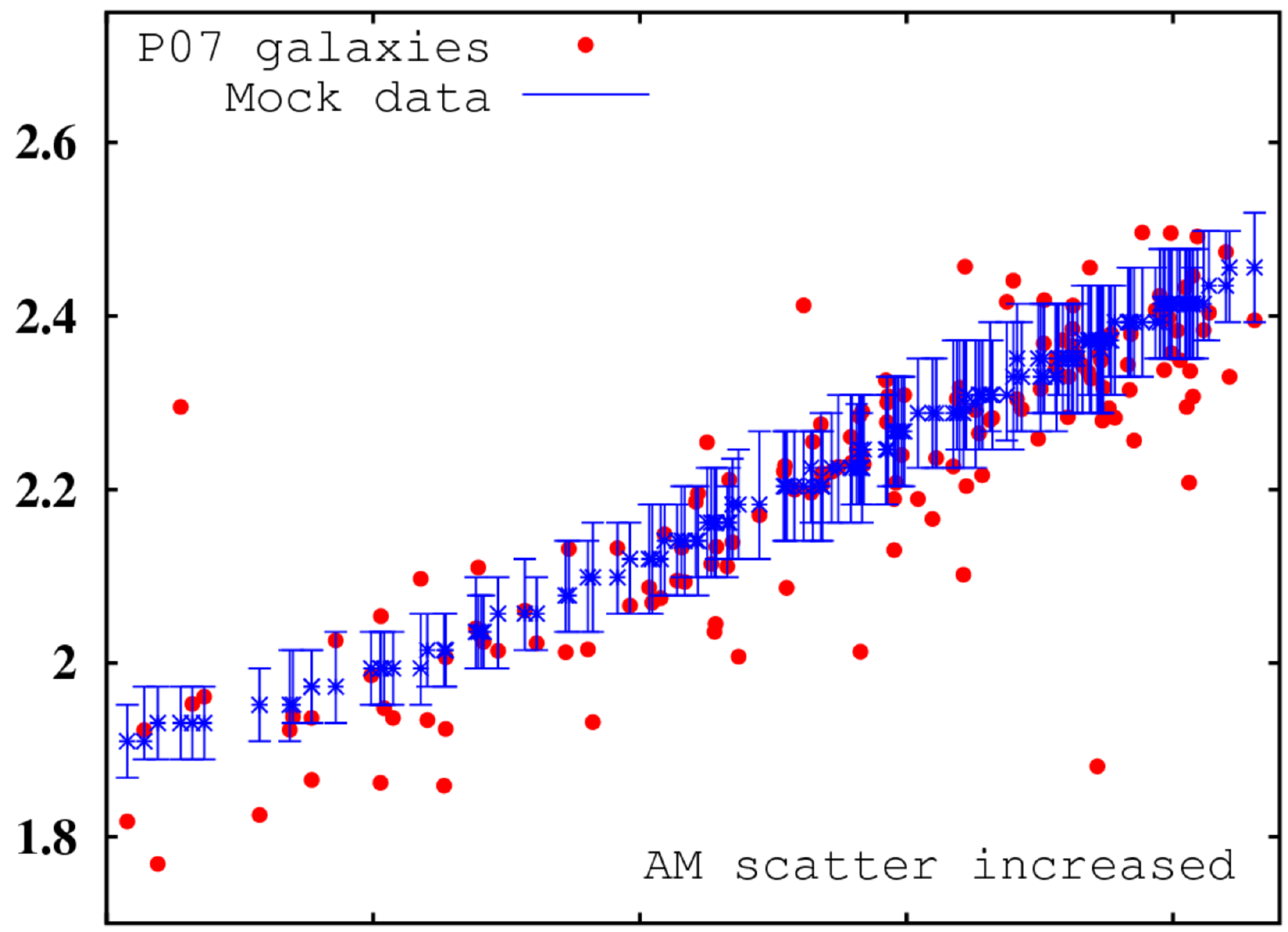}
  \hspace*{-0.9cm}
  \includegraphics[height=5.33cm, width=0.38\textwidth]{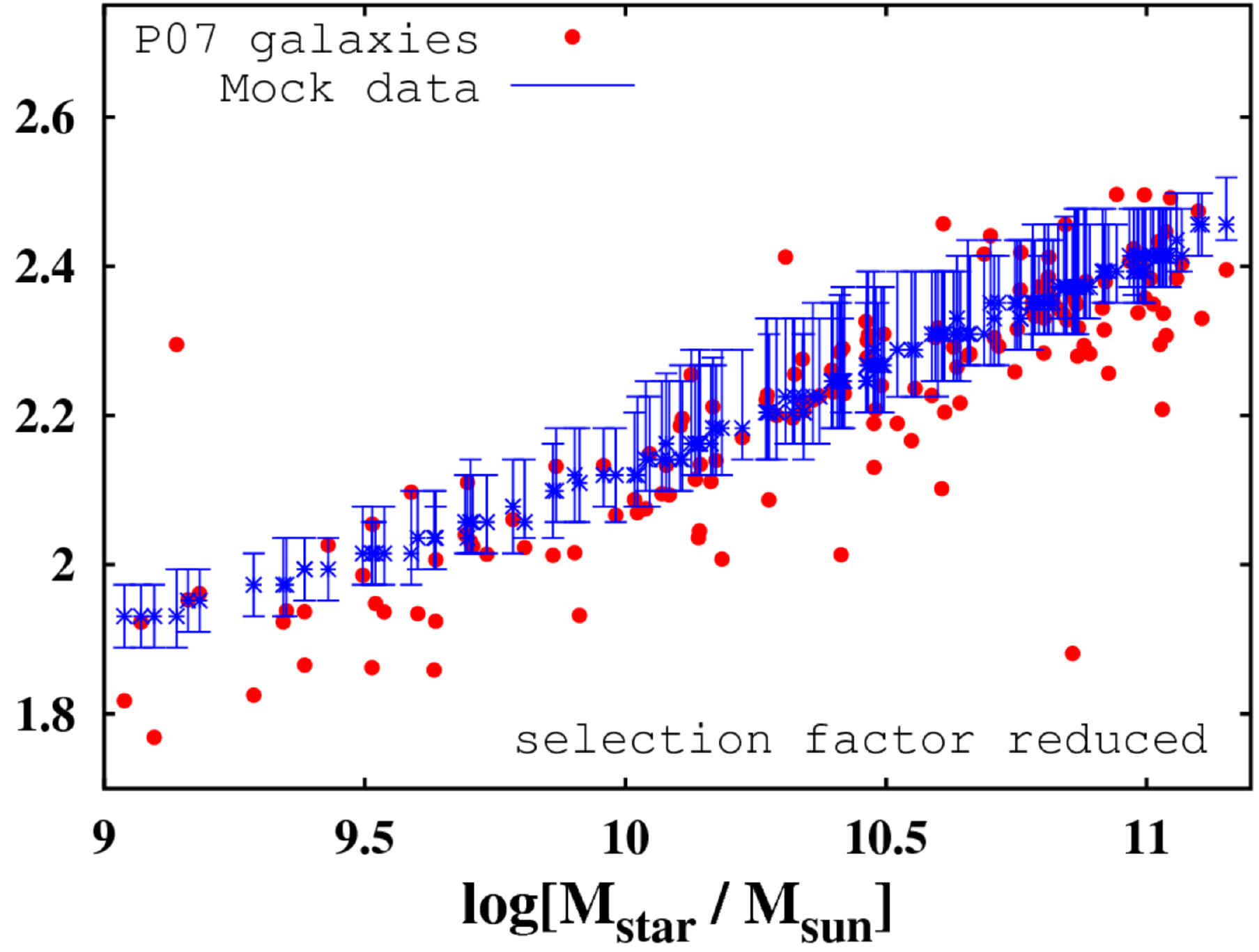}
  \caption{Comparison of the P07 data (red points) to 1000 mock TFR data sets generated using the maximum-likelihood parameter values of the $V_\mathrm{peak}$ model (top panel; see also Table~\ref{tab:ML_Parameters}), and with parameters individually perturbed to illustrate their impact on the TFR (other panels). The blue points are at the median velocities of the mock galaxies at the stellar masses of the P07 galaxies, and the error bars show the interquartile range.}
  \label{fig:PointsPlots}
\end{figure}

\renewcommand{\thefigure}{A2}

\begin{figure}
   \centering
  \hspace*{-0.9cm}
  \includegraphics[height=4.3cm, width=0.38\textwidth]{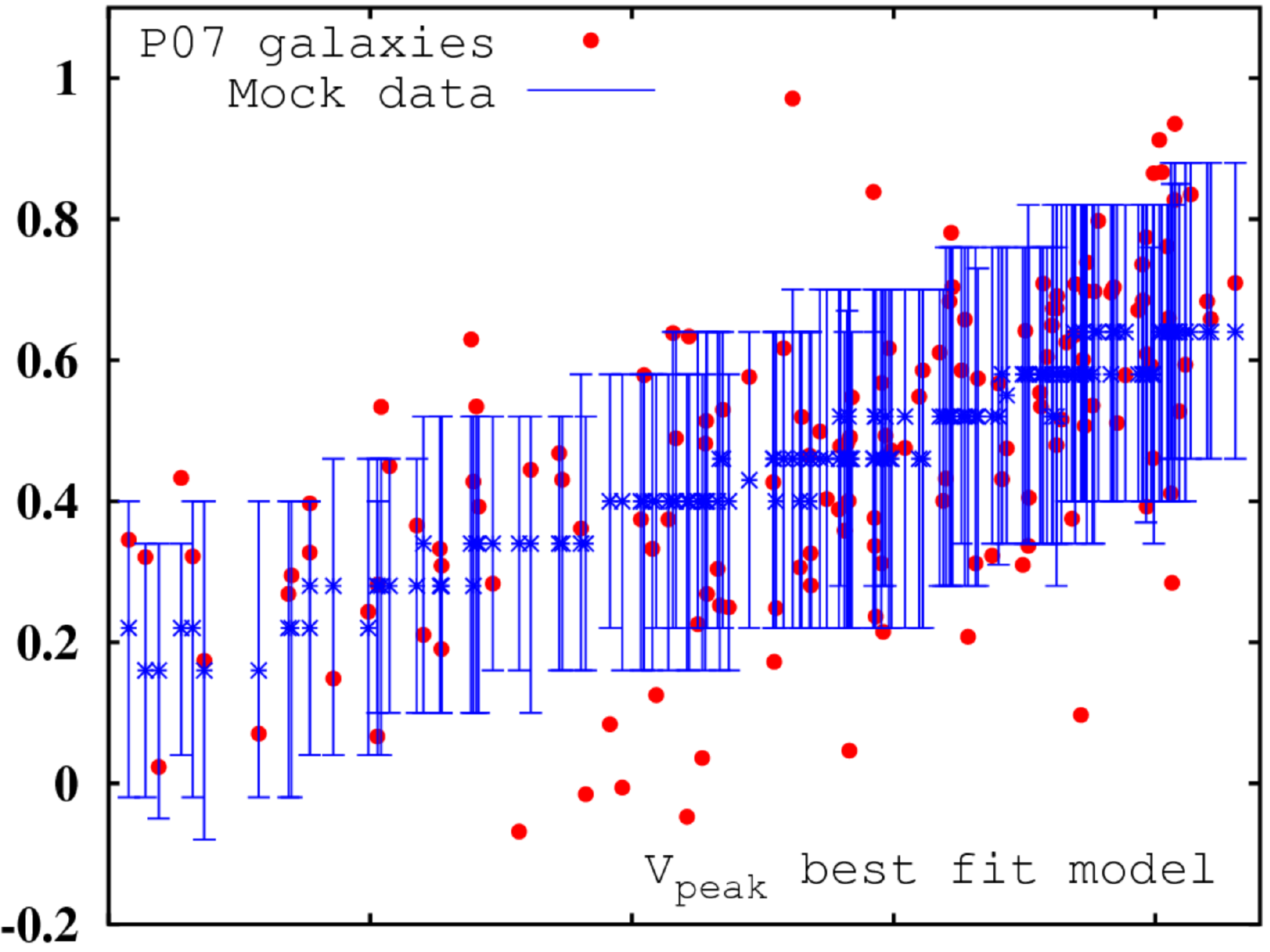}
  \hspace*{-0.9cm}
  \includegraphics[height=4.3cm, width=0.38\textwidth]{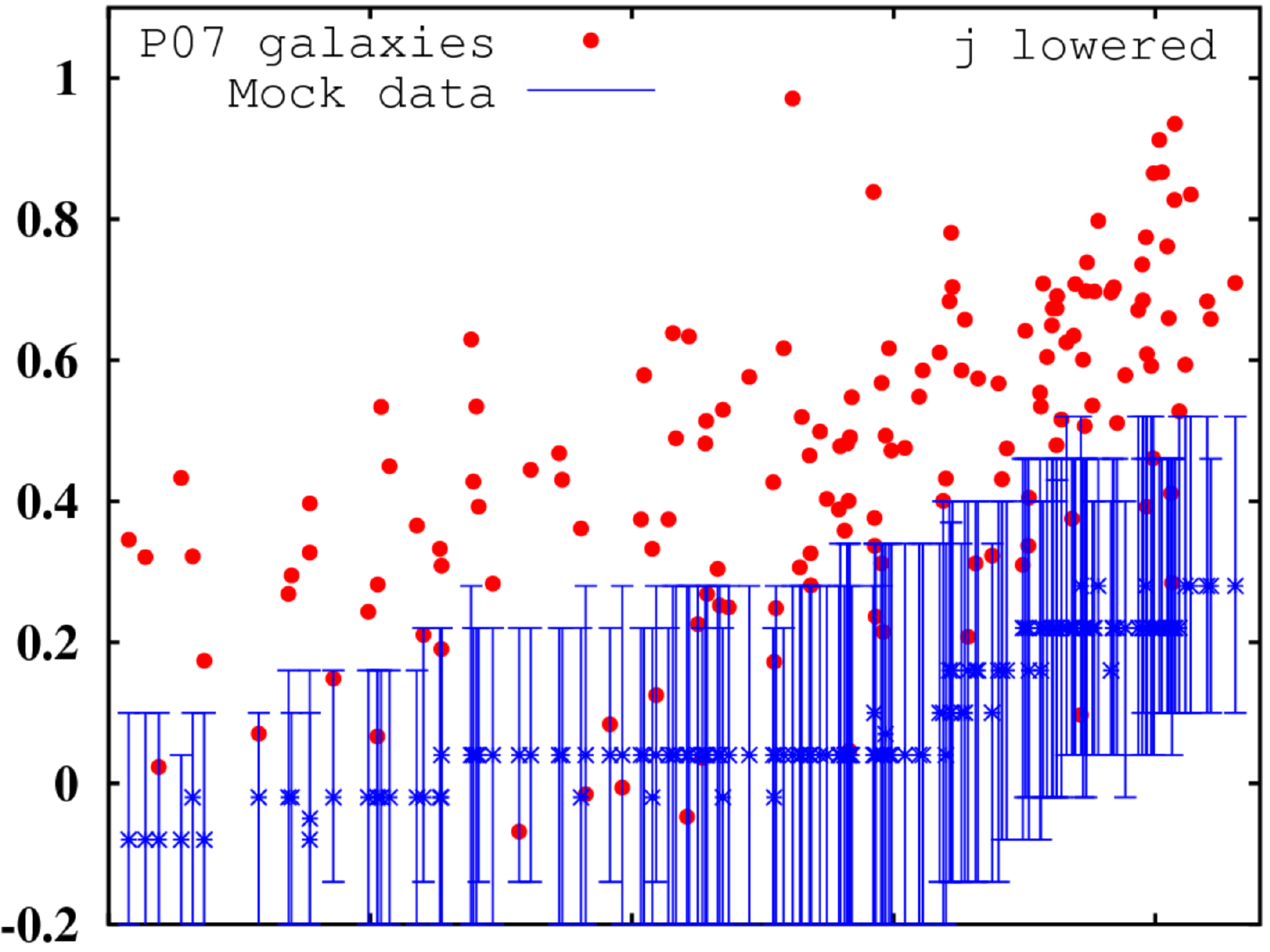}
  \hspace*{-1.63cm}
  \includegraphics[height=4.3cm, width=0.425\textwidth]{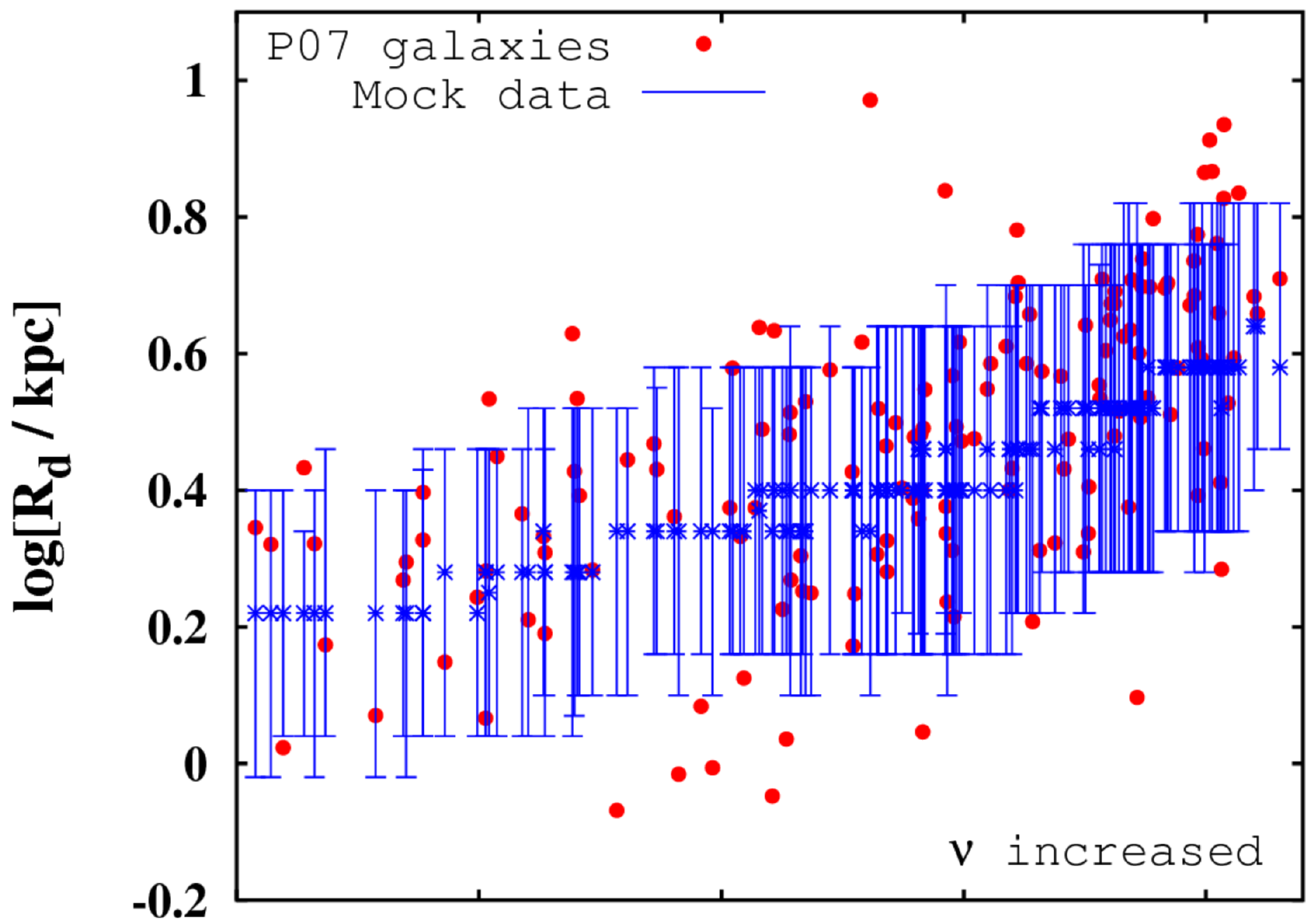}
  \hspace*{-0.9cm}
  \includegraphics[height=4.3cm, width=0.38\textwidth]{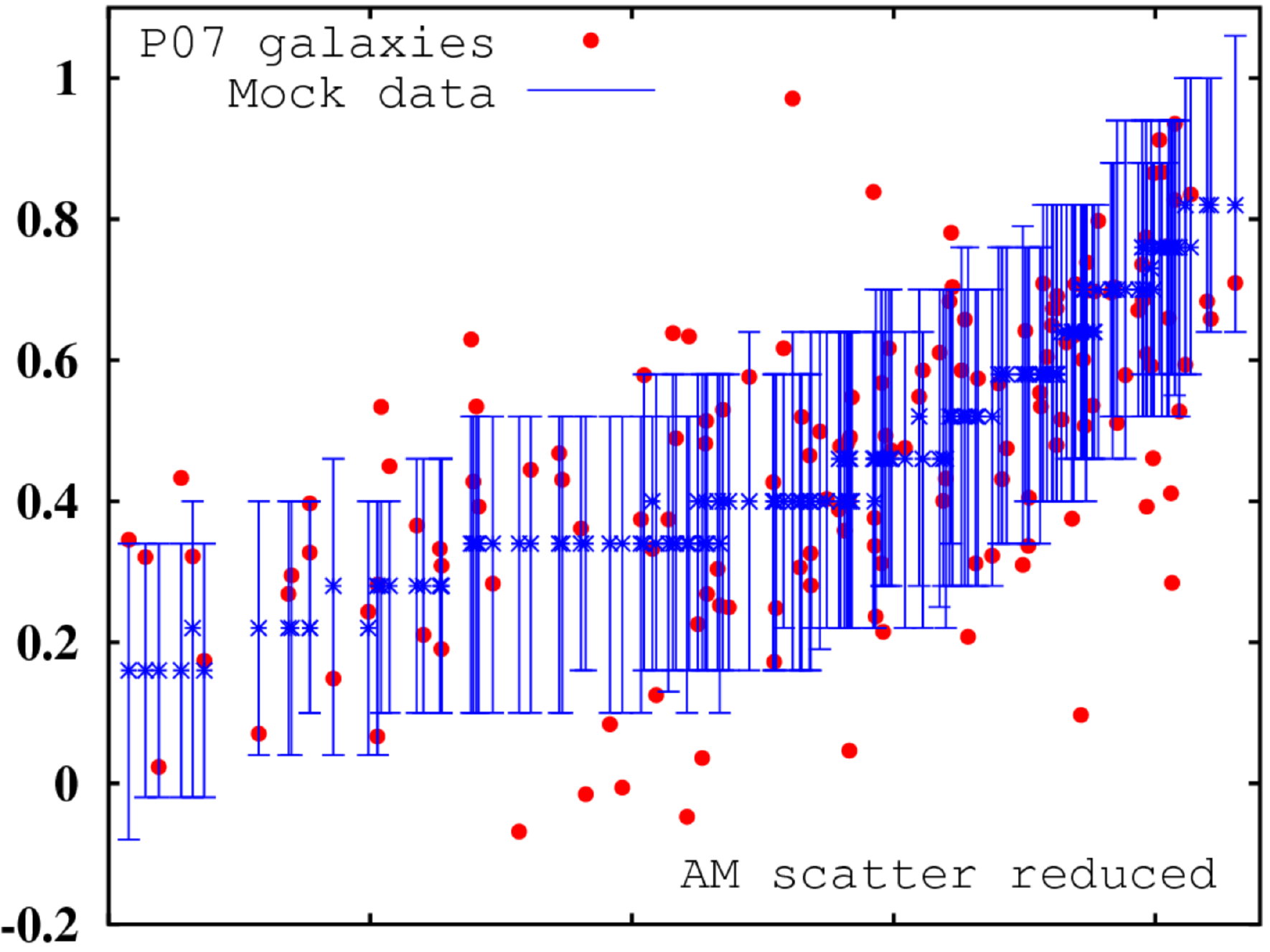}
  \hspace*{-0.9cm}
  \includegraphics[height=5.3cm, width=0.38\textwidth]{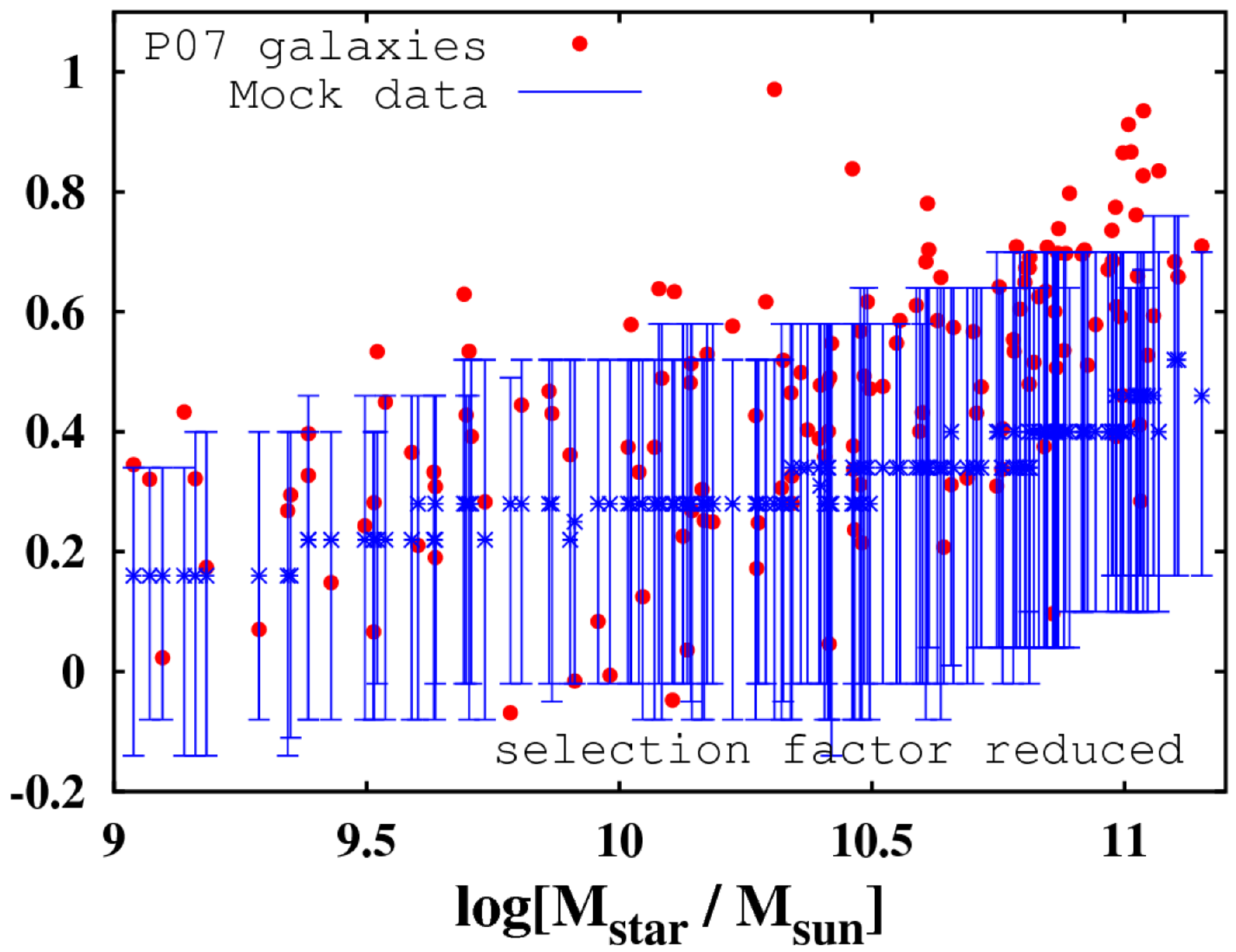}
   \caption{Same as Fig.~\ref{fig:PointsPlots}, but for the MSR.}
  \label{fig:PointsPlots_Rad}
\end{figure}

}

\section{Assumptions and Systematic Uncertainties}
\label{sec:assumptions}

We review here the most significant assumptions in our analysis. We discuss the consequences of reasonable deviations, and, where possible, provide quantitative estimates of the systematic errors that may be induced, in order to demonstrate the robustness of our results to them. Overall, we believe the systematic error to be under control at the 10 per cent level or better.

\vspace{3 mm}

\noindent \emph{Fraction of galaxies removed by selection cuts}: we have assumed that the exclusion from the P07 sample of galaxies with rotation curves poorly fit by an arctan function does not significantly bias the halo population in which these galaxies live, i.e. that the deviation of a galaxy's rotation curve from an arctan function is not strongly correlated with halo properties. In reality, some galaxies likely failed this cut because they were early-type and did not show strong signs of rotation -- as discussed in Section~\ref{sec:method}, early-type galaxies are expected to reside in systematically different haloes than late-type galaxies at fixed galaxy mass. However, we do not expect this to affect our main conclusions since the fraction of galaxies removed as a function of mass is degenerate with our selection factor parameter. Reducing the fraction of galaxies retained during the selection step would merely lower the preferred value of the selection factor.

\vspace{3 mm}

\noindent \emph{Modelling of selection effects}: we have assumed that selection effects can be modelled reliably via a correlation between observed disc axis ratio $b/a$ (the primary quantity used for selection in P07) and $V_{80}$. This is not inconsistent with the more physical assumption of a correlation between a direct morphological proxy (e.g. S\'{e}rsic index or true axis ratio) and any alternative measure of dynamical mass (e.g. $V_\mathrm{max}$ or halo $M_\mathrm{vir}$), which would imply a correlation (albeit weakened) between $b/a$ and $V_{80}$. Our methodology will therefore capture the first order effect of any such correlation. Although the impact of selection is almost certainly more complicated in detail, the absence of additional information about the relation between galaxy morphology and host halo properties makes ours a practical approach from a phenomenological perspective.

\vspace{3 mm}

\noindent \emph{Impact of cosmology}: the {\tt c-125} and {\tt c-250} simulations from which we drew our haloes assume $h=0.7$, $\Omega_\mathrm{m}=0.29$, $\Omega_\mathrm{b}=0.047$, $\sigma_\mathrm{8}=0.82$, and $n_\mathrm{s}=0.96$. These values are slightly different to the best-fitting Planck values ($h=0.67$, $\Omega_\mathrm{m}=0.32$, $\Omega_\mathrm{b}=0.049$, $\sigma_\mathrm{8}=0.83$ and $n_\mathrm{s}=0.96$;~\citealt{Planck}), which themselves have some uncertainty. The underlying cosmology (most significantly $\Omega_\mathrm{m}$, $\sigma_\mathrm{8}$ and $n_\mathrm{s}$) affects the $\log(\mathrm{concentration})$--$\log(M_\mathrm{vir}/M_\odot)$ (c--m) relation of the haloes in the $z=0$ snapshot, and hence the $V_{80}$ values of their galaxies at fixed stellar mass. To investigate the robustness of our results to reasonable variations in cosmological parameter values, we use the model of~\citet{Diemer_Kravtsov} to generate approximate c--m relations over the mass range of interest. In particular, we begin by fitting a power law to the c--m relation generated by the cosmology of our simulations [pivoted at the approximate median $\log(M_\mathrm{vir}/M_\odot)$ of the P07 sample, 11.7], and then calculate the maximum fractional variation in the normalization when cosmological parameters are varied within $2\sigma$ Planck uncertainties. We find that the normalization can be changed by at most $\sim10$ per cent in this way.\footnote{The change in the slope of the c--m relation as cosmological parameters are varied within $2\sigma$ Planck uncertainties causes fractional changes in the concentrations over the mass range of interest that are around 3 times smaller than the normalization difference, so may be safely ignored.} To determine the effect this would have on our parameter constraints, we repeat the TFR MCMC analysis (for the $V_\mathrm{peak}$ model) with all halo concentrations shifted downwards by 10 per cent. We find that the best fitting value of $\nu$ is raised by $\sim2\sigma$ to $-0.1^{+0.1 (+0.3)}_{-0.2 (-0.4)}$ (c.f. Table~\ref{tab:Constraints}, first row), indicating that a small amount of halo contraction is now permitted. The upper limit on the AM scatter is raised slightly to $0.26$ ($1\sigma$). All other constraints are affected at the $\lesssim5$ per cent level.

\vspace{3 mm}

\noindent \emph{Halo concentrations}: estimates of concentration made by halo finders may be unreliable, especially for haloes composed of few particles. We have checked that median concentrations of \textsc{rockstar} haloes in the {\tt c-125} and {\tt c-250} boxes agree to better than $5$ per cent with the model of~\citet{Diemer_Kravtsov}, and that the median fractional deviation between different ways of calculating the concentration is $<5$ per cent (see also Section~\ref{sec:sim_data}). We therefore conclude that there is at most a $\sim5$ per cent systematic error in average concentrations, and have shown above (in the context of the impact of cosmology) that this has relatively little effect on our results.

\vspace{3 mm}

\noindent \emph{Stellar masses}: we have taken stellar masses for the P07 galaxies from the NSA, and the SMF for AM from~\citet{Bernardi}. Several prescriptions exist for mapping observed luminosity to total luminosity to stellar mass, and their relative merits have been discussed extensively in the literature (e.g.~\citealt{Bell, McGaugh_STFR, Bernardi}). However, a fractional shift in stellar mass that is the same for all galaxies of a given mass would not affect our results because the real and mock data would be affected in exactly the same way. Our constraints would only be modified if the masses of the P07 galaxies were affected differently to the average for the entire galaxy population, which sets the shift in the SMF. Since the P07 galaxies are typical in terms of colour and other properties (P07), this is unlikely.

\vspace{3 mm}

\noindent \emph{Impact of disc formation on halo}: we have modelled the effect of disc formation on halo density profile using the generalized adiabatic contraction model of~\citet{D07}. Although this model does allow for a halo expansion (and should therefore capture the leading order effect of associated processes), the true mass dependence of the expansion may not be included. In particular, it is expected from some cosmological simulations that the effect of stellar feedback on the dark matter halo is largest for $\log(M_\star/M_\odot) \approx 8.5$~\citep{diCintio_1, diCintio}, whilst in our model the effect is largest at the peak of the stellar-to-halo mass fraction [$\log(M_\star/M_\odot) \approx 10$]. Changing the stellar mass dependence of the effect of baryons on the halo density profile would alter the shape of the TFR (for example, the~\citealt{diCintio} model would reduce the predicted velocity more at the low end of the P07 mass range, steepening the predicted TFR), but would affect none of our main conclusions. We note, however, that haloes produced in the \textsc{eagle} simulations (which also include the effect of baryons) are most contracted around the peak of the stellar mass fraction, an effect captured to first order by our generalized adiabatic contraction prescription when $\nu>0$~\citep{EAGLE_2, EAGLE_1}.

\vspace{3 mm}

\noindent \emph{Mass-to-light ratios}: we have assumed that the mass-to-light ratios of the bulge and disc are equal. This assumption is required to map the bulge-to-disc $i$-band flux ratios observed in the P07 study to mass ratios required for our rotation curve modelling. It is, however, generally accepted that galactic bulges have higher mass-to-light ratios than discs; thus the bulge should be more massive relative to the disc for a given flux ratio. We have investigated the effect on $V_{80}$ of variations in the ratio of the mass-to-light ratios of the bulge and disc from 0.3 to 3, and find that changes are typically at or below the 1 per cent level, due to the relatively small bulges in these late-type systems.

\vspace{3 mm}

\noindent \emph{Velocity dispersion}: we assume that the model galaxies in the population from which the P07 sample could have been drawn have negligible velocity dispersion relative to their rotation velocity. It is expected that galaxies with properties similar to those in the P07 sample have velocity dispersions of order $20\:\mathrm{km\:s^{-1}}$ (McGaugh, private communication), which would cause a relatively small shift of our model galaxies towards lower $V_{80}$ values. In addition, we assume our model galaxies are thin when we simulate their rotation curve; real discs have finite thickness which reduces their rotation velocity at fixed mass and size. For realistic thicknesses (e.g. $h/R_\mathrm{d}\approx0.2$), this is a $\lesssim5$ per cent effect (e.g.~\citealt{Casertano}). 

\vspace{3 mm}

\noindent \emph{Dynamical effect of gas}: we assume that the cold gas in galaxies contributes negligibly to the total dynamical mass, and hence to $V_{80}$. Using the approximate scaling relations for the gas and stellar masses and disc scalelengths presented in~\citet{D11}, we expect the gas disc to increase typical galaxies' $V_{80}$ values by at most around 2 per cent in the mass range we are considering.

\vspace{3 mm}

\noindent \emph{Uncertainty in stellar masses}: we have assumed no uncertainty in any stellar mass estimate. If this uncertainty were non-zero, each AM scatter value that we quote would become the quadrature sum of this uncertainty and the intrinsic scatter associated with AM. Since it is the \textit{true} rather than \textit{observed} stellar mass which determines a model galaxy's rotation curve, however, one would need to perturb the stellar masses output by AM according to their uncertainty before continuing with the calculation of rotation velocity and disc scalelength. This would increase the intrinsic scatter of the theoretical TFR and MSR. However, the effect would be small because the low stellar-to-halo mass fractions induced by AM (at most $\sim5$ per cent) make a galaxy's own contribution to its rotation curve far less than that of its halo. The slope and normalization of the relations would be on average unaffected, as would the correlation of the TFR and MSR residuals, which is recorded at fixed stellar mass.

\vspace{3 mm}

\noindent \emph{Halo sphericity}: we assume haloes are perfectly spherical. However, cosmological simulations create triaxial haloes with smallest-to-largest axis ratios of the order of 0.5~\citep*{Allgood,Despali}. This is expected to be increased by the process of galaxy formation~\citep*{Triaxiality_2}, and weak lensing observations provide only weak evidence that haloes are not spherical~\citep{Triaxiality_1}. Triaxiality would cause the rotation velocity of the disc to be a function of angle in addition to radius, introducing some thickness to the observed rotation curve and increasing the scatter in the TFR, but the magnitude of the effect is unlikely sufficient to affect our conclusions.

\vspace{3 mm}

\noindent \emph{Halo density profile}: we assume that haloes have a density profile perfectly described by the NFW form. Deviations from pure NFW would cause variations in the circular velocities of the haloes for given $r_\mathrm{s,klypin}$ and $r_\mathrm{vir}$, generating additional scatter in the TFR. However, this approximation is unlikely poor enough to constitute a significant source of systematic error in our analysis.

\vspace{3 mm}

\noindent \emph{Disc intrinsic axis ratio}: we take all discs to have an intrinsic axis ratio of 0.2, as assumed by P07. It is possible, however, that these could be significantly higher, especially at low mass (Geha, private communication). Underestimating the axis ratios would cause the inclinations to be underestimated, and hence the inclination-corrected $V_{80}$ to be overestimated. To explore the impact this could have on our parameter estimation, we repeat the MCMC analysis of Section~\ref{sec:method} with two alternative assumptions. In the first, the intrinsic axis ratios of all the P07 galaxies are set to 0.6. In the second, we make the axis ratio a linear function of $\log(M_\star/M_\odot)$, falling from 0.6 at $\log(M_{\star}/M_\odot)$ = 9 to 0.2 at $\log(M_{\star}/M_\odot)$ = 11.2. (This modifies the slope of the observed TFR as well as its normalization.) In both cases, we find that our parameter constraints are affected at the $\lesssim5$ per cent level.

\section{Technical Details of the Method}
\label{sec:technical}

In order for the likelihood evaluations to be sufficiently rapid for a full MCMC analysis to be feasible, the halo AM step cannot be performed each iteration. We therefore began by creating and storing arrays of the $M_{\star}$ values associated with each halo for each AM parameter and scatter, in addition to the haloes' $M_\mathrm{vir}$, concentration and $\lambda$ values, and (for Section~\ref{sec:environment}) the \textsc{rockstar} `pid' variable that distinguishes satellites from centrals. We then created a look-up table to map from a point in \{$M_\mathrm{vir}$, concentration, $M_{\star}$, $\lambda$\} space to a point in \{$R_\mathrm{d}$, $V_{80}$\} space using the method described in Section~\ref{sec:method}, partitioning each space into a grid with the properties described below. This was done separately for each value of the $\nu$ parameter, which affects this map. Then, in the likelihood function and for a given set of parameter values, we determined the number of haloes in each \{$M_\mathrm{vir}$, concentration, $M_{\star}$, $\lambda$\} grid cell, and incremented the counts in the TFR \{$M_{\star}$, $V_{80}$\} or MSR \{$M_{\star}$, $R_\mathrm{d}$\} arrays by the corresponding amount. These arrays were then used to generate mock data sets. This procedure allowed each likelihood evaluation to take only 25 s. More quantitative information concerning the convergence conditions and resolution of our algorithm is given below.

\begin{itemize}

\item{} Convergence requirement for iterative calculation of disc scalelength (see Section~\ref{sec:method}): 2.5 per cent.

\item{} Convergence requirement of generalized adiabatic contraction: 5 per cent.

\item{} Resolution of grid of $\log(M_\mathrm{vir}/M_\odot)$, $\log(\mathrm{concentration})$, $\log(M_{\star}/M_{\odot})$, $\log(\lambda)$ values: 50 cells in ranges $9.5:13.0$, $0.4:2.1$, $9.0:11.2$, $-2.3:-0.4$ respectively. These limits were chosen to enclose $\ge 99$ per cent of model galaxies for all AM parameters.  

\item{} Resolution of $\log(V_{80}/\mathrm{km\:s^{-1}})$ grid: 50 cells in range $1.70:2.75$.

\item{} Resolution of $\log(R_\mathrm{d}/\mathrm{kpc})$ grid: 50 cells in range $-0.8:2.2$.

\item{} Rotation curve sampled at 500 logarithmically uniform intervals over $10^{-5} \le r/R_\mathrm{d} \le 30$.

\item{} Number of mock data sets for likelihood evaluations: 200.

\item{} AM scatter resolution: 0.02 dex.

\item{} $\nu$ resolution: 0.1.

\end{itemize}

\end{document}